\documentclass[traditabstract]{aa}
\usepackage{amsmath} 
\usepackage{longtable}
\usepackage{array}
\usepackage{natbib}
\usepackage{bm}
\usepackage{multirow}
\usepackage{graphicx}
\usepackage{booktabs,supertabular,array}
\usepackage{rotating}
\usepackage{lipsum}

\newcommand\blfootnote[1]{%
  \begingroup
  \renewcommand\thefootnote{}\footnote{#1}%
  \addtocounter{footnote}{-1}%
  \endgroup
}
\usepackage{lscape}
\usepackage[caption=false]{subfig}
\usepackage{xcolor}
\usepackage[english]{babel}     
\usepackage{multicol}
\usepackage{enumerate}
\usepackage{url}

\usepackage{ulem}
\usepackage{xcolor}
\usepackage[utf8]{inputenc}

\usepackage{soul}

\begin{document}

\title{Spectroscopic observations of the machine-learning selected anomaly catalogue from the AllWISE Sky Survey}

\author{A. Solarz\inst{\ref{inst1}}, R. Thomas\inst{\ref{inst1}}, F. M.  Montenegro-Montes\inst{\ref{inst1}},   M. Gromadzki\inst{\ref{inst2}}, E. Donoso\inst{\ref{inst3},\ref{inst4}}, M. Koprowski\inst{\ref{inst5}}, L. Wyrzykowski\inst{\ref{inst2}}, C. G. Diaz\inst{\ref{inst3},\ref{inst6}}, E. Sani\inst{\ref{inst1}},  M. Bilicki\inst{\ref{inst7}} }

\institute{European Southern Observatory, Ave. Alonso de C\'ordova 3107, Vitacura, Santiago, Chile\label{inst1}\\ 
         \email{asolarz@eso.org}
\and
Astronomical Observatory, University of Warsaw, Al. Ujazdowskie 4, 00-478 Warszawa, Poland\label{inst2}
\and
Instituto de Ciencias Astron\'omicas, de la Tierra, y del Espacio (ICATE), CONICET, San Juan, Argentina\label{inst3}
\and
Universidad Nacional de San Juan (UNSJ), Mitre Oeste 396, J5402 San Juan, Argentina \label{inst4}
  \and  
Institute of Astronomy, Faculty of Physics, Astronomy and Informatics, Nicolaus Copernicus University, Toru\'n, Poland\label{inst5}
   \and
Gemini Observatory, Southern Operations Center, La Serena, Chile \label{inst6}
\and
Center for Theoretical Physics, Polish Academy of Sciences, al. Lotnik\'ow 32/46, 02-668, Warsaw, Poland\label{inst7}
        }

\date{Received <date>/ Accepted <date>}

\abstract{

We present the results of a programme to search and identify the nature of unusual sources within the All-sky Wide-field Infrared Survey Explorer (WISE) that is based on a machine-learning algorithm for anomaly detection, namely one-class support vector machines (OCSVM).
 Designed to detect sources deviating from a training set composed of known classes, this algorithm was used to create a model for the expected data based on WISE objects with spectroscopic identifications in the Sloan Digital Sky Survey (SDSS).
 Subsequently, it marked as anomalous those sources whose WISE photometry was shown to be inconsistent with this model.
We report the results from optical and near-infrared spectroscopy follow-up observations of a subset of 36 bright ($g_{\rm AB}<$19.5) objects marked as 'anomalous' by the OCSVM code to verify its performance. 
Among the observed objects, we identified three main types of sources: i) low redshift ($z\sim 0.03-0.15$)
galaxies containing large amounts of hot dust (53\%), including three Wolf-Rayet galaxies; ii) broad-line quasi-stellar objects  (QSOs) (33\%) including low-ionisation broad absorption line (LoBAL) quasars and a rare QSO with strong and narrow ultraviolet iron emission;
iii) Galactic objects in dusty phases of their evolution (3\%). The nature of four of these objects (11\%) remains undetermined due to low signal-to-noise or featureless spectra. The current data show that the algorithm works well at detecting rare but not necessarily unknown objects among the brightest candidates. They mostly represent peculiar sub-types of otherwise well-known sources.
To search for even more unusual sources, a more complete and balanced training set should be created after including these rare sub-species of otherwise abundant source classes, such as LoBALs. 
Such an iterative approach will ideally bring us closer to improving the strategy design for the detection of rarer sources contained within the vast data store of the AllWISE survey.
}

\titlerunning{Spectroscopic observations of anomaly catalogue from the AllWISE Sky Survey}
\authorrunning{A. Solarz, R. Thomas, F. Montenegro et al.}
\maketitle 

\blfootnote{Based on observations collected at the European Southern Observatory, Chile. Programme IDs 0101.A-0539  and  0102.A-0305.}

\section{Introduction}

In recent years, astronomy has seen rapid growth in the amount and complexity of the information collected by multi-wavelength surveys. Currently available databases of astronomical observations already contain vast amounts of data. 
The Sloan Digital Sky Survey (SDSS, \citealt{york00}) has provided the community with the photometry and spectra of over 4 million sources covering almost 30\% of the sky totalling to 273 TB  \footnote{\url{https://www.sdss.org/dr16/data_access/volume/}}.  Wide Field Survey Explorer (WISE, \citealt{wright10}), which performed a photometric all-sky survey in near- and mid-infrared passbands delivered over 23 TB \footnote{\url{https://irsa.ipac.caltech.edu/holdings/catalogs.html}} (excluding multi-epoch and reject catalogues).
Soon, facilities that  are currently under development, such as the Square Kilometre Array (SKA; \citealt{dewdney09}) and the Large Synoptic Survey Telescope (LSST; \citealt{ivezic08}) will provide even larger volumes of data: The LSST is expected to deliver in total 60 PB of raw images\footnote{\url{https://www.lsst.org/scientists/keynumbers}}, while the SKA would give over 5 ZB \citep{ska}.
Such an exponential growth in the quantity of data has compelled astronomers to develop automatic tools for extracting knowledge about known objects as well as to discover new information.
A new possibility for efficient data extraction from surveys has been made possible thanks to the advent of machine learning (ML) algorithms, a tool of artificial intelligence that is designed to learn from the provided data set itself. There are two main branches of ML. On the one hand, \textit{} 'supervised' 
learning algorithms are designed to create a model that can recognise specific patterns within the data based on  training examples provided by the user (supervisor). 
On the other hand, 'unsupervised' learning algorithms search for similarities between input data points without any a priori input knowledge from the user.
From a practical point of view, both approaches have different applications and suffer from various handicaps. Supervised ML can only recognise data with properties that are similar to the training set. For that reason, any rare or unseen data structures would be lost and the output catalogues would lack purity (e.g. \citealt{kurcz2016}). 
Unsupervised ML, on the other hand, produces an output which has to be verified by the user, as the similarity of the data chosen by the algorithm must be of physical significance to interpret the results. 
This verification process makes unsupervised ML applications a time-consuming procedure of trial and error.
A fusion of the two approaches, known as 'semi-supervised' learning, combines the best aspects of both methods.
Semi-supervised algorithms are designed to recognise typical data structures based on training examples. However, they have the freedom to pinpoint all those data points which are inconsistent with the known patterns in the data. 
Thanks to the increasing depth and sensitivity of present and future surveys, we should expect to encounter lesser known or rare phenomena more frequently.
Therefore, the semi-supervised approach works well for detecting 'anomalies' in the data, especially in large photometric surveys which lack spectroscopic follow-up observations.
This approach is of crucial importance for statistical studies, as we want our samples to be as pure and as complete as possible. 

Currently, more and more attention is drawn towards data mining for unique objects. 
\citet{unusualqso} used Kohonen self-organising maps \citep{som} to search for the most unusual quasi-stellar objects  (QSOs), within the SDSS survey. \citet{baron17} used unsupervised machine learning methods, such as random forest (RF, \citealt{Breiman2001}), to search for peculiar objects in the SDSS database by finding abnormal spectral features. \citet{reis18} used RF similarity matrix for detection of stellar spectra, which otherwise would remain hidden when using classical modelling methods. \citet{hocking18} used an assembly of unsupervised ML algorithms to label galaxies in astronomical imaging surveys using only pixel data. This technique can also facilitate the search for rarer objects in the data sets. Other automated anomaly detection algorithms are applied to other tasks, for example, photometric redshift estimation meant to reject any bad data that may be present within the training set \citep{hoyle15}.

\citet{solarz17}, henceforth  S17, created a catalogue of AllWISE objects \citep{cutri13}, which exhibit unusual mid-IR behaviour with respect to the optical source population. The catalogue is a result of the application of a semi-supervised anomaly detection algorithm called the one-class support vector machine (OCSVM, \citealt{ocsvm}), designed to find objects deviating from user-defined known sources. The model for the expected data was created based on AllWISE objects with spectroscopic identifications found in the SDSS DR13 \citep{sdss13}. 
In this study, we aim to characterise the bright end of the S17 catalogue by performing a pilot spectroscopic programme. The observations are performed using the ESO Faint Object Spectrograph and Camera (EFOSC2, \citealt{efosc}) and Son of ISAAC (SofI, \citealt{sofi}). Both instruments are mounted on the New Technologies Telescope (NTT) at the La Silla Observatory.
The full-scale project intends to 'close the loop' of the machine learning process, where we study the nature of the underrepresented objects flagged as 'anomalies' by the algorithm. Ultimately it will lead to improving underrepresented populations within the sample of known objects.
The physical characterisation of the anomaly detection algorithm output is a crucial step in an iterative search of truly abnormal or perhaps even new phenomena registered in wide-angle photometric surveys.

This paper is organised as follows. Section 2 presents the description of the original catalogue of S17. Details about the spectroscopic observations and data reduction are included in Section 3, while Section 4 presents the results. We summarise the paper in Section 5.

\section{Anomaly selection}

We searched for abnormal sources present within the AllWISE catalogue.
The AllWISE contains more than 747 million sources detected over the whole sky, measured in four passbands ({\it W1}, {\it W2}, {\it W3}, {\it W4}) covering near- and mid-IR wavelengths centred at 3.4, 4.6, 12, and 22~$\mu$m with an angular resolution of the filters of 6.1", 6.4", 6.5", and 12.0", respectively.  The sensitivity to point sources at the 5$\sigma$ detection limit is estimated to be at least (depending on sky position) 0.054, 0.071, 0.73, and 5 mJy for the {\it W1}, {\it W2}, {\it W3,} and {\it W4} bands (equivalent to 16.6, 15.6, 11.3, and 8.0 Vega mag). 
The sensitivity of AllWISE allows for the construction of extragalactic catalogues that extend at least twice as deep as earlier all-sky datasets (e.g. \citealt{bilicki16}, \citealt{jarrett17}), such as the ones provided by Infrared Astronomical Satellite (IRAS) or Two Micron All-Sky Survey (2MASS). The WISE satellite provided a database of significant depth but without any spectroscopic information for most of its sources. 
As an infrared photometric all-sky survey, AllWISE extends the spectral information of objects enabling to study their dust properties.
Also, due to its sky coverage and depth, the AllWISE database may contain rare or unusual objects which have been missed by optical surveys or can reveal an unexpected behaviour of otherwise regular source.
When searching for unusual sources, it is preferable to predict the appearance of the known types of objects. When the 'ordinary' object types can be distinguished successfully, we can find unusual objects as those that are inconsistent with all known sorts of objects. However, sometimes a unique source can mimic the appearance of a regular object and remain unrecognised as a novelty.

To find outliers in the AllWISE data, we used the OCSVM algorithm, a semi-supervised method designed to find outlying data points (the anomalies) based on a sample of a user-defined training set of objects with known properties. Each training object must contain an {\it N}-dimensional feature vector, composed from {\it N} observables, such as flux, colour, morphology, etc. 
Then OCSVM projects the data through a nonlinear function, $\phi,$ to space with a higher dimension, called the feature space, $F$. 
Based on $F$, the OCSVM attempts to find the best separation boundary that will enclose the training sample (in a two-dimensional space, this boundary would be a circle or an ellipse, which would encompass the majority of the training set).
In other words, the algorithm estimates a probability distribution function which makes most of the known data more likely than the rest and a decision rule that creates the largest possible margin to separate these known observations from the outliers.
Once obtained, the boundary serves as a decision function and new objects are classified according to their location: if the point lies within the boundary, it is classified as an object with features that resemble those of the known objects. If it falls outside of the frontier, then it is classified as an outlier: an object with characteristics that differ from the training sample.
The outcome of the decision function relies on the dot-product of the vectors in the {\it F} (i.e. all the pairwise distances for the vectors).
The distance between the boundary and the outlying points can be treated as a measure of how anomalous a given source is: the further the object is from the decision boundary, the more uncharacteristic it is with respect to the training set.
S17 aimed for the detection of such sources in the AllWISE survey provided that they differ from those observed by the SDSS.
 The SDSS was chosen as a database for known sources, as it is the most comprehensive wide-angle survey to spectroscopically label various source types (stars, galaxies, QSOs). 
The training set was based on SDSS DR13 and any source with a measured redshift (velocity for stars) was included in the training sample.
The input parameter space was constructed using the following descriptive characteristics: $W1$ magnitude, $W1-W2$ colour and $w1mag1-w1mag3$ concentration parameter (defined as the difference between flux measurements in two circular apertures in the $W1$ passband in radii equal to 5.5" and 11.0";  \citealt{bilicki14}). 
We used the $W1-W2$ colour instead of $W2$ magnitude alone to ensure the maximum coverage of the parameter space by known sources. The concentration parameter $w1mag1-w1mag3$ was chosen as an input parameter as it serves as a proxy for morphological information: point-like  sources typically  have smaller $w1mag1-w1mag3$ values than extended ones.
For more details, see S17.

Once applied to the full catalogue of AllWISE data, the OCSVM algorithm selected $\sim 33,000$ objects with very 'red' mid-IR colours, namely, with very large $W1-W2$ ($>1$). 
This result is somewhat expected. As the training sample was based on the sources observed in the optical survey, the more dusty or higher redshift objects are bound to be flagged by the algorithm as outliers.
The high IR colour could be a result of a foreground or background source near the observed object. However, all the detected sources are well-isolated and no nearby sources contaminate the $W1$ and $W2$ flux, as, by design, S17 cleaned the data from the blends. Additionally, we checked the number of point spread function (PSF) components used simultaneously in the profile-fitting for these sources, called {\it nb} flag in the AllWISE database\footnote{\url{http://wise2.ipac.caltech.edu/docs/release/allwise/expsup/sec2_1a.html}}. If the {\it nb} flag is equal to 1, then only the central source is fitted. If {\it nb} is greater than 1, then the source is deblended. WISE deblends sources either in a passive way by fitting the source with other nearby detections or in an active way by splitting it into two components during the fitting process. All selected objects have {\it nb} flag equal to 1, and therefore their AllWISE photometry is free of any additional flux contamination. It is worth noting that 51\% of the sources flagged by the algorithm as anomalies are absent from optical catalogues such as the Pan-STARRS DR1 \citep[e.g.][]{kaiser10, chambers2016}, Gaia DR2 \citep{gaia2}, SuperCOSMOS Sky Survey \citep{supercosmos}, or SkyMapper DR1 \citep{skymapper}.\footnote{We find no counterparts in 2DFGRS \citep{2dfgrs}, 6DFGS \citep{6dfgs} or WiggleZ \citep{wigglez}.}
For this reason, it was necessary to perform follow-up spectroscopic observations to determine the nature of the anomalies and verify the performance of the algorithm.

\begin{figure}
    \centering
    \includegraphics[scale=0.14]{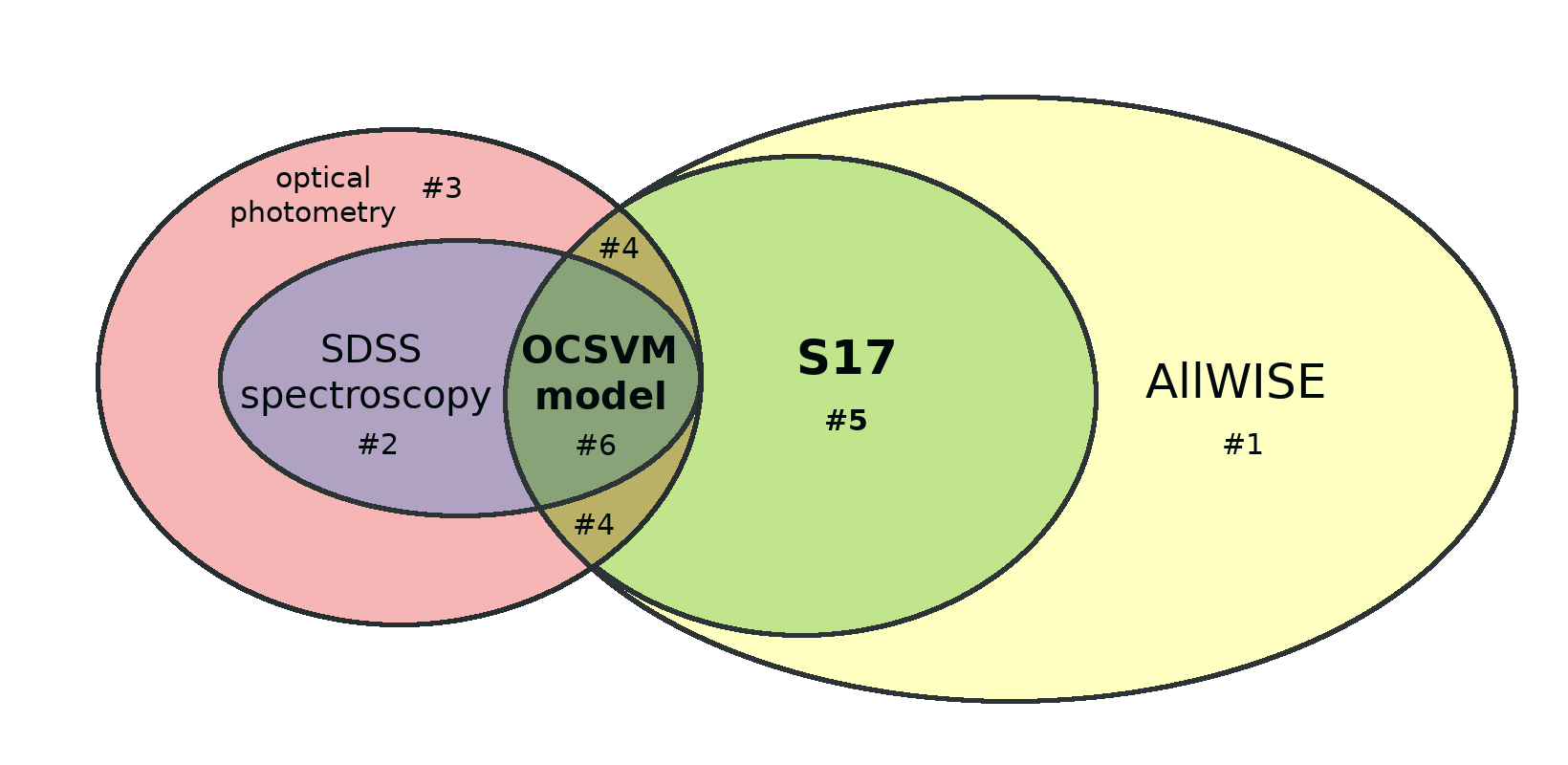}
    \caption{Schematic representation of the S17 catalogue creation. Set 1 represents the AllWISE database; set 2 the SDSS spectroscopic observations; set 3 objects with optical photometry (not limited to SDSS); set 4 sources present both in AllWISE and optical photometric catalogues but which do not have spectra; set 5 sources included in S17 anomaly catalogue; set 6 sources used as a training sample for the OCSVM algorithm. The ellipses do not reflect the sizes of the databases. }
    \label{vann}
\end{figure}

A schematic representation of the catalogue creation is shown in Fig.~\ref{vann}. 
According to the notation in Fig.~\ref{vann}, set 6 is the one used to train the ML model, which contains the unique class of so-called known objects. According to the same representation, we apply such a model to find outliers in group 4, namely, objects that have been detected by AllWISE with optical photometry. 
Therefore, the anomalies should be those sources which are in groups 4 and 5 but have different characteristics from those of group 6. The spectroscopic follow-up programme was designed to determine the nature of the 'anomalous' objects which have optical photometry (i.e. group 4 in Fig.~\ref{vann}).
For that reason we expect these anomalies to be: 1. objects that have photometry from SDSS but had not been selected as targets for spectroscopy (majority); 2. objects with SDSS photometry which, despite having been selected for spectroscopy, could not be successfully classified from their spectra (minority); 3. exotic objects that had not been covered at all by the SDSS and therefore not yet discovered (very rare).
 
This means that the S17 catalogue is actually probing the bias of SDSS spectroscopic target selection criteria, along with a few rare objects that the SDSS could not classify. These 'rare' objects are therefore not necessarily new kinds of exotic objects or very low in absolute numbers, but instead, they are mostly objects that are very under-represented in optical surveys. 
Nevertheless, these anomalies could represent potentially overlooked sources which are not considered in the statistical studies. 
To estimate the contamination ($c$) of the S17 catalogue, we counted how  many  times  an  object  whose character was  known   was, indeed,  correctly classified by the OCSVM (true positive; TP) and how many  times  a  known  object  was  classified  as  an  outlier (false negative; FN) during the training process. 
Based on these counts we found $c_{train}=FN/(TP+FN)=0.01\%$.
Furthermore, the accuracy estimation was followed by counting TP and FN in a validation set - a set of objects with known nature, but which were not used for training the algorithm (1\% of the full SDSS catalogue). The contamination of the validation set is $c_{valid}=0.02\%$.

\section{Spectroscopic follow-up observations}
\label{Sec:Data}

We designed a pilot spectroscopic programme targeting objects from the anomaly catalogue which have an optical counterpart within a radius of 2 arcsec from the WISE position and with apparent magnitude $g_{\rm AB} < 19.5$ mag. This magnitude limit is required for spectroscopy with a high enough signal-to-noise ratio (S/N) on a 4-m class telescope.  
We find 4,543 objects with optical counterparts for $g_{AB}<19.5$~[mag] (see Table~\ref{optcount}).

\begin{table}
\begin{footnotesize}
\begin{center}
\caption{Summary of the search for optical counterparts in photometric surveys within 2'' matching radius. $N_{g_{\rm AB}<19.5}$ denotes how many objects from all counterparts have $g_{\rm AB}$ brighter than limiting 19.5 mag; $N_{unique}$ shows how many objects are unique to a given survey; $N_{obs}$ denotes how many objects from a given survey we have observed.}
\label{optcount}
\setlength{\tabcolsep}{1.5 mm} 
\begin{tabular}{ccccc}
\hline
\hline
& 2'' x-match & $N_{g_{\rm AB}<19.5}$ & $N_{unique}$ & $N_{obs}$ \\
\hline
PannSTARRs DR1 & 13881 & 1446 & 1446 & 15 \\
Gaia DR2       & 8370  & 2926 & 1603 & 6  \\
SkyMapper DR1  & 702   & 171  & 37   & 15 \\
\hline
\end{tabular}
\end{center}
\end{footnotesize}
\end{table}

As the nature of the sources in the S17 catalogue is unverified, the selection of the observed targets was based on observing conditions and visibility: airmass $<1.5$, the maximum distance from the Moon and weather.
The observations were performed on the NTT in August-December 2018 in visitor mode (Programme ID: 0101.A-0539 and 0102.A-0305). During these runs, we observed the total of 36 objects using the EFOSC2 and SofI instruments.
The EFOSC2 was operating in long-slit mode (width equal to 1 arcsec),  with the Grism \#13 delivering spectra from 3650 to 9250~$\text{\AA}$ with average resolving power
$R=\lambda/\Delta\lambda \sim 355$. No blocking filter was used. The SofI was operating in long-slit mode (width equal to 1 arcsec), with the blue grism delivering spectra from  9350 to 16450~$\text{\AA}$ with spectral resolving power $R=\lambda/\Delta\lambda \sim 550$. 
The first run in August 2018 was conducted during the bright time with a fraction of lunar illumination (FLI) $\sim 0.97$, but the average seeing was below 1 arcsec.
The run in December was conducted during the 'dark time', with FLI$\sim$ 0.06 but seeing was higher ($\sim 1.45"$). The details of the observed sources are summarised in Table~2. 
 For the EFOSC2 data, we took a set of 11 bias frames before the start of each observing night, which is used to create a nightly master bias. Then spectroscopic dome flatfields and arc frames were taken before the beginning of the night to reduce the overheads. Each night we observed a spectrophotometric standard star to calibrate the flux.
Exposure times of science targets vary between 1800 and 2700 seconds.

 All sources observed with SofI were obtained using a standard ABBA sequence, including a small random offset in the A and B positions between exposures. This strategy is necessary due to a high flux level of the sky, which is usually higher than the flux of the target itself. It can vary on minute timescales and, therefore, it must be measured at similar times as the target observations. For that reason, the sky-subtraction was performed by subtracting consecutive frames from the ABBA nodding cycles from each other.
 The bias was subtracted together with the sky background, as recommended in the SofI handbook. Subsequently, the frames were shifted on top of each other and a median frame was calculated.
Spectroscopic flat fields were taken once per run (one in August, one in December) and are composed of flat field pairs taken with and without lamp illuminating the dome. Then the flats with lamp-off are subtracted from the ones with lamp-on removing the thermal background of the instrument\footnote{\url{https://www.eso.org/sci/facilities/lasilla/instruments/sofi/doc/manual/sofiman_2p40.pdf}}.
To calibrate the flux each night, we observed a spectrophotometric standard star, just like in the case of optical observations.
Each SofI target was followed by observations of bright Vega-type telluric standard stars, at similar air masses to the target object.
Wavelength calibration was made using the lines of a xenon lamp observed before every night.

Exposure times for SofI targets are a sum of detector integration time (DIT), the number of DITs (NDIT), the number of jitter positions A and B (NJITTER), and the number of offset positions (NOFFSETS). The typical DIT for our targets varied between 90 and 110 s.
\subsection*{Data reduction and redshift estimation}

The data reduction was carried out with the software written for the \textit{ePESSTO} survey by Stefano Valenti \citep{epessto} dedicated for EFOSC2 and SofI instruments \footnote{The PESSTO Pipeline for EFOSC2 and SOFI reductions  \url{http://wiki.pessto.org/pessto-operation-groups/data-reduction-and-quality-control-team}}. 

The S/N varies across the spectra, but typically it is $\sim 10-30$ at 6000-6200 $\text{\AA}$ and $\sim 10-30$ at 12000-12200 $\text{\AA}$ for EFOSC2 and SofI, respectively\footnote{With the exception of four objects described further in the following sections}.
 For telluric absorption correction, the ePESSTO data reduction pipeline uses a model of the atmospheric absorption to correct for the H$_2$O and O$_2$ absorption using the Line  By  Line  Radiative Transfer Model (LBLRTM; \citealt{clough05}). A detailed description of the model and available parameters can be found in \cite{patat11}. 
The observing details, including starting times, exposure times, and starting and ending airmasses for each exposure are listed in Appendix~\ref{app1} (tables~\ref{listaobiektowexp} and ~\ref{listaobiektowexpsofi}).

The redshift determination for the observed objects was based on visual inspection of the spectral features. Then wavelength measurements of the features via Gaussian profile fitting was performed, using the IRAF\footnote{The Image Reduction and Analysis Facility is distributed by the
National Optical Astronomy Observatory, which is operated by the
Association of Universities for Research in Astronomy (AURA) under
a cooperative agreement with the National Science Foundation.} task \textit{splot}. 
We did not include emission lines which were located near the edge of the wavelength range. Similarly, the spectral features contaminated by sky emission lines and significant residuals left by the sky subtraction were excluded. For most line centres, the typical formal statistical errors are $\sim 1 \mbox{\AA}$ and they translate into redshift errors of less than 0.001, ignoring the wavelength calibration error.
To evaluate the redshift uncertainties, we compared the redshifts derived from different lines of the same object. For objects with multiple lines, we adopted the average difference as redshift error.

\begin{table*}
\begin{footnotesize}
\begin{center}
\caption{List of targets observed
and the features identified in their spectra (column 4),
which were used for redshift measurement (column 5) and
object classification (column 6).}
\label{sum}
\setlength{\tabcolsep}{3.5 mm} 
\begin{tabular}{cccccc}
\hline
\hline
Object ID & RA & Dec & Spectral features & Redshift & Classification \\
\hline
J232450.80 & 23:24:50.90 & -35:37:15.80 & \ion{Si}{iv}$+$\ion{O}{iv}], \ion{C}{iv}, \ion{C}{iii}]     & $1.962\pm0.010$ & Broadline QSO \\
J201747.34 & 20:17:47.30 & -50:45:32.30 & \ion{C}{iv}, \ion{C}{iii}], \ion{Mg}{ii}     & $1.639\pm0.004$ & Broadline QSO \\
J191725.48 & 19:17:25.50 & -69:42:27.70 & \ion{C}{iv}, \ion{C}{iii}]         & $1.586\pm0.001$ & Broadline QSO \\
J214658.64 & 21:46:58.60 & -66:36:31.60 & \ion{C}{iv}, \ion{C}{iii}], \ion{Mg}{ii}     & $1.452\pm0.006$ & Broadline QSO \\
J204635.56 & 20:46:35.60 & -57:08:30.60 & Ly$\alpha$, \ion{Si}{iv}$+$\ion{O}{iv}], \ion{C}{iv} & $2.605\pm0.022$ & Broadline QSO \\
J190137.97 & 19:01:38.00 & -62:08:02.90 & \ion{C}{iii}], \ion{Mg}{ii}           & $1.278\pm0.001$ & Broadline QSO \\
J231638.59 & 23:16:38.60 & -44:48:56.70 & \ion{C}{iii}], \ion{Mg}{ii}           & $1.310\pm0.005$ & Broadline QSO \\
J072450.77 & 07:24:50.77 & -68:49:49.90 & \ion{C}{iv}, \ion{C}{iii}]  & $1.578\pm0.002$ & Broadline QSO \\
J061858.03 & 06:18:58.04 & -03:33:44.90 & \ion{C}{iii}], \ion{Mg}{ii}           & $1.409\pm0.003$ & Broadline QSO \\
J084059.57 & 08:40:59.57 & -22:10:00.50 & \ion{Si}{iv}$+$\ion{O}{iv}], \ion{C}{iv}, \ion{C}{iii}], \ion{Mg}{ii} & $1.755\pm0.009$ & BAL QSO    \\
J040840.58 & 04:08:40.59 & -17:40:41.50 & \ion{Si}{iv}$+$\ion{O}{iv}], \ion{C}{iv}, \ion{C}{iii}], \ion{Mg}{ii} & $1.687\pm0.007$ & BAL QSO    \\
J040754.75 & 04:07:54.76 & -65:22:35.00 & \ion{Mg}{ii}, H$\beta$, [\ion{O}{iii}], H$\alpha$ & $1.356\pm0.006$ & Fe rich QSO \\
J204544.00 & 20:45:44.00 & -35:39:41.30 & [\ion{O}{ii}], [\ion{O}{iii}], H$\alpha$ & $0.108\pm0.006$ & Seyfert 2     \\
J195131.19 & 19:51:31.20 & -67:22:09.70 & [\ion{O}{ii}], H$\beta$, [\ion{O}{iii}], H$\alpha$ & $0.067\pm0.001$ & SF \\
J062035.99 & 06:20:35.99 & -56:57:56.50 & [\ion{O}{ii}], H$\beta$, [\ion{O}{iii}], H$\alpha$ & $0.080\pm0.001$ & SF  \\
J070647.77 & 07:06:47.77 & -39:34:14.00 & [\ion{O}{ii}], H$\beta$, [\ion{O}{iii}], H$\alpha$ & $0.078\pm0.001$ & SF  \\
J205821.02 & 20:58:21.02 & -14:44:22.00 & [\ion{O}{ii}], H$\beta$, [\ion{O}{iii}], H$\alpha$ & $0.114\pm0.001$ & SF  \\
J052522.78 & 05:25:22.79 & -07:04:19.10 & [\ion{O}{ii}], H$\beta$, [\ion{O}{iii}], H$\alpha$ & $0.148\pm0.001$ & SF  \\
J022908.52 & 02:29:08.52 & -00:38:19.60 & [\ion{O}{ii}], H$\alpha$        & $0.069\pm0.001$ & SF            \\
J053520.86 & 05:35:20.87 & -25:34:42.40 & [\ion{O}{ii}], H$\beta$, [\ion{O}{iii}], H$\alpha$ & $0.073\pm0.001$ & SF  \\
J040051.83 & 04:00:51.83 &  06:50:34.30 & H$\beta$, [\ion{O}{iii}], H$\alpha$ & $0.032\pm0.001$ & SF        \\
J022718.11 & 02:27:18.11 &  07:06:14.30 & [\ion{O}{ii}], H$\beta$, [\ion{O}{iii}], H$\alpha$ & $0.105\pm0.001$ & SF  \\
J164737.38 & 16:47:37.39 & -00:57:52.80 & \ion{Ca} K \& H, H$\alpha$  & $0.088\pm0.001$ & S0            \\
J010409.21 & 01:04:09.22 & -43:17:36.80 & \ion{Ca} K \& H, H$\alpha$  & $0.030\pm0.001$ & S0            \\
J064043.96 & 06:40:43.96 & -08:32:57.40 & \ion{Ca} K \& H, H$\alpha$  & $0.138\pm0.002$ & S0            \\
J060127.78 & 06:01:27.79 & -53:22:06.70 & \ion{Ca} K \& H, H$\alpha$  & $0.033\pm0.002$ & S0            \\
J071734.84 & 07:17:34.84 & -54:23:03.00 & \ion{Ca} K \& H, H$\alpha$  & $0.077\pm0.001$ & S0            \\
J211637.73 & 21:16:37.74 & -35:38:51.00 & \ion{Ca} K \& H, H$\alpha$  & $0.073\pm0.001$ & S0            \\
J191825.52 & 19:18:25.53 & -71:53:13.70 & \ion{Ca} K \& H, H$\alpha$  & $0.041\pm0.001$ & S0            \\
J185020.05 & 18:50:20.05 & -51:23:13.70 & \ion{Ca} K \& H              & $0.078\pm0.001$ & E             \\
J141606.17 & 14:16:06.18 & -36:09:34.70 & \ion{Ca} K \& H             & $0.041\pm0.002$ & E             \\
J155603.92 & 15:56:03.92 & -23:23:09.40 & -                      & -               & YSO           \\
J062245.36 & 06:22:45.37 & -45:10:24.30 & -                      & -               & unknown       \\
J185103.11 & 18:51:03.12 & -17:46:45.00 & -                      & -               & unknown       \\
J183949.54 & 18:39:49.40 & -50:49:16.70 & -                      & -               & unknown       \\
J173122.32 & 17:31:22.33 & -22:25:16.60 & -                      & -               & unknown       \\
\hline
\end{tabular}
\end{center}
\end{footnotesize}
\end{table*}

\section{Results}

The sample of 36 targets observed in this campaign includes the following objects:
\begin{itemize}
\item Nineteen galaxies at $z$ $<$ 0.15
\item Twelve broad-line QSOs located at redshifts $1.3<z<2.6$ including two broad absorption line (BAL) QSOs and one with strong and narrow iron emission lines
\item one young stellar object (YSO)
\item four unidentified objects due to a very low S/N for the spectra
\end{itemize}
Table~\ref{sum} lists all observed objects with identified features in their spectra, which we used to determine their redshifts and classification.
\begin{figure*}[ht]
\centering
\captionsetup[subfloat]{labelformat=empty}
    \subfloat[]{\includegraphics[scale=0.55]{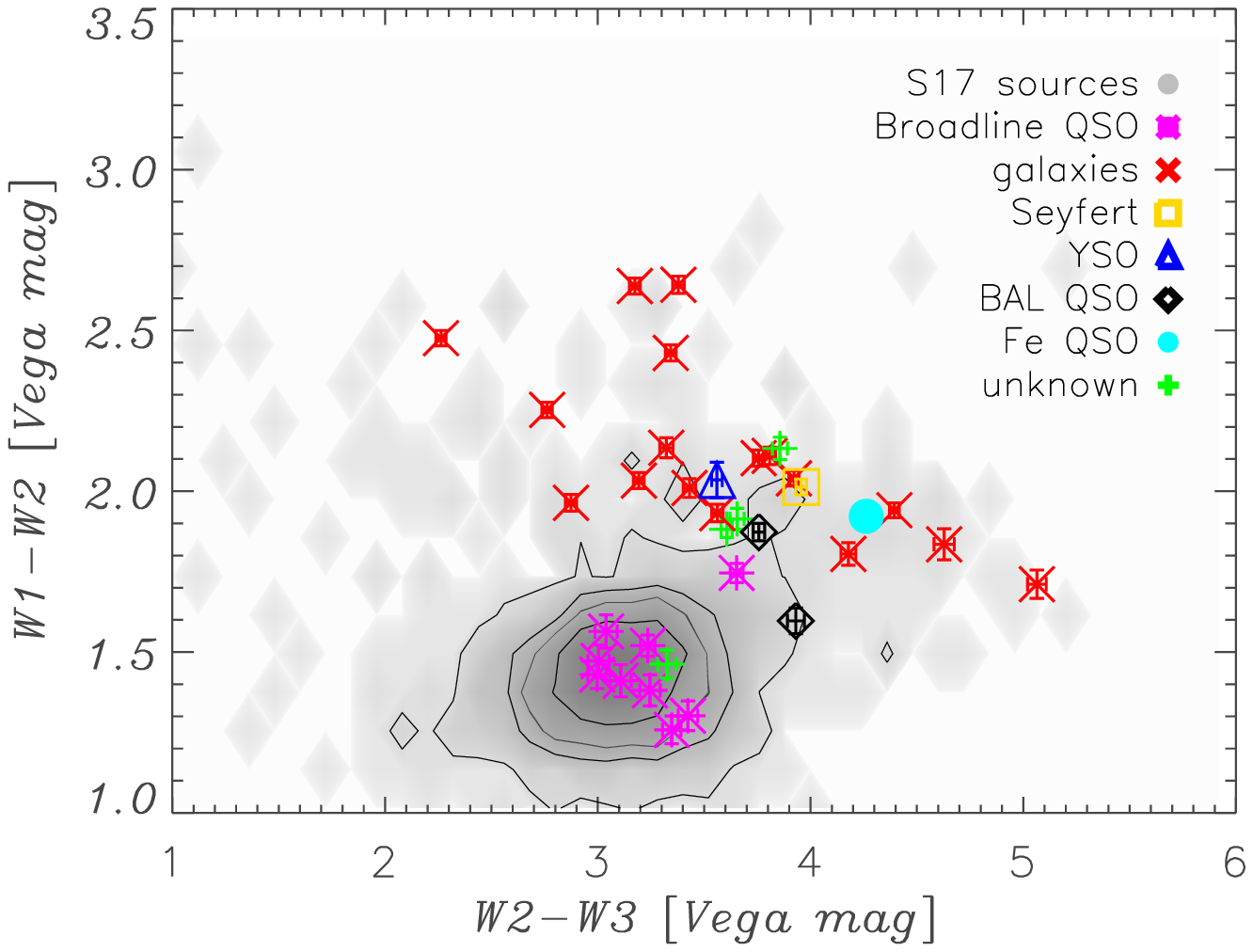}}
   \quad
    \subfloat[]{\includegraphics[scale=0.55]{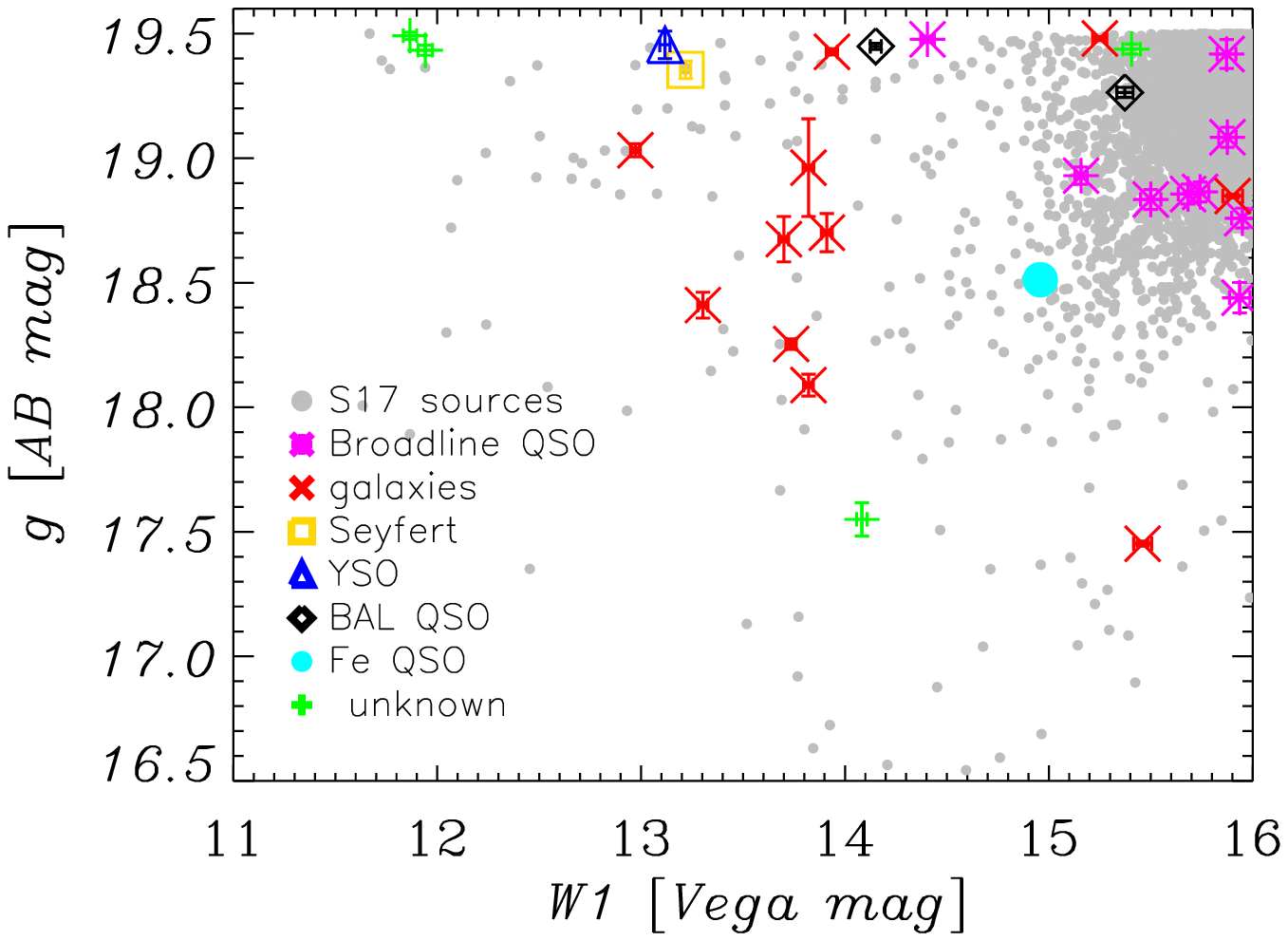}}
  \caption{ Left panel: Colour-colour diagram with the classes of observed anomalies marked as different symbols: broadline QSOs are shown as asterisks, BAL QSOs as diamonds, a YSO as a triangle, galaxies as crosses, a Seyfert galaxy as a  square, and unidentified objects as plus signs.  Right panel: {\it W1} magnitude in Vega plotted against $g_{\rm AB}$ magnitude for the targeted sources (same colour and symbol coded as for left panel). Grey contours and points mark the position of all OCSVM anomalies contained in the S17 catalogue which have optical detection with $g_{\rm AB}<19.5$.}
  \label{nq}
\end{figure*}%
The left panel of Fig.~\ref{nq} shows the positions of the observed objects on a colour-colour diagram for WISE passbands. The right panel shows the distribution of the apparent optical brightness in $g_{\rm AB}$-band relative to the {\it W1}. 
\begin{figure}[ht]
\centering
\includegraphics[width=\hsize]{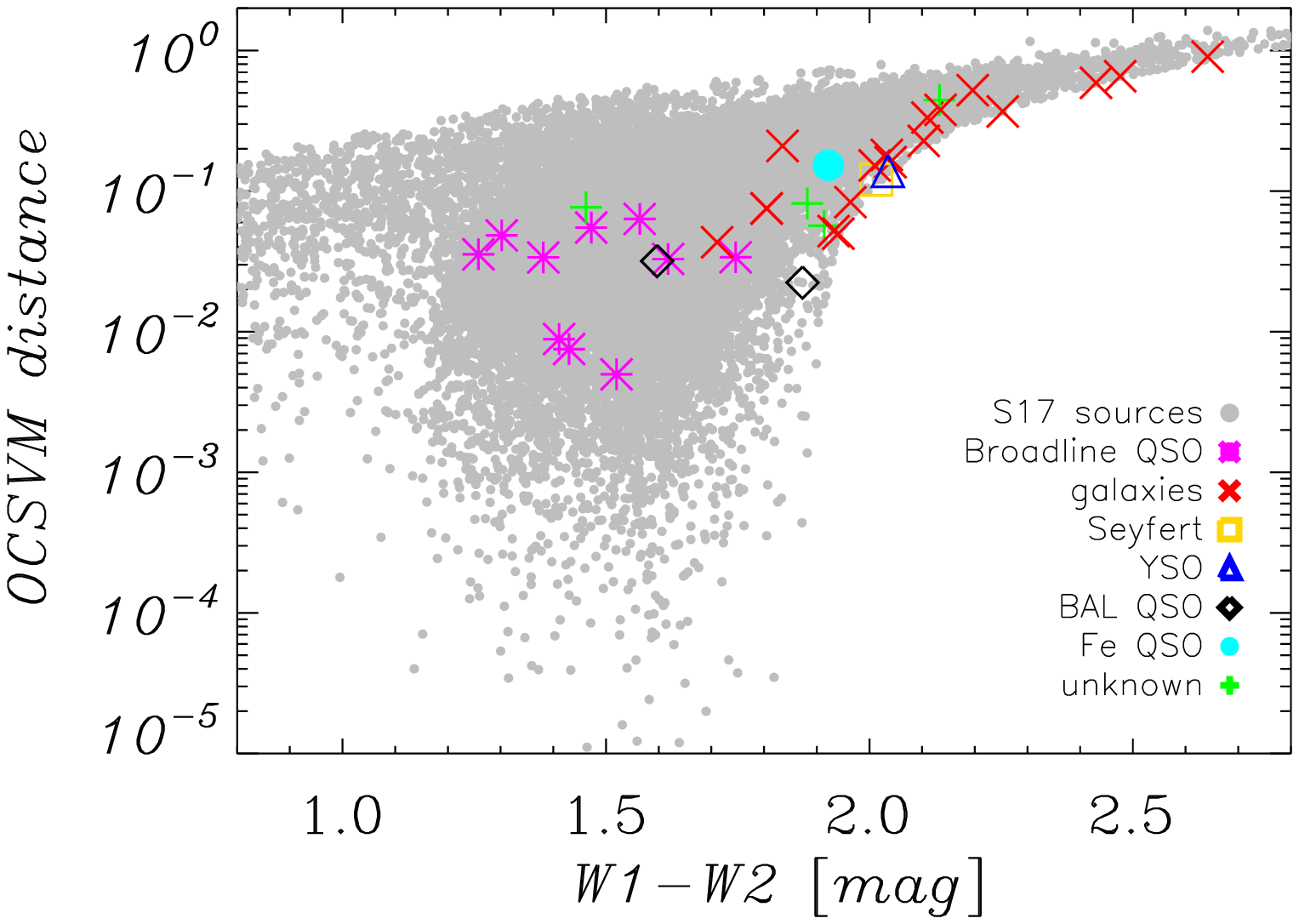}
\caption{$W1-W2$ colour distribution as a function of the distance from the OCSVM decision boundary. The symbols are the same as in Fig.~\ref{nq}. }
\label{dist}
\end{figure}
Fig.~\ref{dist} presents the distribution of the $W1-W2$ colour as a function of their respective distance from the OCSVM decision boundary for all observed objects. The distance measurement indicates how uncharacteristic the outlier is with respect to the training objects.  We can see that there is a correlation between these two values, especially for the observed galaxies, which are both the reddest and the most uncharacteristic. What is more, the QSO with strong iron emission is also more anomalous than other observed QSOs and BAL QSOs.
In the following subsections, we describe in more detail the different classes of the observed objects.

\subsection{Low redshift red galaxies}

\begin{figure*}[ht]
\centering
\includegraphics[width=\hsize]{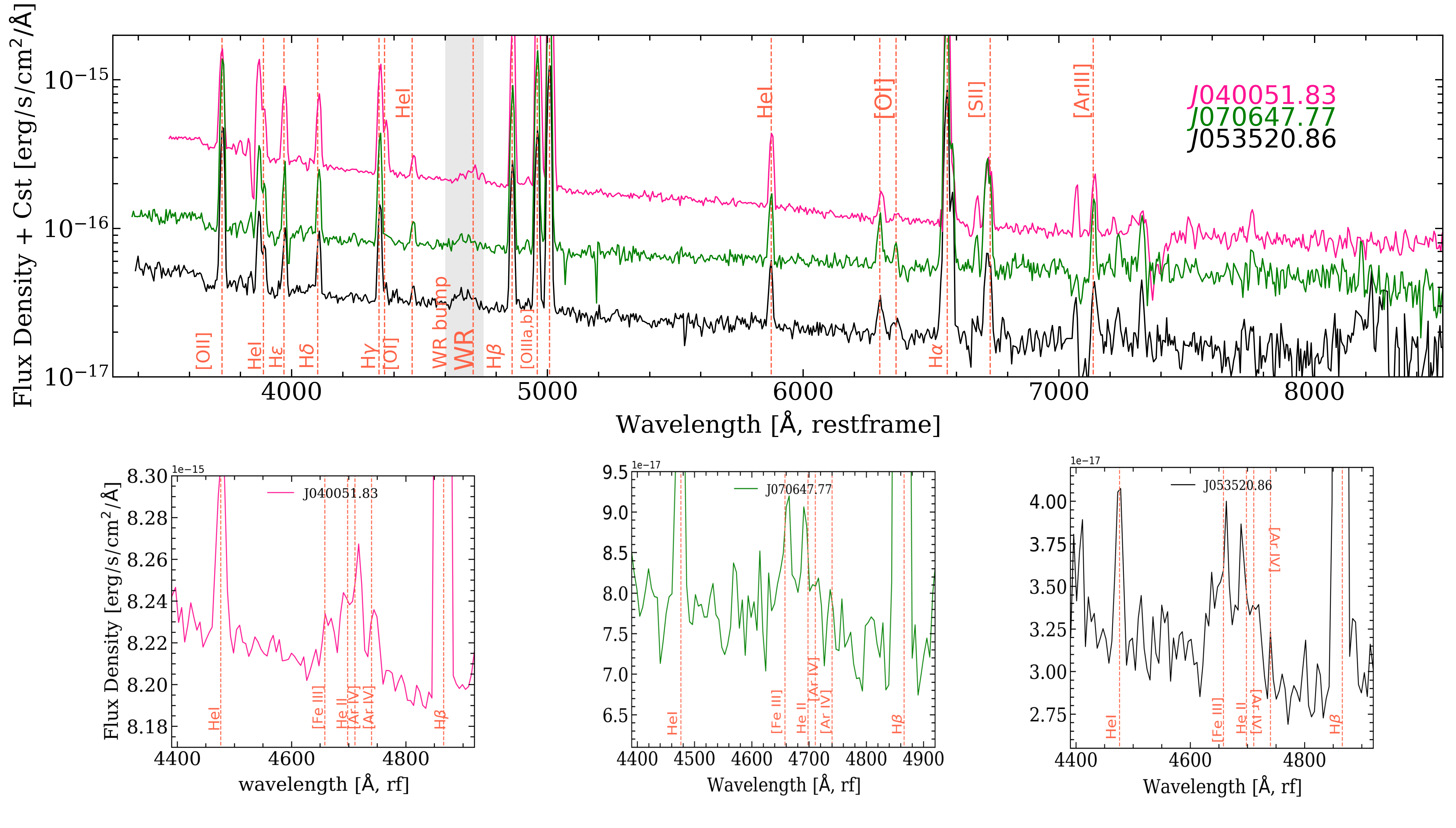}
\caption{Observed targets with Wolf-Rayet signature (WR). \textit{Top:} All observed objects with line identifications. The WR signature is indicated by the grey span. \textit{Bottom:} Close-up on the WR region for each object.}
\label{sb}
\end{figure*}

\begin{figure}[ht]
\centering
\includegraphics[width=\hsize]{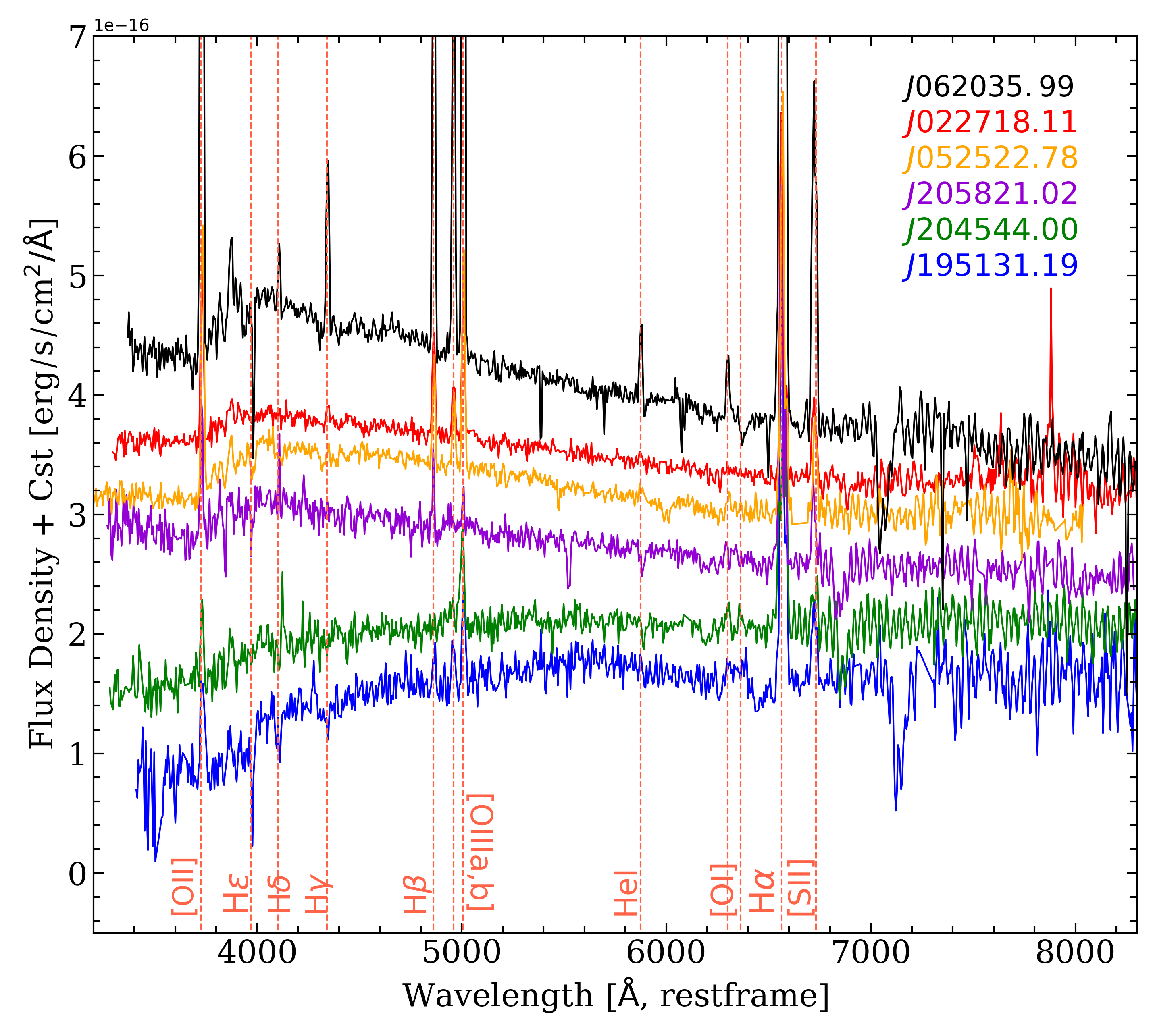}
\caption{Observed local star-forming galaxies.  }
\label{localdusty}
\end{figure}%

\begin{figure}[ht]
\centering
\includegraphics[width=\hsize]{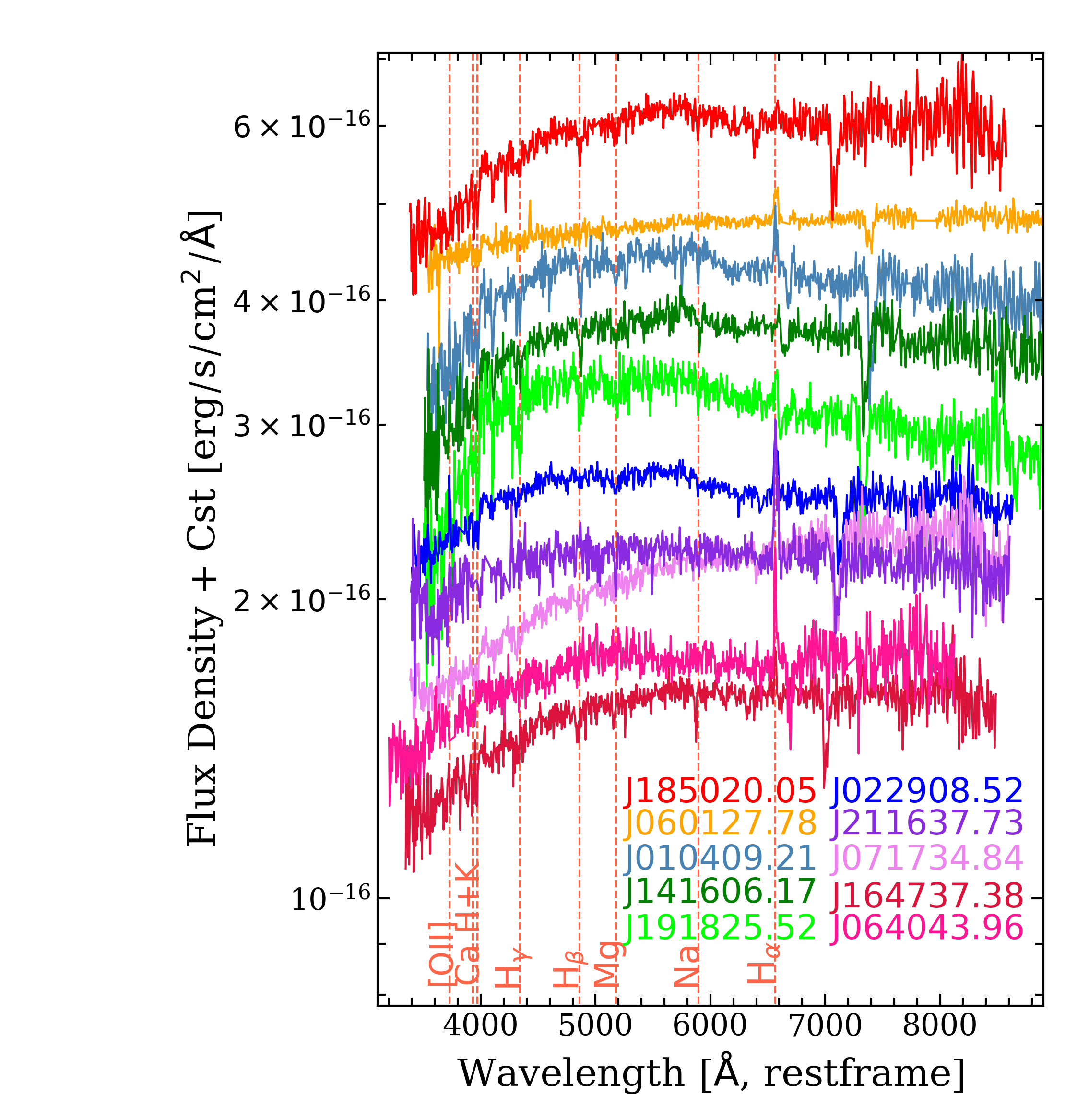}
\caption{Observed local early-type galaxies. The only detected emission line is $H_{\alpha}$}. 
\label{regular}
\end{figure}%

Nineteen of the observed objects are galaxies located at $z$ $<$ $0.15$ and displaying very red $W1-W2$ colours ($>1.7$), which can indicate the presence of hot dust, which is heated during an active galactic nucleus (AGN) phase or a violent star-forming (SF) phase, or combination of the two phenomena.
The observed spectra are divided into specific classes depending on their spectral features and are presented in Figs.~\ref{sb}, \ref{localdusty}, and \ref{regular}. 
Local galaxies with extremely red mid-IR colours are a particularly interesting source population. Usually, such galaxy colours are thought to be indicative of AGN activity in massive galaxies (e.g. \citealt{stern12}). 
However, for low-mass galaxies observational evidence reports a lack of AGN signatures in either broad or narrow optical emission lines \citep[e.g.][]{izotov11,sartori15,hainline16,oconnor16,kauffmann18}.
\begin{figure*}[ht]
\centering
\includegraphics[scale=0.4]{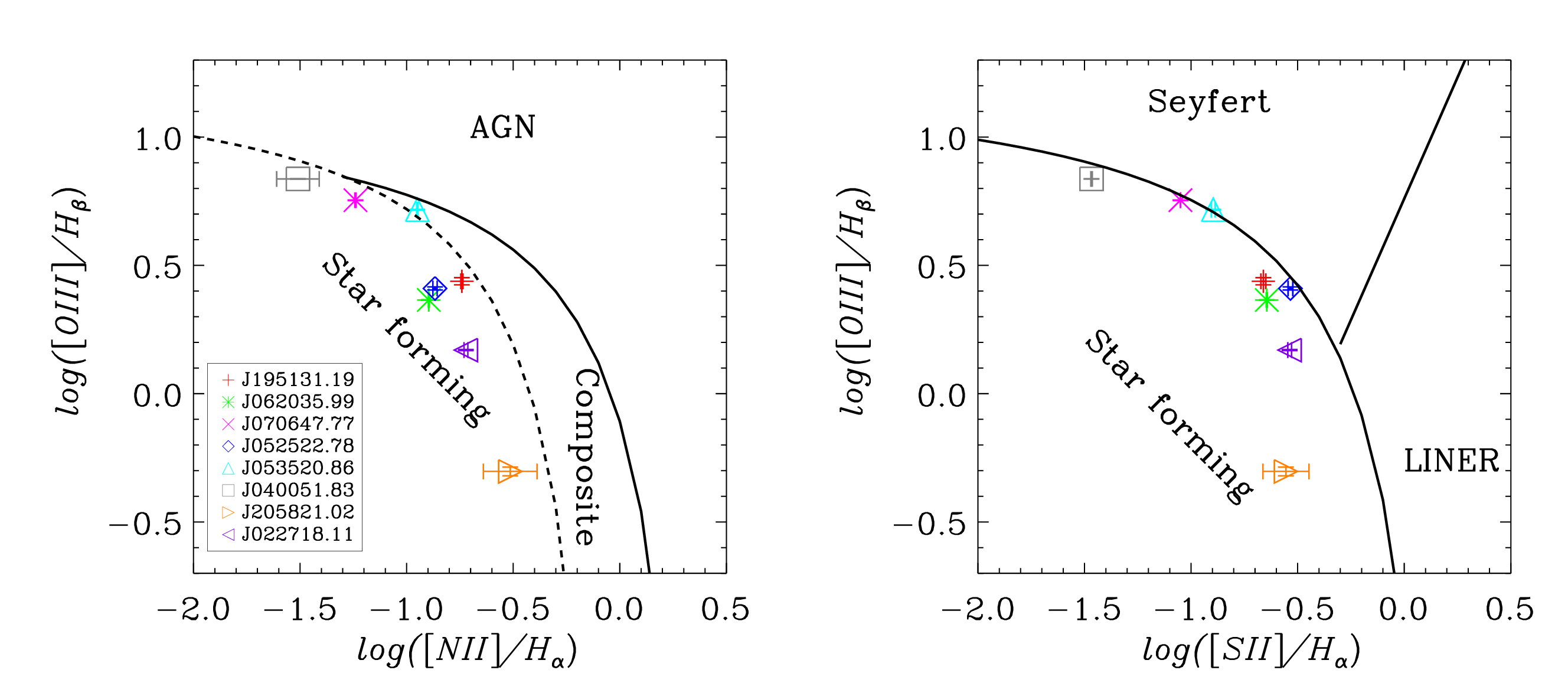}
\caption{Two BPT diagrams used to classify the emission-line galaxies as: Seyfert, LINER, composite, and ambiguous galaxies. Left: Solid line shows the \citet{kauff03} selection criteria. \citet{kewley06} classification is shown as the dashed line. Right: Solid lines separate the SF galaxies from the active galaxies and the dashed lines represent the Seyfert-LINER demarcation.  }
\label{bpt}
\end{figure*}
In nine of these sources, we detect multiple emission lines (see Table~\ref{sum}), such as H$\beta$, [\ion{O}{iii}] $\lambda$5008.24, H$\alpha$, [\ion{N}{ii}] $\lambda$6585.27, and [\ion{S}{ii}]  $\lambda$6718.29. 
We employ the BPT diagram \citep{BPT}, along with the selection criteria of \cite{kauff03} and \cite{kewley01, kewley06}, to identify the class of these sources (see Fig.~\ref{bpt}). 
All of the observed galaxies lie just below the separation line between SF and composite galaxies. 

As we do not detect any apparent signatures of AGN activity within these sources (i.e. broad emission lines or corresponding emission line ratios), we consider two possible explanations for the extreme dust heating in these galaxies. 
On the one hand,  these galaxies could contain AGNs which are heavily obscured and optically invisible. 
On the other hand, they might not contain AGN and the dust heating that results in high $W1-W2$ colour could be caused by stars alone.

In looking for evidence for the presence of an AGN within our objects, we first searched for counterparts in publicly available astronomical databases of soft and hard X-rays: Chandra \citep{chandra}, XMM Newton \citep{xmm}, Swift/BAT \citep{swift}, and INTEGRAL/IBIS \citep{integral}. We used a maximum search radius of 30''.
X-ray emission traces the accretion history of the Universe, offering a highly efficient method of detecting growing black holes in galaxies over a wide range of redshifts. X-ray surveys with XMM-Newton and Chandra at energies $<10$ keV are sensitive to all but the most heavily obscured AGN (e.g. \citealt{dc08}). Hard X-ray emission is thought to be the most unbiased AGN selection technique  (e.g. \citealt{swift}), as it is less affected by the obscuring material (up to $N_{H}\sim10^{23.5}-10^{24}$ cm$^{-2}$, \citealt{ricci15}).
Also, a large number of X-ray sources have been confirmed to be regular galaxies without any obvious optical signatures of the AGN presence \citep[e.g.][]{caccianiga07, smith14}. 
We find that our 'red' galaxies have no X-ray counterparts. We note, however, that all-sky surveys have a lower sensitivity, so the X-ray emission might be still present in these systems. 
Additionally, we searched for radio counterparts of our objects in The NRAO VLA Sky Survey (NVSS, \citealt{condon98}) and Faint Images of the Radio Sky at Twenty-cm (FIRST, \citealt{first}) surveys, with no success. For this search, we adopted a 60'' for the maximum matching radius.
Ten out of the 19 objects appear in the area of the sky covered by these radio surveys; however, no radio emission is detected. We cannot determine the radio emission for the remaining objects, as they fall outside of the coverage of these surveys.
Finally, in the work of \citet{satyapal18}, a colour cut was derived to distinguish the SF contribution from that of an AGN in red galaxies. Objects fulfilling the following criteria are considered to have a significant AGN contribution:
$W 1 - W 2 \ge 0.52$ and $W 1 - W 2 \ge 5.78 \times (W 2 - W 3) - 24.50.$
All but three of our galaxies fulfil this AGN criterion.
Objects J070647.77, J040051.83 and J053520.86, could, therefore, turn out to be purely SF galaxies. Their optical spectra are shown in Fig.~\ref{sb}. They show clear signatures of Wolf-Rayet (WR) stars
manifested by series of broad emission lines in the optical wavelength range such as the broad \ion{He}{ii} $\lambda$4686, \ion{N}{iii} $\lambda$4640, or \ion{C}{iii/iv} $\lambda$4650, features originating in the expanding atmospheres of the most massive stars \citep{crowther07}. This spectral region manifesting the existence of the WR population within a galaxy is often referred to as 'WR bump' (bottom panels of Fig.~\ref{sb}).
Such WR stars, 
initially identified by Wolf \& Rayet (1867), evolve from O-type stars with an initial mass of 25~$M_{\odot}$ or larger.
Galaxies hosting living high-mass stars during their WR phase are reported to be very rare, as WR stars present a small fraction of the stellar population in a galaxy and have a short lifetime (e.g. \citealt{liang20}).
The high dust temperatures, as indicated by the large $W1-W2$ colours could
arise from very young, hot stars hosted in low-metallicity environments (\citealt{griffith11}, \citealt{izotov14}). To estimate the oxygen abundance, where possible, we use the 
$R_{23}$ index introduced by \citet{pagel}:

\begin{equation}
    R_{23}=\frac{\mathrm{[O II]} \lambda3727+ \mathrm{[O III]} \lambda4959 + \mathrm{[OIII]} \lambda5007}{\mathrm{H}\beta}
.\end{equation}

The $R_{23}$ abundance  relation has two solutions, a low and high metallicity estimate for most values of $R_{23}$ (e.g. \citealt{kk04}).
For that reason, we use an additional line ratio,  [\ion{N}{ii}] $\lambda$6583 / [\ion{O}{ii}] $\lambda$3727, to break this degeneracy.
The upper and lower branches of the $R_{23}$ calibration bifurcates at $\log$([\ion{N}{ii}]/[\ion{O}{ii}])$\sim -1.2$ for the SDSS galaxies \citep{ke08}, which corresponds to a metallicity of $12+\log(\mathrm{O/H})\sim8.4$.
To estimate the $12+\log(\mathrm{O/H})$ we use the empirical calibration method for $R_{23}$ presented by \citet{pt05} based on electron temperature ($T_{e}$) metallicities for a sample of \ion{H}{ii} regions. 
They derive the relationship between $R_{23}$ and $T_{e}$-metallicities that includes an excitation parameter $P$ that corrects for the effect of the ionisation parameter, defined as:   $P=$(([\ion{O}{iii}] $\lambda$4959 + [\ion{O}{iii}] $\lambda$5007)/H$\beta$) /$R_{23}$.
Their calibration has an upper branch that is applicable to metallicities $12 + \log (\mathrm{O/H})>8.25$, and a lower branch that is valid for metallicities $12 + \log (\mathrm{O/H})<8.0$. We use the [\ion{N}{ii}]/[\ion{O}{ii}] ratio to discriminate between the upper and lower branches, and we apply the appropriate upper and lower-branch calibrations: 
\begin{equation}
   12+\log(\mathrm{O/H})_{high}=\frac{R_{23}+726.1+842.2P+337.5P^{2}}{85.96+82.76P+43.98P^{2}+1.793 R_{23}}
\end{equation}
and
\begin{equation}
    12+\log(\mathrm{O/H})_{low}=\frac{R_{23}+106.4+106.8 P-3.4 P^{2}}{17.72+6.60P+6.95P^{2}-0.302R_{23}}.
\end{equation}

The metallicity estimates for our galaxies are summarised in Table~\ref{met}.
All three WR galaxies have low metallicities, as their average is $12 + \log (\mathrm{O/H})\sim 8.28$. Their low measured metallicities coupled with the presence of WR bump could indicate that the extreme MIR colours could be a result of purely SF processes.
The metallicity of the remaining SF galaxies is measured to be slightly higher (average $12+\log(O/H)\sim 8.5$).
As we do not detect an $H\beta$ line in object J04544.00, we cannot estimate its metallicity.

\begin{table}
     \centering
          \caption{Values of $12+\log (\mathrm{O/H})$ determined from the \citet{pt05} calibrations together with $\log$([\ion{N}{ii}]/[\ion{O}{ii}]) values used to break the $R_{23}$ metallicity degeneration. Asterisks denote the WR galaxies in the sample.}
     \begin{tabular}{c|c|c}
     \hline\hline
        OBID&$\log([\ion{N}{ii}]/[\ion{O}{ii}])$&$12+\log(\mathrm{O/H})$\\\hline
        J040051.83$^*$&$-1.12\pm0.06$&$8.29\pm0.07$\\
        J053520.86$^*$&$-0.68\pm0.02$&$8.28\pm0.09$\\
        J070647.77$^*$&$-0.81\pm0.03$&$8.28\pm0.03$\\
        J062035.99&$-0.63\pm0.02$&$8.36\pm0.04$\\
        J022718.11&$-0.32\pm0.03$&$8.54\pm 0.04$\\
        J052522.78&$-0.99\pm0.05$&$8.61\pm0.03$\\
        J205821.02&$-0.66\pm0.03$&$8.41\pm0.05$\\
        J195131.19&$-1.13\pm 0.04$&$8.45\pm0.03$\\\hline
     \end{tabular}

     \label{met}
 \end{table}

In the next step, we study the physical properties of the observed galaxies by performing a spectral energy distribution (SED) modelling.
For this purpose we use the Code Investigating GALaxy Emission (CIGALE)\footnote{For a detailed description of the code, please refer to the CIGALE webpage \url{cigale.lam.fr}}. 
We combine all available photometric information from public databases to fit the SEDs: using Pan-STARRS and  SkyMapper Southern Survey \citep{skymapper} with WISE photometry and the AKARI Far-infrared Surveyor (FIS) AllSky Survey \citep{kawada07} catalogue. Only nine of the observed galaxies have a counterpart in AKARI FIS within 15 arcsec search radius (corresponding to the PSF of AKARI, \citealt{pepcus}), providing far-infrared photometric data points at 65, 90, 140, and 160~$\mu$m. Only nine of the observed galaxies have an FIR counterpart, so to get reliable star formation rates (SFR) and dust estimates, we must focus on those objects alone.
CIGALE uses a $\chi^{2}$ minimisation technique over a broad array of model templates covering the wavelength range from ultraviolet (UV) to FIR  \citep{noll09}.
 To perform SED fitting for the observed 'red' galaxies, we combined \citet{bc03} models  with assumed solar metallicities\footnote{CIGALE has a fixed set of discrete metallicities. The closest metallicity value available for our galaxies, where we could determine the oxygen abundances, is solar.}. 
We adopt a Chabrier initial mass function (IFM, \citealt{chabrier03}) 
 and account for the dust reddening using Calzetti's law \citep{calzetti00}. Star formation history is modelled as exponentially delayed: $SFR(t)\propto te^{-t/\tau_{main}}$, where $t$ is time and $\tau_{main}$ is the time since the onset of SF to the peak of the history.  We use values of the old stellar ages covering the range between 100 and 5000 Myr and e-folding time of the main stellar population to vary from  100-2000 Myr. 
We also include the dust reddening parametrised by the colour excess {\it  E(B-V)}, set to vary between 0 and 1 with a 0.05 step.
The CIGALE code includes a possibility to model a dusty torus emission heated by a central AGN component \citep{fritz06}. The model is described by six parameters related to the geometrical configuration of the dusty torus, properties of the dust and radiative transfer equation. To avoid the degeneracy of the SED fitting due to 
the limited photometric data we currently have, we fix several parameters of the AGN contribution to average values found by \citet{hatziminaoglou08}. We set the outer-to-inner radius ratio to 60, dust density parameters to -0.5 and 0, and opening angle to $100^\circ$, and we allow the optical depth at 9.7 $\mu$m to vary between two values: 1 and 6, corresponding to low and high optical depth. 
 CIGALE allows to fit templates for different values of the angle between the line of sight and the axis of the torus, denoted by $\psi$.
 However, \citet{ciesla15} proved that only extreme values of the angle between the AGN axis and the line of sight could provide reliable results. Therefore we allow the $\psi$ parameter to assume only two values: $0^\circ$ and $90^\circ$, corresponding to Type II and Type I AGNs, respectively. 
We let the contribution of the AGN to the total infrared (IR) luminosity ($\mathrm{frac_{AGN}}$) to assume the following values: 0.001\%, 0.1\%, 10\%, 15\%, 30\%, 40\%, and 49\%. 
Table~\ref{ajaj} summarises the range of parameters used for the SED fitting.

\begin{table*}
\begin{footnotesize}
\begin{center}
\caption{Input parameter ranges for the SED fitting with CIGALE for the spectroscopically confirmed galaxies.}
\label{ajaj}
\setlength{\tabcolsep}{3.5 mm} 
\begin{tabular}{cc}
\hline
\hline
Parameter & Range \\
\hline
$\tau$ e-folding time of main stellar population / Myr               & 100-2000         \\
Age (old stellar population) / Myr                                   & 100-5000         \\
metallicity                                                          & 0.02             \\
colour excess of the stellar continuum light of the young population & 0.1-1            \\
Ratio of the maximum to minimum radii of the dust torus              & 60               \\
Optical depth at 9.7$\mu$m                                           & 1.0, 6.0         \\
radial dust distribution of the torus                                & -0.5             \\
angular distribution of dust in the torus                            & 0.0              \\
Full opening angle of the dust torus                                 & 100.0            \\
angle between equatorial axis and the line of sight [$^{\circ}$]     & 0.001 and 89.990 \\
fractional contribution of AGN                                       & 0.001-0.49       \\
Mass fraction of PAH                                                 & 0.47-4.58        \\
\hline
\end{tabular}
\end{center}
\end{footnotesize}
\end{table*}

\begin{table*}
\begin{footnotesize}
\begin{center}
\caption{Summary of the SED modelling for the observed galaxies at $z <0.15$ with FIR detection.}
\label{fritz}
\setlength{\tabcolsep}{3.5 mm} 
\begin{tabular}{ccccccc}
\hline
\hline
Name & $\chi^{2}$ & AGN$_{\rm frac}$ & $E(B-V)$ & log$(L_{\rm dust}/{\rm L_{\odot}})$ & SFR/${\rm M_{\odot} yr^{-1}}$ & log$(M_{*}/{\rm M_{\odot}})$\\
\hline
J195131.19 & 0.80 & $0.40\pm0.04$ & $0.10\pm0.01$ & $10.89\pm0.03$ & $20.19\pm 2.57$ & $8.37\pm 0.06$\\
J204544.00 & 0.72 & $0.15\pm0.08$ & $0.60\pm0.05$ & $11.56\pm0.04$ & $45.80\pm 8.23$ & $9.27\pm 0.04$\\
J141606.17 & 0.73 & $0.30\pm0.04$ & $0.15\pm0.01$ & $10.23\pm0.06$ & \phantom{1}$3.12\pm0.72$ & $8.82\pm 0.03$\\
J164737.38 & 0.63 & $0.30\pm0.05$ & $0.45\pm0.16$ & $10.51\pm0.10$ & \phantom{1}$2.29\pm0.27$ & $9.86\pm 0.08$\\
J010409.21 & 0.60 & $0.45\pm0.20$ & $0.50\pm0.04$ & $10.76\pm0.02$ & \phantom{1}$8.21\pm1.49$ & $8.40\pm 0.01$\\
J071734.84 & 1.50 & $0.15\pm0.05$ & $0.45\pm0.07$ & $11.01\pm0.03$ & \phantom{1}$6.53\pm 0.32$ & $9.48\pm 0.06$\\
J052522.78 & 0.72 & $0.49\pm0.06$ & $0.30\pm0.14$ & $11.49\pm0.08$ & $58.57\pm 9.28$ & $8.83\pm 0.64$\\
J205821.02 & 1.23 & $0.30\pm0.05$ & $0

.50\pm0.17$ & \phantom{1}$9.36\pm 0.06$ & \phantom{1}$0.29\pm 0.05$ & $7.08\pm 0.13$\\
J022718.11 & 0.64 & $0.30\pm0.06$ & $1.00\pm0.05$ & $11.26\pm0.03$ & $15.72\pm1.15$ & $9.34\pm0.05$\\
\hline
\end{tabular}
\end{center}
\end{footnotesize}
\end{table*}

\begin{figure*}[ht]
\centering
\subfloat{
\includegraphics[width=85mm]{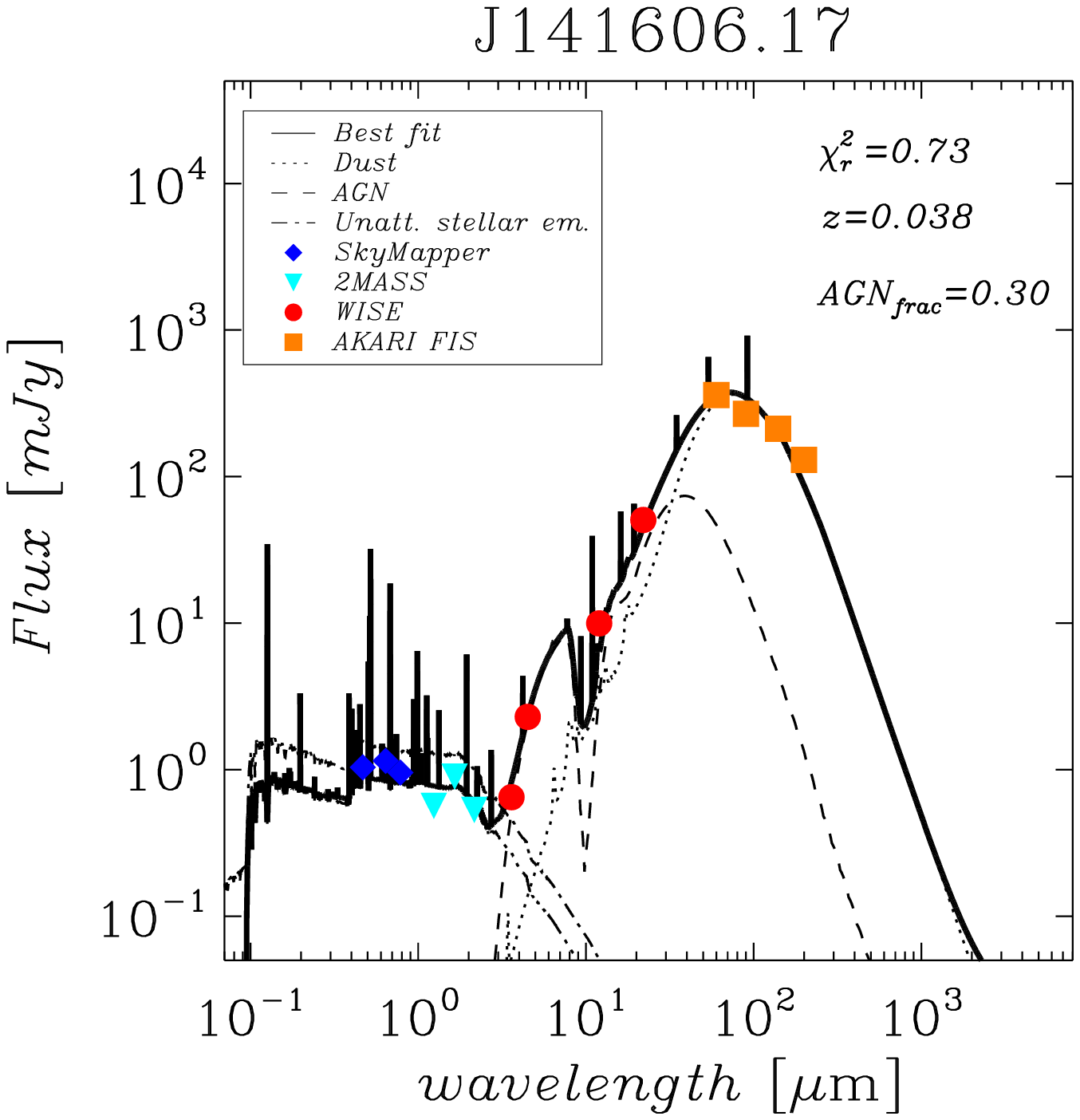}
}
\subfloat{
\includegraphics[width=85mm]{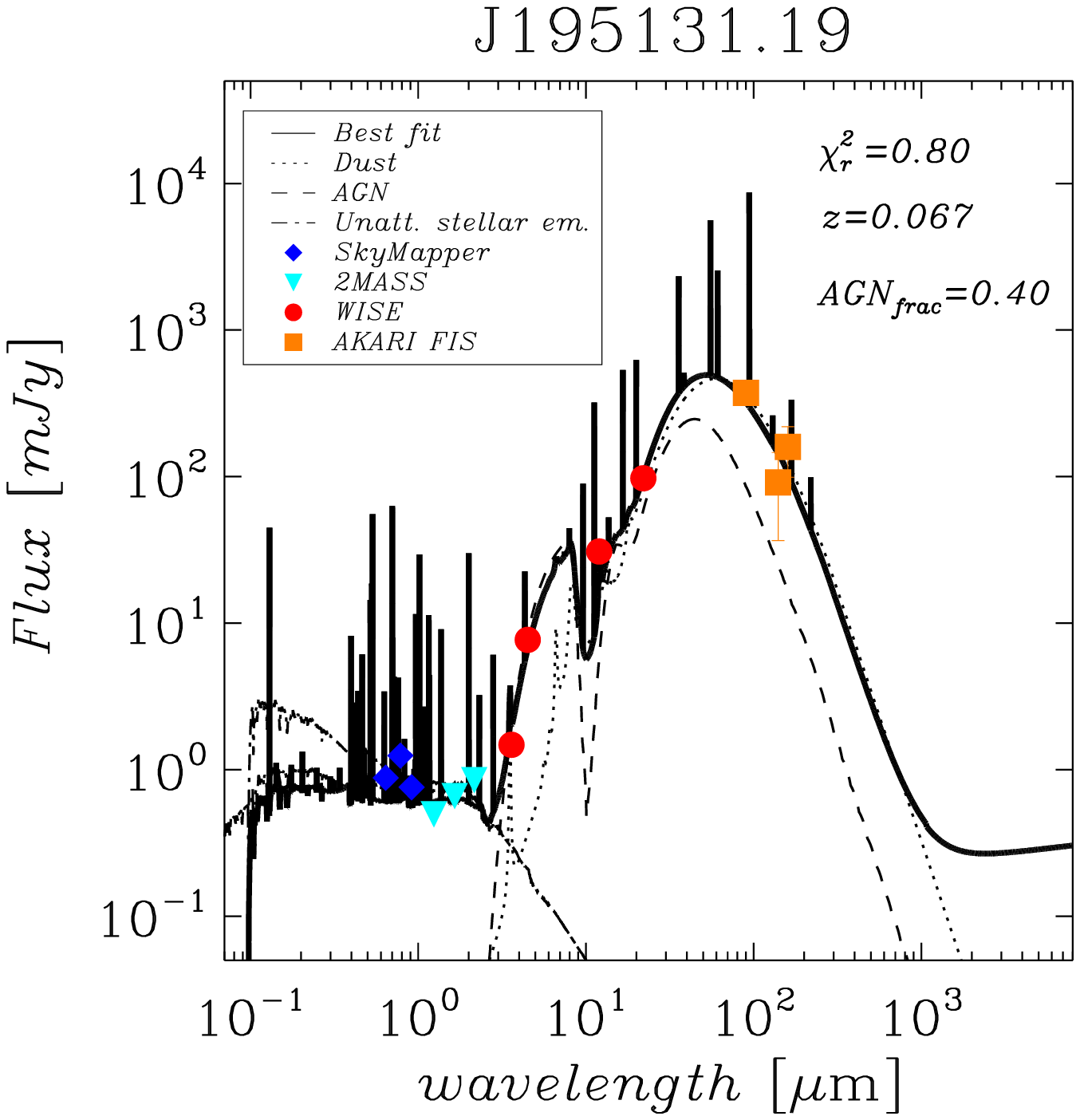}
}
\hspace{0mm}
\subfloat{
\includegraphics[width=85mm]{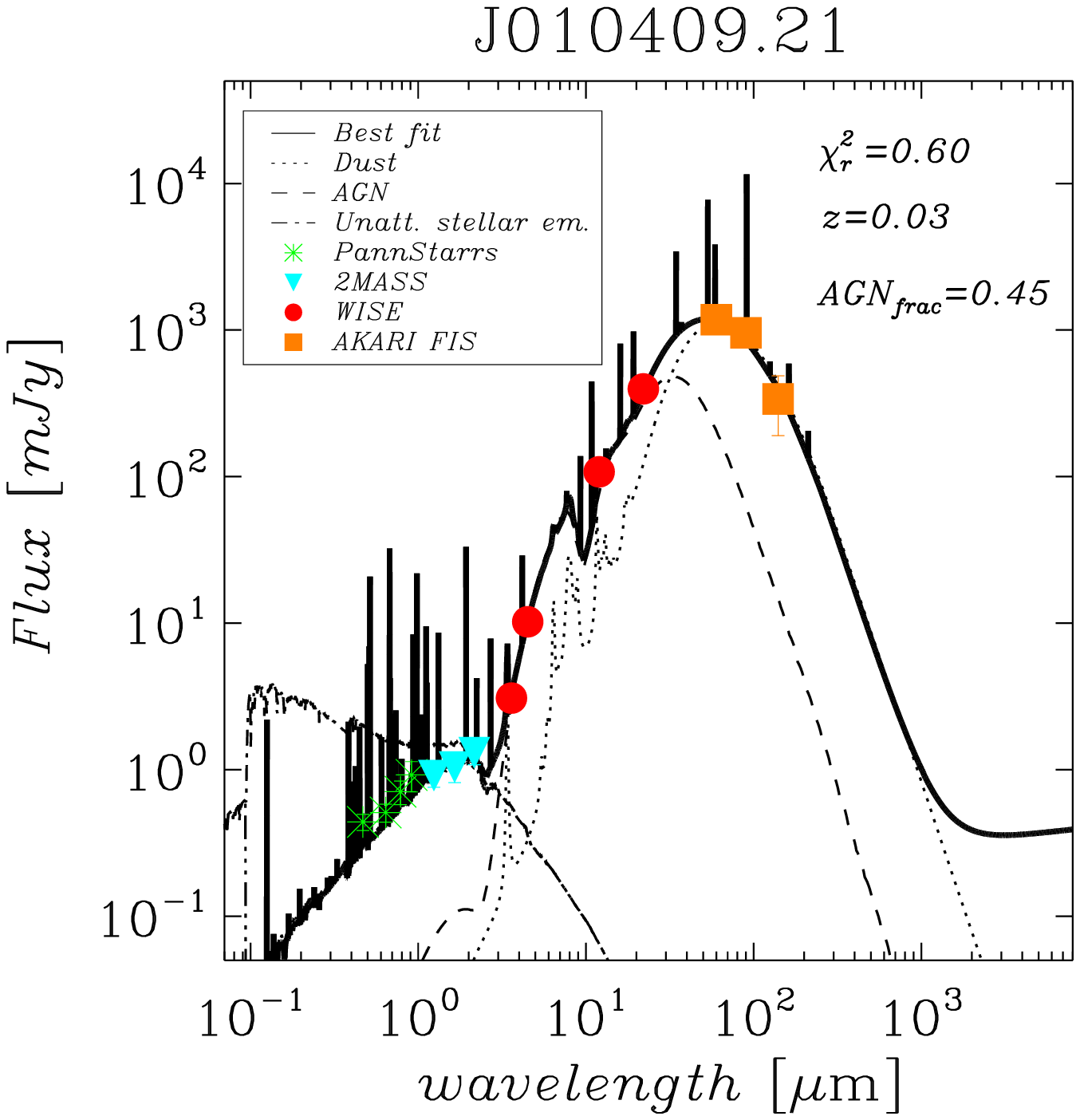}
}
\subfloat{
\includegraphics[width=85mm]{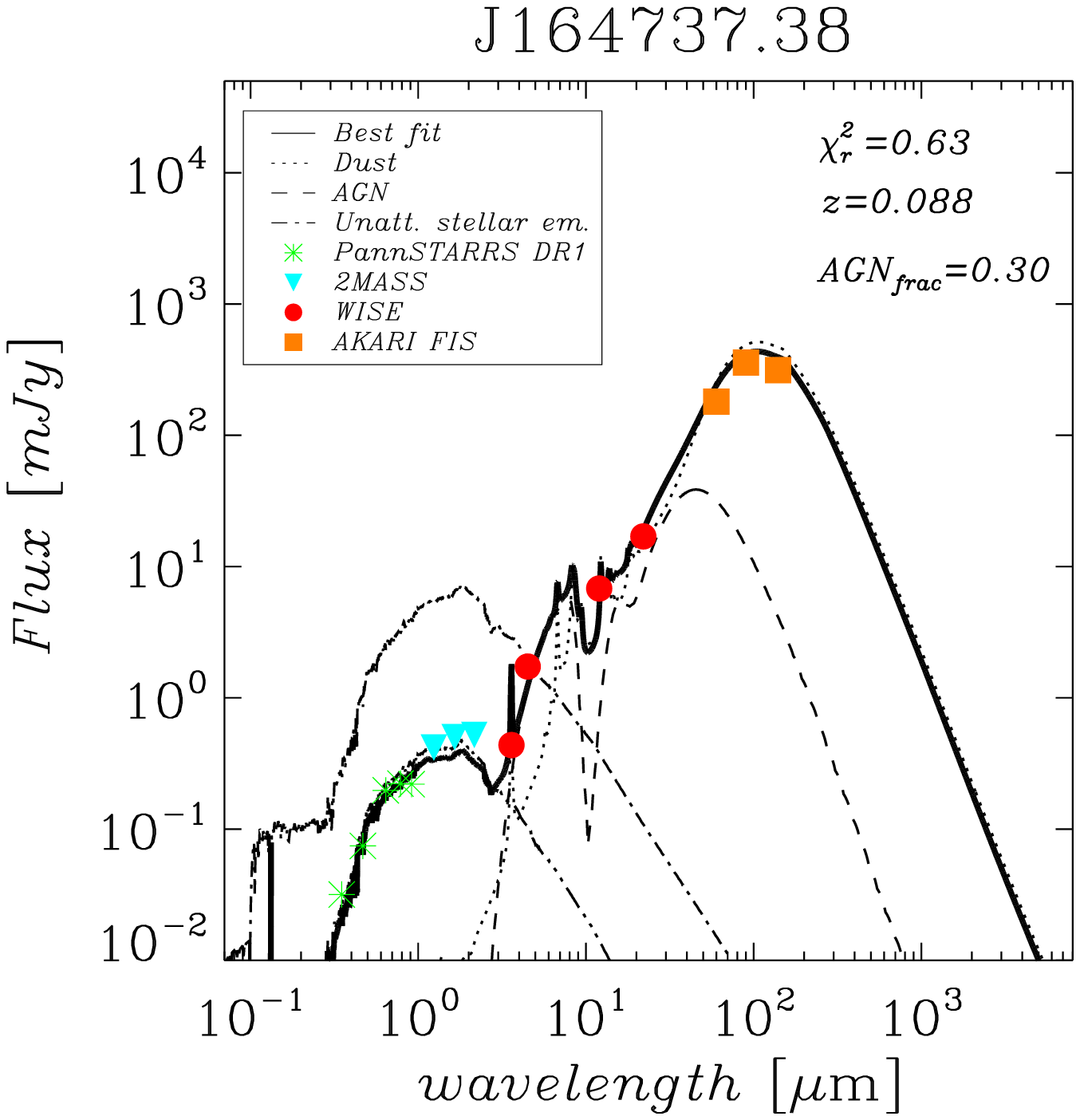}
}
\caption{Examples of SED fitting for four of the observed galaxies, with detected FIR emission in AKARI survey. Error bars are smaller than the data points.}
\label{seds}
\end{figure*}

Table~\ref{fritz} shows the physical parameters derived from the fitting procedure and Fig.~\ref{seds} shows SED fit plots for four red galaxies from the OCSVM sample. SEDs for the remaining five objects are shown in Appendix~\ref{app2} in Fig.~\ref{cdf}.
 SED fitting shows that the AGN component dominates on short wavelengths 3.4 and 4.5 $\mu$m, while cold dust models explain the emission at longer wavelengths well.
 We find a significant AGN fraction (15\% and higher) for most of the sources  despite the lack of typical AGN features in their optical spectra.
In addition, we find that eight galaxies (J195131.19, J204544.00, J141606.17, J010409.21, J071734.84, J052522.78, J205821.02, and J022718.11) have stellar masses estimated as lower than that of the Large Magellanic Cloud ($M_{*}<3 \times 10^{9} M_{\odot}$, \citealt{lmc}).
As the possible AGNs may be heavily obscured, we tested whether assuming average values for the AGN contribution is justified. To this aim, we released the geometrical parameters of the AGN model (outer-to-inner radius ratio, radial and angular dust distribution, and the opening angle) while retaining the previously defined grid of other parameters for each object. 
We find that while the best fit SED models with free AGN parameters tend to have different values than the fixed ones, the resultant physical properties do not differ in a significant way.
The differences in the SFR, $L_{Dust}$ and $M_{*}$ are $\sim 0.091$ dex, $\sim 0.053$ dex, and $\sim 0.320$ dex, respectively. 
Releasing the AGN parameters has the most significant impact on the stellar mass estimation. However, these results do not impact the final classification.

\subsection{Broad Absorption Line QSOs}

\begin{figure}[ht]
\centering
\includegraphics[width=\hsize]{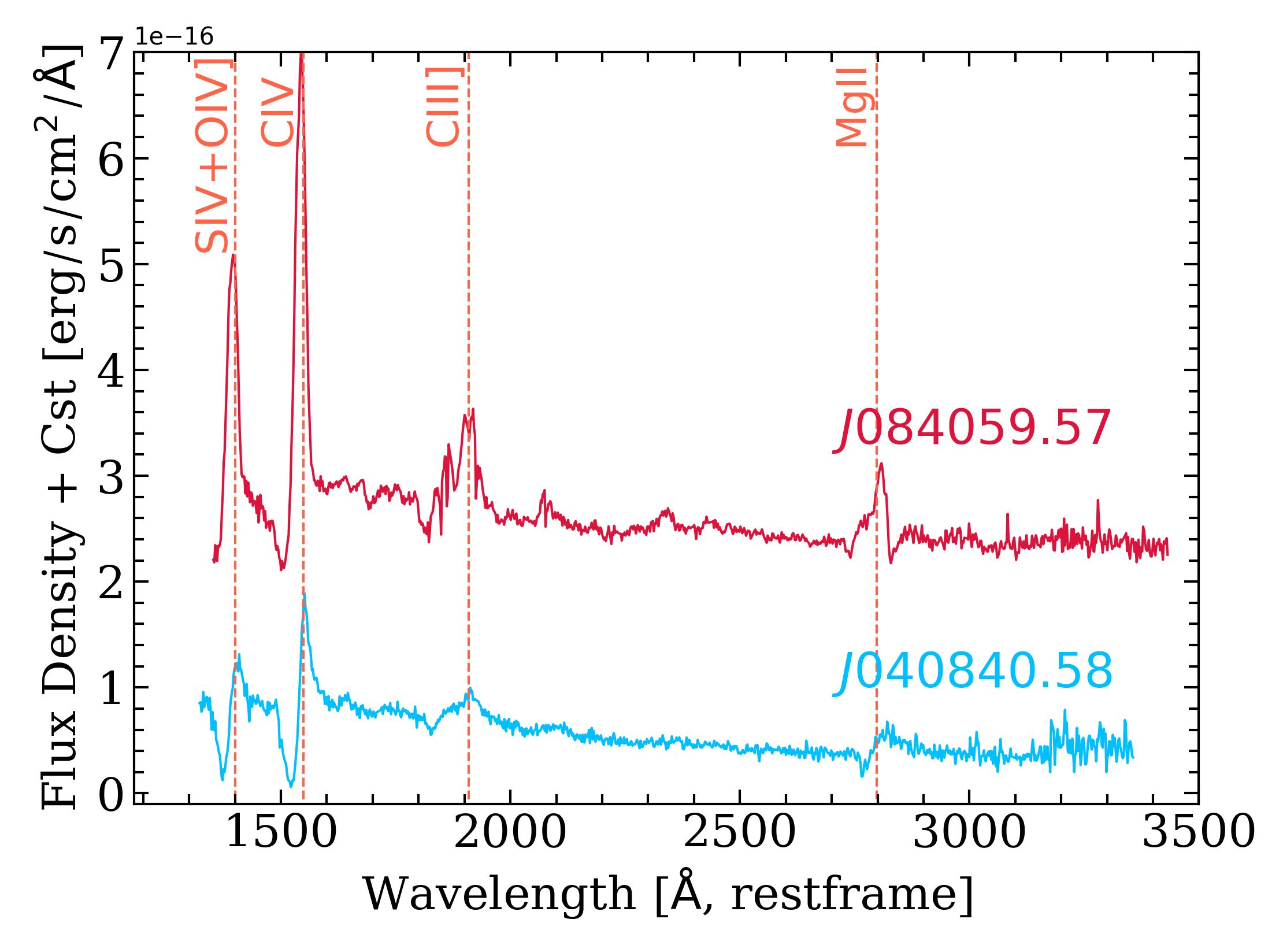}
\caption{Two LoBAL QSOs detected during the observations: J040840.58 and J084059.57}
\label{bals}
\end{figure}%

\begin{figure}[ht]
  \centering
\includegraphics[scale=0.55]{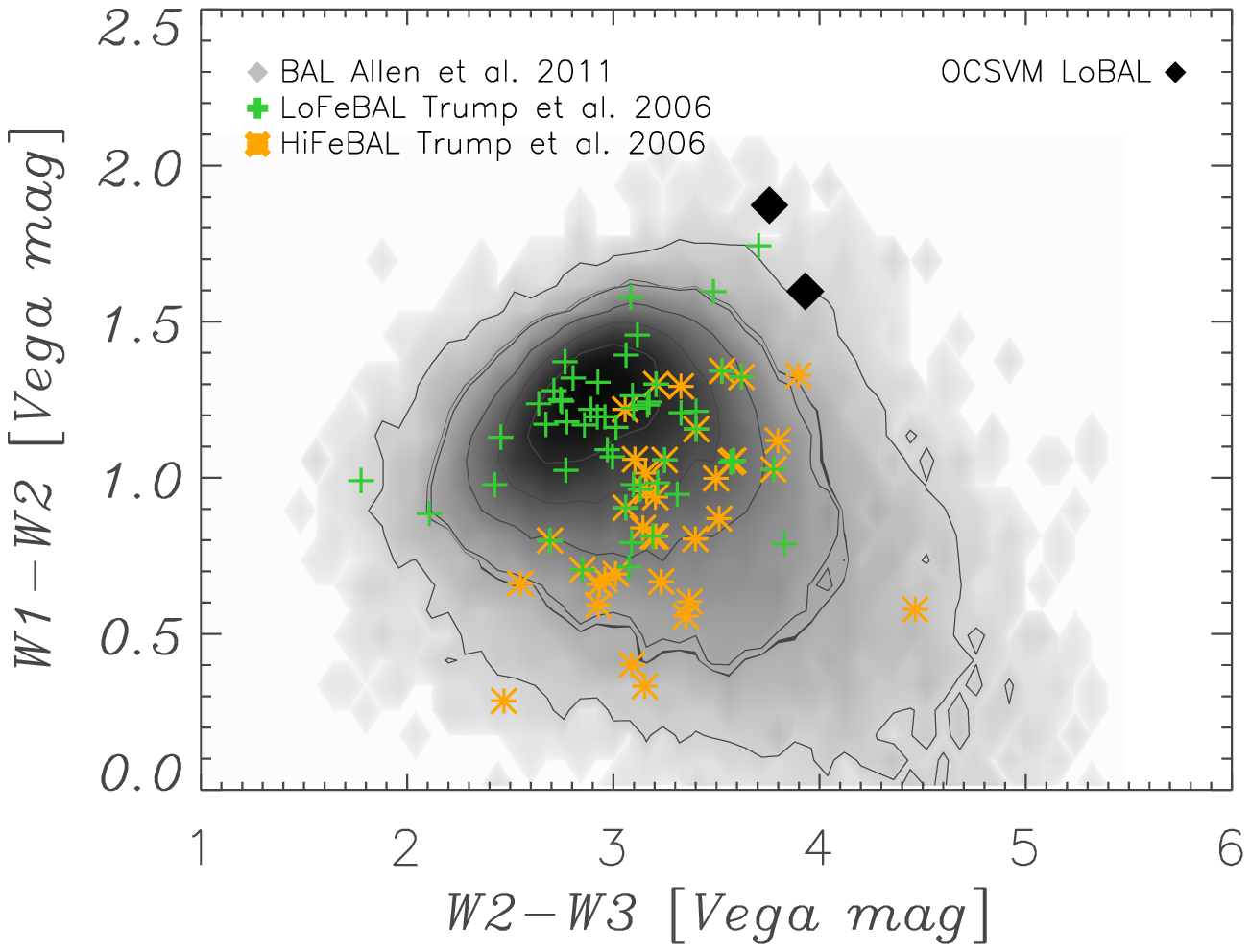}

  \caption{Colour-colour diagram of observed LoBAL objects from the OCSVM catalogue (LoBALs marked as filled diamonds) compared to the BAL catalogue of \citet{allen11} marked by grey contours and samples of HiBAL and LoBAL from \citet{trump06}.}
\label{ccbal}
\end{figure}

Among the observed OCVSM anomalies, we identified two BAL QSOs (Fig.~\ref{bals}).
Overall, BAL QSOs are observed in 15\% of the optically selected QSOs (e.g. \citealt{hewett03}, \citealt{gibson09}), but the intrinsic fraction might be as high as $\sim$ 40\% \citep{allen11}. Also, 15\% of all BAL QSOs exhibit absorption lines from low-ionisation species such as \ion{Al}{iii} and \ion{Mg}{ii} in addition to absorption lines from more highly ionised species like \ion{C}{iv} and \ion{Si}{iv}$+$\ion{O}{iv}]. These are referred to as LoBAL QSOs.
Two observed objects, J040840.58 and J084059.57, show absorption features from low-ionisation species. 
We compare the WISE colour distribution of the observed BAL QSOs to the publicly available catalogues (see Fig.~\ref{ccbal}) of such sources from \citet{allen11} and \citet{trump06} selected from the SDSS survey. 
SDSS QSO spectroscopic selection algorithm does not include the reddest QSOs \citep{allen11}.
We can see that the BAL QSOs selected by the anomaly catalogue have much redder MIR colours than the majority of the objects of this class reported in the literature.
This could indicate higher dust masses contained within these QSOs; however, without additional FIR data, we cannot verify this assumption. 
 
\subsection{Type I QSOs}

\begin{figure}[ht]
  \centering
\includegraphics[width=9cm]{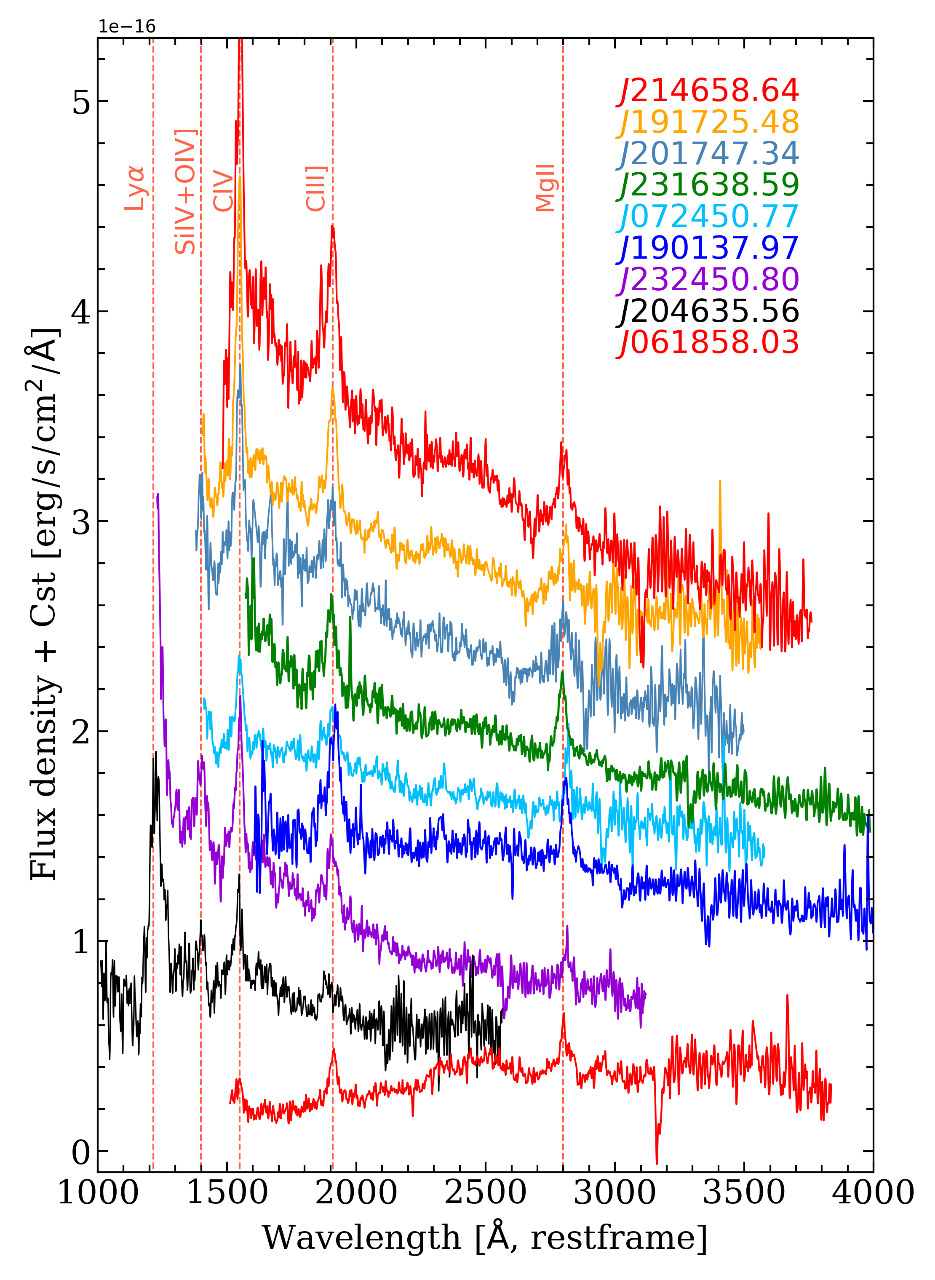}

  \caption{Spectra of the observed Type 1 QSOs shifted to rest-frame wavelength. The spectra were shifted vertically by small offsets for clarity.}
\label{type1}
\end{figure}%
Figure~\ref{type1} presents the spectra of 9
QSOs observed from the OCSVM sample, classified as Type 1 based on
the broad emission lines seen in their spectroscopy. Where measured, the mean line widths\footnote{all FWHMs values are given in rest-frame.} are FWHM(\ion{Mg}{ii}) $\sim$ 2243.31 $\pm $ 76.79 km$\rm s^{-1}$ and FWHM(\ion{C}{iv}) $\sim$ 2430.98 $\pm$ 42.33  km$\rm s^{-1}$.
The spectra of eight objects show blue continua while one (J061858.03) is reddened and shows strong iron emission.
Rest-frame UV spectra for these objects do not display any abnormal properties. The average $W1-W2$ colour of these sources is 1.3 - a typical value of the SDSS Type-I QSOs detected by AllWISE at similar redshifts (e.g. \citealt{assef10}, \citealt{stern12}). 

The reason why the algorithm failed while filtering out unobscured QSOs, which are present in the training sample, stems from its initial construction.
The initial training sample in S17 was created to include all objects found in both AllWISE and SDSS databases, even though SDSS contains 59\% galaxies, 18\% stars and 23\% of QSOs. Additionally, a $W1$ magnitude cut at $16$ [Vega mag] was applied to the training (matched databases of SDSS and WISE) and test (AllWISE full-sky only) sets, as the completeness of the galaxy sample drops off beyond this brightness limit. This action was performed to ensure uniform parameter space coverage by the known sources. It resulted in the exclusion of $\sim 61\%$ of all SDSS QSO, which severely underrepresented broadline QSOs in the training dataset. 
In the future, we will increase the weight of this particular class to compensate for the lower number of training examples of sources belonging to this class. 
\subsection{QSO with strong and narrow UV Fe emission}
\begin{figure*}[ht]
\centering

\includegraphics[width=\hsize]{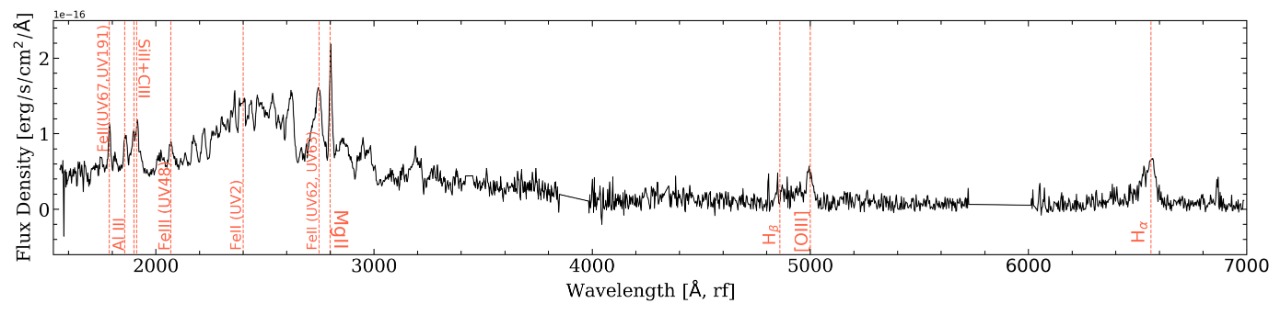}
\caption{Combined EFOSC2 and SofI spectra for J040754.75 QSO. 
} 
\label{lofebal}
\end{figure*}

We did not identify any new type of unusual QSO spectra. However, we detected one peculiar QSO, J040754.75, showing strong and narrow UV \ion{Fe} emission lines (Fig.~\ref{lofebal}). 
It exhibits emission from the multiplets \ion{Fe}{ii} $\lambda$1785 (UV67,UV191), \ion{Fe}{iii} $\lambda$1926 (UV34), \ion{Fe}{iii} $\lambda$2070 (UV48), \ion{Fe}{ii} $\lambda$2400 (UV2), \ion{Fe}{ii} $\lambda$2600 (UV1), and \ion{Fe}{ii} $\lambda$2750 (UV62,UV63). The positions of these lines, as well as the positions of the typical QSO emission lines (\ion{Al}{iii} $\lambda\lambda$1854.7,1862.8, \ion{C}{iii}] $\lambda$1908.7, \ion{Mg}{ii} $\lambda\lambda$2796.3,2803.5), are marked by the vertical lines.

\ion{The Fe}{ii} emission strongly influences the energy balance of broad emission line regions in AGNs. Currently, theoretical models are unable to fully reproduce the strength and shape of the observed iron complexes.
For that reason, the identification of new systems with both strong and narrow \ion{Fe}{ii} is important for improving the models and defining the areas of parameter space occupied by broad emission line regions.
Similar objects were found by \citet{hall02} and \citet{unusualqso} in the SDSS survey to a limited extent.

\subsection{Galactic objects}

\begin{figure*}[ht]
    \centering
    \includegraphics[scale=0.4]{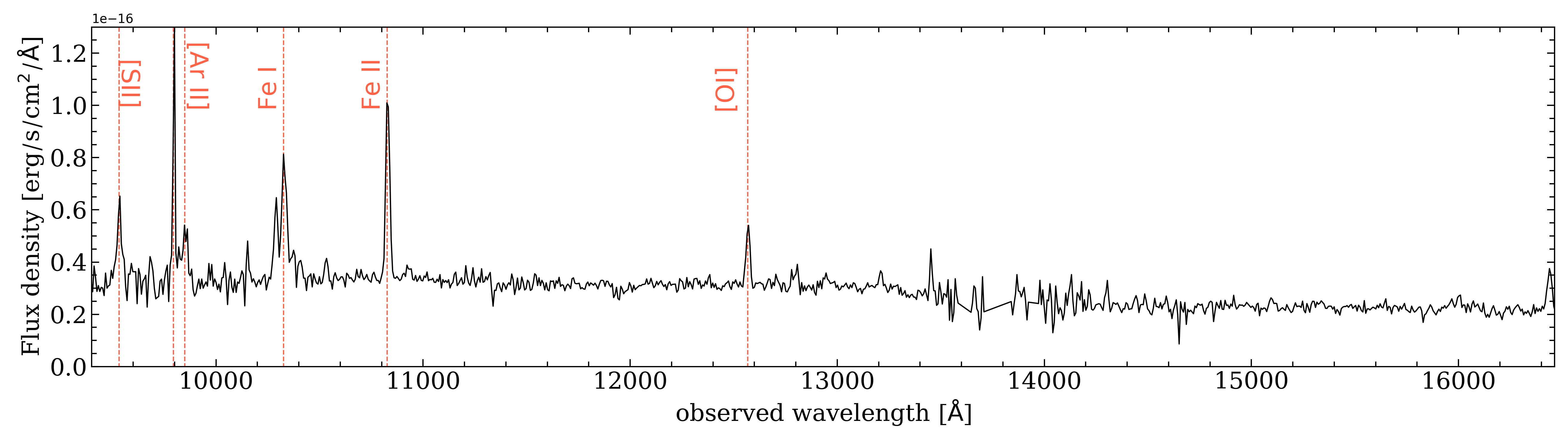}
    \caption{ Object J155603.92 was observed by SofI instrument only. No optical spectrum was obtained as the object is too faint to be targeted by EFOSC2. }
    \label{yso}
\end{figure*}

Object J155603.92 was observed with the SofI instrument only, as the optical flux is too small to be observed by EFOSC2. The NIR spectrum is shown in Fig.~\ref{yso}.
This particular object has a counterpart in Gaia DR2 (\citealt{gaia1}, \citealt{gaia2}) where its parallax is measured to be $\pi=4.2705\pm0.7170$ mas and with relatively high proper motion 
($\mu_{\alpha}cos \delta \sim 14$~mas yr$^{-1}$ and $\mu_{\delta} \sim 29$~ mas yr$^{-1}$), 
placing this object at a very close distance of 246$^{+64}_{-42}$ pc, according to \cite{bailer-jones18}.
Based on these measurements, we find that this object has a 60\% probability of being a part of Upper Scorpius OB association \citep{scorpius}.

The OCSVM code is sensitive to red galactic objects like YSOs, which often have extremely red $W1-W2$ colours due to the dusty cocoons in which stars are born and because the training sample was lacking a representative amount of objects of this type. SDSS DR13 does not cover the Galactic plane regions, and therefore no early stages of stellar evolution are present in the training set. This is an important issue, which must be addressed in the future creation of the training sample based solely on optical data.
To differentiate the galactic sources from the extragalactic ones based on photometry alone is to either manually remove the data falling inside the areas of the Galactic plane and bulge or to include a statistically significant sample of such objects from other spectroscopic surveys \citep[e.g.][]{fischer16,jones17}. 

\subsection{Unknown}
\begin{figure}[ht]
    \centering
    \includegraphics[scale=0.4]{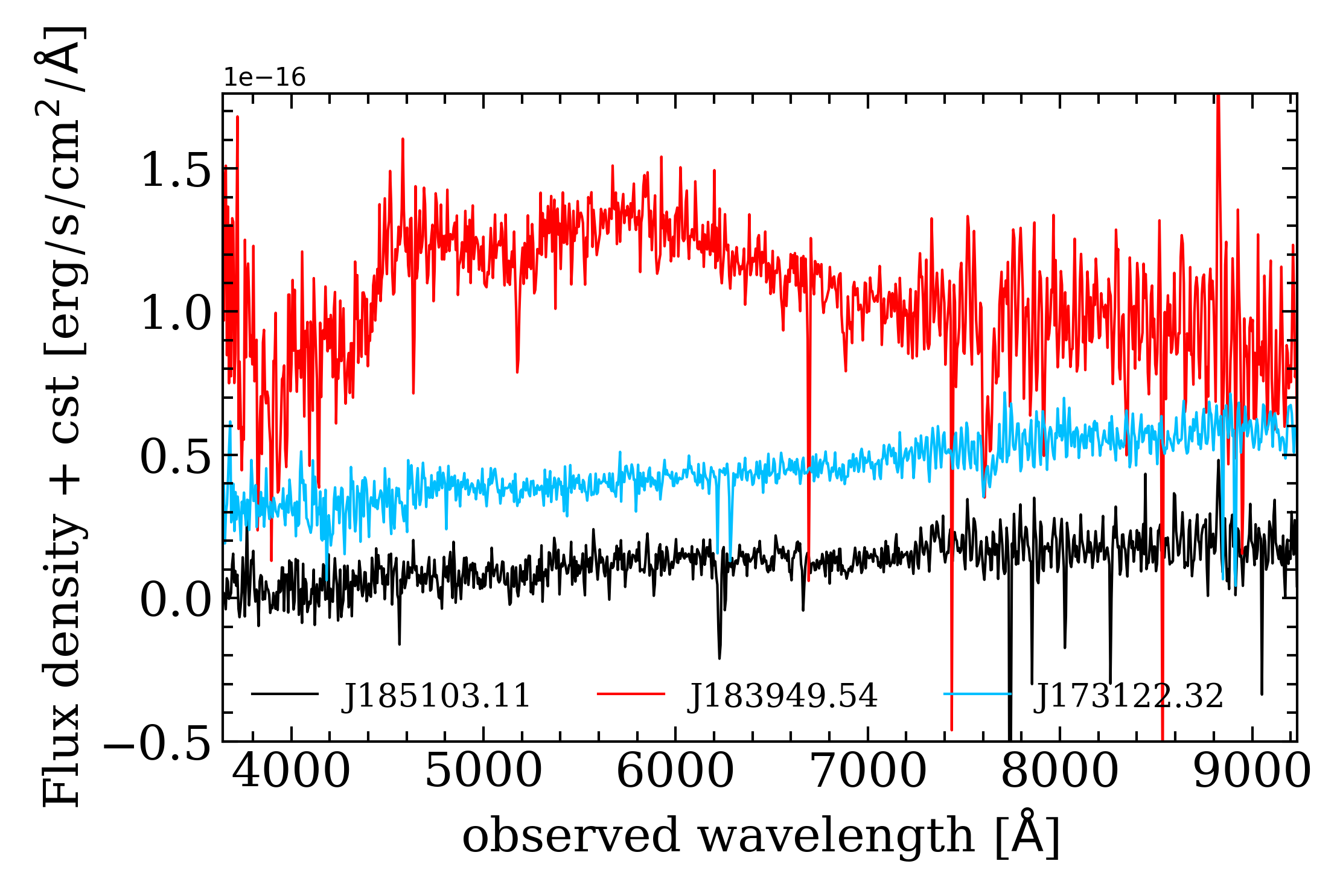}
    \caption{Spectra of objects with unidentified nature observed with EFOSC2: J183949.54, J173122.32, and J185103.11. }
    \label{unknownopt}
\end{figure}

\begin{figure}[ht]
    \centering
    \includegraphics[scale=0.4]{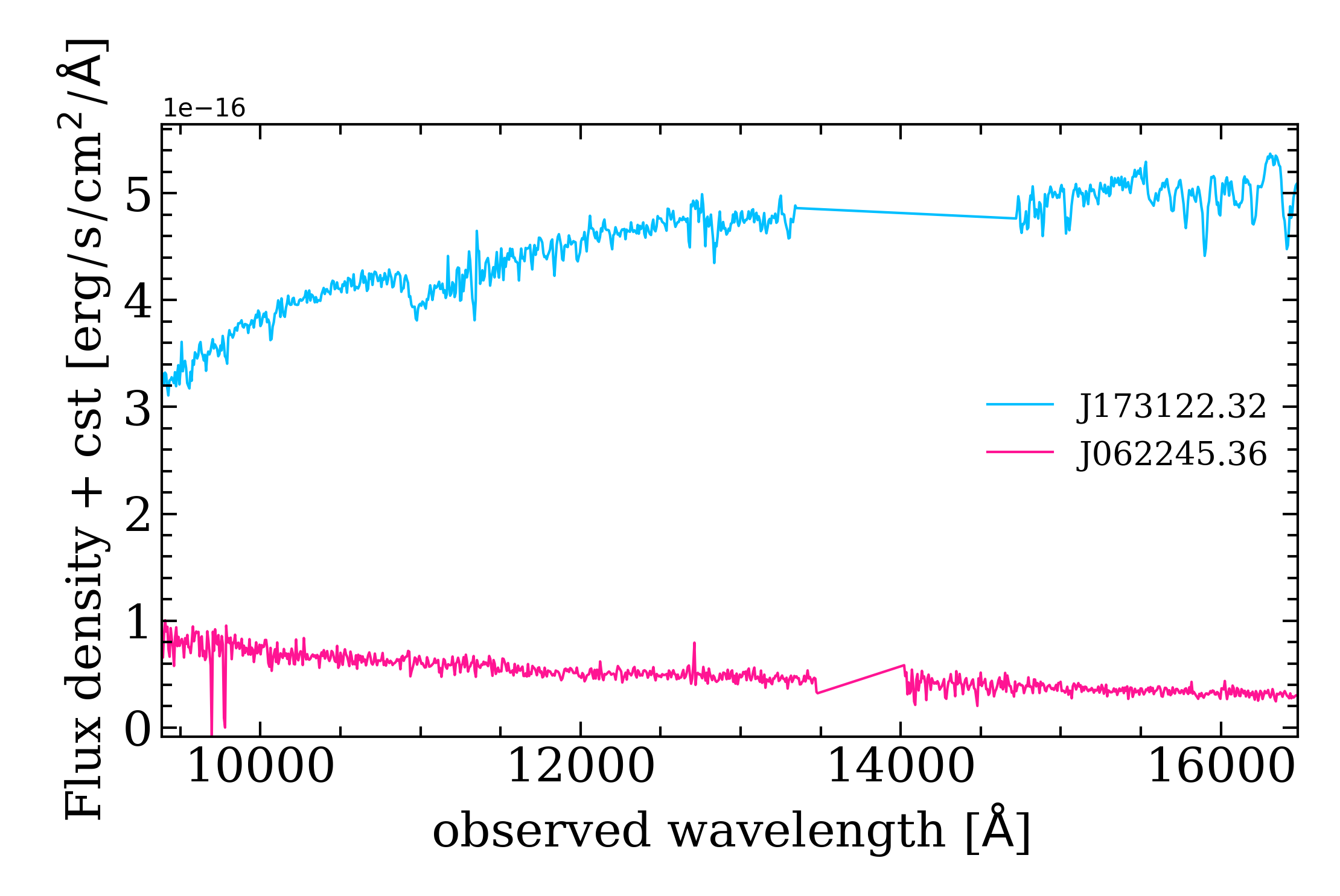}
    \caption{Spectra of two unidentified objects observed with SofI: J173122.32 and J062245.36.}
    \label{unknownir}
\end{figure}

Figures~\ref{unknownopt} and \ref{unknownir} show the four objects (J062245.36, J183949.54, J173122.32, and J185103.11) which we were not able to identify either due to  insufficient exposure time that resulted in low S/N or due to lack of any lines to identify these sources.
The optical spectrum of J173122.32 has a S/N of$~3$ and J185103.11 has a S/N of$~2$, which makes it impossible to identify them. 

Object J173122.32 was followed-up with the SofI instrument (see Fig.~\ref{unknownir}); however, the spectrum shows no additional clues of what the nature of this source is. 
This source is detected in Gaia DR2 with proper motions ($\mu_{\alpha}cos \delta \sim 4.8$~mas  yr$^{-1}$, $\mu_{\delta} \sim 2.4$~ mas yr$^{-1}$) as well as the distance estimate from \cite{bailer-jones18} (5.5$^{+4.0}_{-2.5}$ kpc) that is consistent with the Galactic bulge. It is located in (l$\sim4$ deg, b$\sim6$ deg).

Object J183949.54 has a $\mathrm{S/N }\sim 10$, however, there are no apparent clues about its nature visible in the spectrum. The 
Gaia DR2 catalogue matches two sources to this object, separated by about three arcsec, hence probably not resolved in AllWISE.
The first Gaia source (ID=6655455826956741248 of G=18.9 mag) is probably a Galactic disk source (b$\sim$19 deg) with 
proper motions of $\mu_{\alpha}cos \delta \sim 4.3$~mas yr$^{-1}$ and $\mu_{\delta} \sim 3.0$~ mas yr$^{-1}$ and distance estimate of 4.0$^{+3.4}_{-1.9}$ kpc from \cite{bailer-jones18}.
The second Gaia source, (ID=6655455831252327424) is fainter in Gaia (G=20 mag); however, its parallax is large, $\pi=$1.48$\pm$0.87 mas, indicating possibly a nearby source (2.7$^{+3.3}_{-1.8}$ kpc) with the lower bound for a distance of only 850 pc \citep{bailer-jones18}. This could indicate a possible nearby brown dwarf, however, given the faintness and possible blending, it is too early to conclude on its true nature. 

Object J062245.36 was observed with SofI only, as the lunar phase and proximity of the object to the Moon did not allow for optical observations.
The spectrum shows a blue continuum slope and no emission/absorption lines.
This may be another low-redshift galaxy, as in NIR we do not observe any spectral lines beyond H$\alpha$.
On the other hand, its Gaia DR2 proper motions ($\mu_{\alpha}cos \delta \sim 3.6$~mas yr$^{-1}$, $\mu_{\delta} \sim 3.1$~ mas yr$^{-1}$) and parallax ($\pi=0.44\pm0.09$) are very well constrained and indicate a Galactic source at 2.0$^{+0.4}_{-0.3}$ kpc. 

These objects could appear in the anomaly catalogue because the training sample was constructed without including signatures of Galactic objects. We expect a significant contamination of the anomaly catalogue with the Galactic objects in dusty phases of their evolution.
\section{Summary}

We  performed pilot spectroscopic follow-up observations of objects catalogued as 'anomalous' by the OCSVM algorithm applied to the IR photometry from the AllWISE survey. 
We observed 36 objects with a limiting brightness of $g_{\rm AB}<19.5$ [mag] and found several different source populations. We describe our observations in the following points: 

    \begin{enumerate}
    \item Nineteen 'red' ($W1-W2>1.7$) local galaxies containing hot dust (due to either an AGN or extreme starburst contribution):
        \begin{itemize}
            \item three WR galaxies
            \item six regular SF galaxies
            \item ten galaxies showing no or only one emission line (H$\alpha$)
        \end{itemize}
    \item Two LoBALs 
    \item one QSO with strong and narrow UV Fe emission    \item one YSO
    \item nine regular broad-line type I QSO    \item four sources with S/N too low to disentangle their nature    \end{enumerate}

Among the nineteen galaxies with extreme mid-infrared colour excess, we found seven low-mass objects which do not display any optical signatures of AGN activity. SED modelling shows that mid-infrared colour excess can be well explained by hot dust emission from an AGN. This hypothesis does not exclude the existence of extreme starbursts, which could produce similar colours in low-metallicity environments. 
Such systems remain elusive and of significant astrophysical interest, as
pinpointing AGNs in low-mass galaxies can give us insights about the primordial black holes that were formed in the early Universe. 
 'Seed' black holes must exist at high redshifts; however, finding such objects is a challenge. Nearby dwarf galaxies are thought to be analogues of low-mass black hole hosts \citep{reines16} and can be used as proxies for studying seed black hole creation.
Hence, anomaly detection algorithms could prove useful for finding such rare objects for future statistical studies. 
Many properties of the rest of the observed 'red' galaxies remain unresolved, as without FIR data we cannot reliably perform the SED fitting for this purpose. However, optically regular galaxies with extreme NIR colours are a rare population: according to \citet{oconnor16}, 0.2\% of all SDSS galaxies display this property. We found three galaxies with emission lines produced by Wolf-Rayet stars - a brief and violent stage of galaxy evolution. These low-metallicity objects can produce red MIR colours due to violent SF processes alone.
Additionally, we have found one truly peculiar, but not unknown object: a QSO with strong UV iron emission.

In conclusion, the OCSVM code selected rare objects, but several underrepresented, yet well-known, classes of sources were still included in the output catalogue. Due to the lack of sufficiently representative quantities and parameter space coverage, both unobscured QSOs (red and blue in terms of the shape of the continuum) and Galactic objects were flagged as anomalies.
Training data preparation is one of the most critical steps in supervised and semi-supervised machine learning schemes. The construction of a 'one-class' training set must include more detailed treatment of underrepresented source populations in comparison to the other classes (i.e. QSOs vs stars and galaxies in case of SDSS). Otherwise, some regions of the parameter space which are occupied by these more sparse objects might be treated as areas containing outliers. 
Moreover, when creating a training set of objects within an IR-based survey based on the optical identifications, 'regular' IR source populations are bound to appear as outliers. Prominent examples of such sources are asymptotic giant branch (AGB) stars, YSOs, \ion{H}{ii} regions among  Galactic sources with IR excess. 
Depending on its future application, the training sample should be created by either excluding the low Galactic latitudes or by including more substantial amounts of such objects in the training sample. 
Such measures were not taken in S17, which led to the detection of these classes in the spectroscopic follow-up observations.
It is currently impossible to determine if such anomalies are propagated throughout the whole catalogue, as the spectroscopic observations were performed for just a small fraction of the full anomaly catalogue.
In the next step of this project, we will focus on an iterative search for truly rare objects. For this purpose, the training sample needs to be recreated by including (i) rare but not unknown objects (i.e. LoBALs); (ii) Galactic objects in the dusty phase of their evolution; and (iii) a weighted selection of broadline QSOs to increase the impact of this class on the algorithm learning process.
The study of the behaviour of the code with respect to training sample composition will be addressed in Solarz et al. (in prep).

\begin{appendix}

\section{Observing logs for the spectroscopic observations.}
\label{app1}
Here, we provide the observing logs for EFOSC2 (Table~\ref{listaobiektowexp}) and SofI (Table~\ref{listaobiektowexpsofi}) observations.

\begin{table*}
\begin{footnotesize}
\begin{center}
\caption{EFOSC2 observing log for the spectroscopic observations. Columns contain data as follows: starting times, exposure times, starting and ending airmasses, and slit position angles for each exposure are listed on separate successive lines.}
\label{listaobiektowexp}
\setlength{\tabcolsep}{3.5 mm} 
\begin{tabular}{cccc}
\hline
\hline
Object ID & UT at start of obs. / yyyy-mm-ddThh:mm:ss & Exp. / s & airmass \\
\hline
J232450.80&2018-08-29T07:28:47.984&2700&1.094-1.194\\
J201747.34&2018-08-30T02:37:33.719&2700&1.075-1.092\\
J191725.48&2018-08-30T23:31:23.723&2700&1.384-1.339\\
J214658.64&2018-08-30T04:33:11.646&2700&1.262-1.290\\
J204635.56&2018-08-30T04:22:25.801&2700&1.139-1.173\\
J190137.97&2018-08-28T03:10:06.460&2700&1.257-1.331\\
J231638.59&2018-08-29T05:45:39.234&2700&1.039-1.061\\
J084059.57&2018-12-06T07:06:57.812&2535&1.057-1.017\\
J040840.58&2018-12-06T04:13:56.953&2700&1.026-1.025\\
J061858.03&2018-12-06T04:21:09.302&2700&1.225-1.143\\
J040754.75&2018-12-05T03:52:24.945&2700&1.236-1.246\\
J072450.77&2018-12-05T05:43:42.439&2700&1.337-1.306\\
J211637.73&2018-08-29T01:13:13.994&1800&1.152-1.091\\
J195131.19&2018-08-30T00:22:37.455&1800&1.323-1.295\\
J204544.00&2018-08-28T04:04:35.377&2700&1.031-1.087\\
J141606.17&2018-08-28T23:38:21.992&1800&1.295-1.428\\
J164737.38& 2018-08-29T00:20:54.295&2700&1.203-1.320\\
J010409.21&2018-08-29T06:47:03.921&1800&1.037-1.030\\
J185020.05&2018-08-30T01:01:54.922&1800&1.079-1.085\\
J064043.96&2018-12-06T07:58:01.625&1800&1.156-1.231\\
J071734.84&2018-12-05T07:29:32.988&2700&1.108-1.134\\
J060127.78&2018-08-29T09:13:08.802&2500&1.319-1.220\\
J022908.52&2018-12-06T01:20:51.724&2700&1.252-1.170\\
J191825.52&2018-08-29T01:51:39.432&1800&1.358-1.368\\
J053520.86&2018-12-05T07:21:27.381&2700&1.044-1.118\\
J040051.83&2018-12-06T02:36:06.998&2700&1.301-1.245\\
J062035.99&2018-12-05T03:00:31.713&2700&1.361-1.255\\
J070647.77&2018-12-05T04:45:21.550&2700&1.137-1.064\\
J052522.78&2018-12-06T05:12:53.551&2700&1.080-1.103\\
J183949.54&2018-08-30T01:44:41.830&2700&1.092-1.136\\
J205821.02&2018-08-28T01:57:08.425&2700&1.091-1.044\\
J022718.11&2018-12-06T01:45:06.816&2700&1.252-1.247\\
J083104.69&2018-12-06T06:35:28.730&1800&1.138-1.077\\
J185103.11&2018-08-27T23:52:39.108&1800&1.072-1.039\\
\hline
\end{tabular}
\end{center}
\end{footnotesize}
\end{table*}

\begin{table*}
\begin{footnotesize}
\begin{center}
\caption{SofI observing log for the spectroscopic observations. Columns show starting times, exposure times, and starting and ending airmasses. Slit position angles for each exposure are listed on separate successive lines.}
\label{listaobiektowexpsofi}
\setlength{\tabcolsep}{3.5 mm} 
\begin{tabular}{cccc}
\hline
\hline
Object ID & UT at start of obs. / yyyy-mm-ddThh:mm:ss & Exp. /s & airmass \\
\hline
J155603.92&2018-08-27T00:04:37.9206&4450&1.096-1.424\\
J173122.32&2018-08-27T01:55:27.9576&4450&1.132-1.347\\%
J040754.75&2018-12-07T00:40:51.6056&4450&1.441-1.293\\
J062245.36&2018-08-27T08:49:35.3248&4450&1.518-1.195\\
\hline
\end{tabular}
\end{center}
\end{footnotesize}
\end{table*}

\section{SED fitting results and reliability check}
\label{app2}

CIGALE is using reduced $\chi^2$ ($\chi^2$/$N_{data}$, hereafter $\chi^2_{r}$). Linear models can be used to estimate the number of degrees of freedom. With nonlinear models, the number of degrees of freedom is nontrivial and whether or not it can be calculated properly is questionable (e.g. Charlot 2016). The best model can be selected by the $\chi^2_{r}$ value for the galaxy from the grid of all models created by the input parameters. However, due to the varied number of observed fluxes and an unknown number of the free parameters, the $\chi^2_{r}$ alone is not the only estimator of the best fits. 
To test the reliability of the parameters, we create a mock catalogue of galaxies, where each object is generated from the best-fit model to the real objects \citep[e.g.][]{giovannoli11, lofaro17}. The mock fluxes are created by deviating the flux of the best-fit model in each passband with a random Gaussian error. Finally, we run CIGALE on the simulated sample using the same set of input parameters as for the original catalogue and compare the physical output parameters of the artificial catalogue with the real ones. 

\begin{figure*}
\caption{Mock catalogue SED fitting results for each parameter evaluated in this work. The x-axis shows the model values, while the y-axis shows the SED fitting of the mock data. Standard error given by the Bayesian analysis is over-plotted as an error bar for each value.}
  \includegraphics[width=.6\columnwidth]{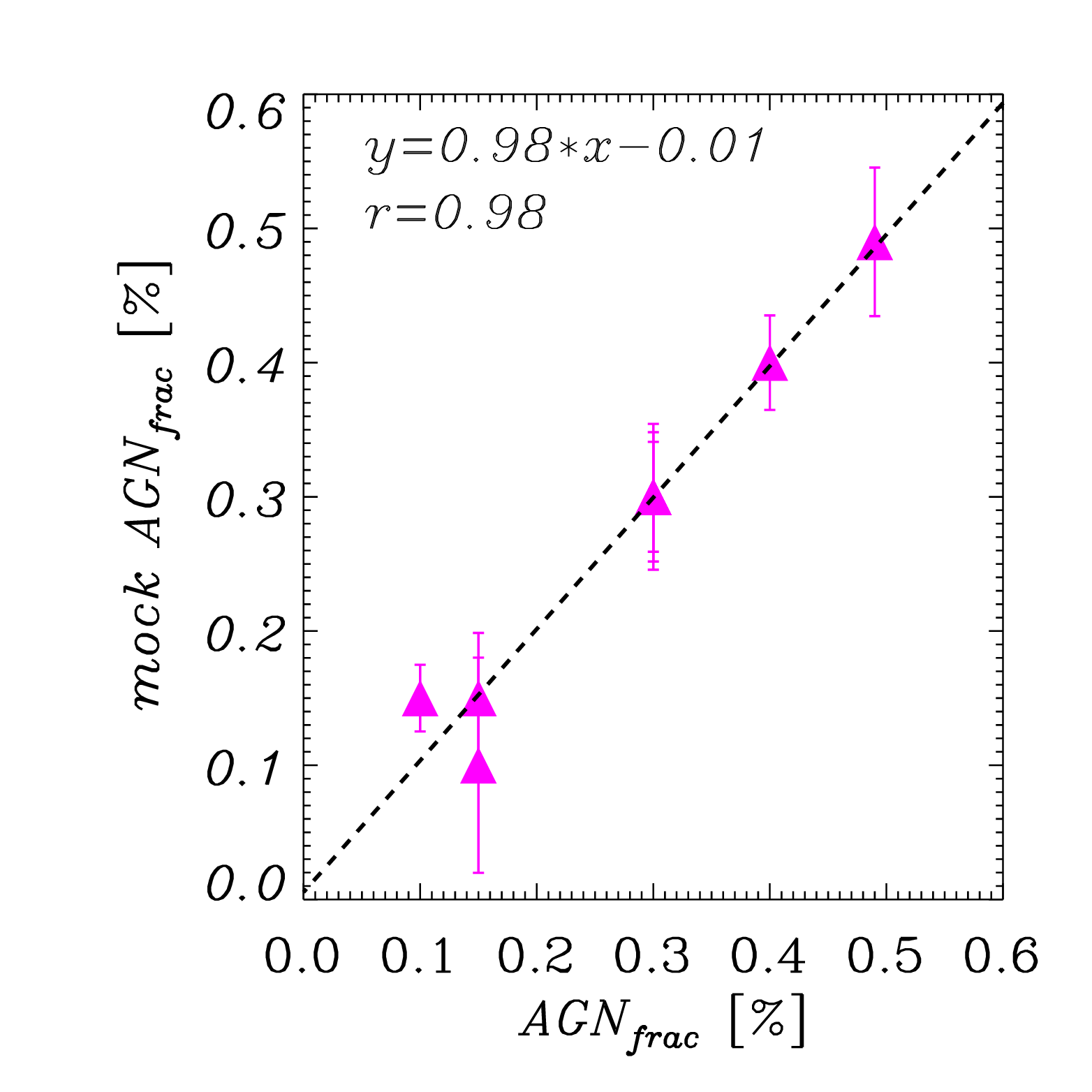}
  \includegraphics[width=.6\columnwidth]{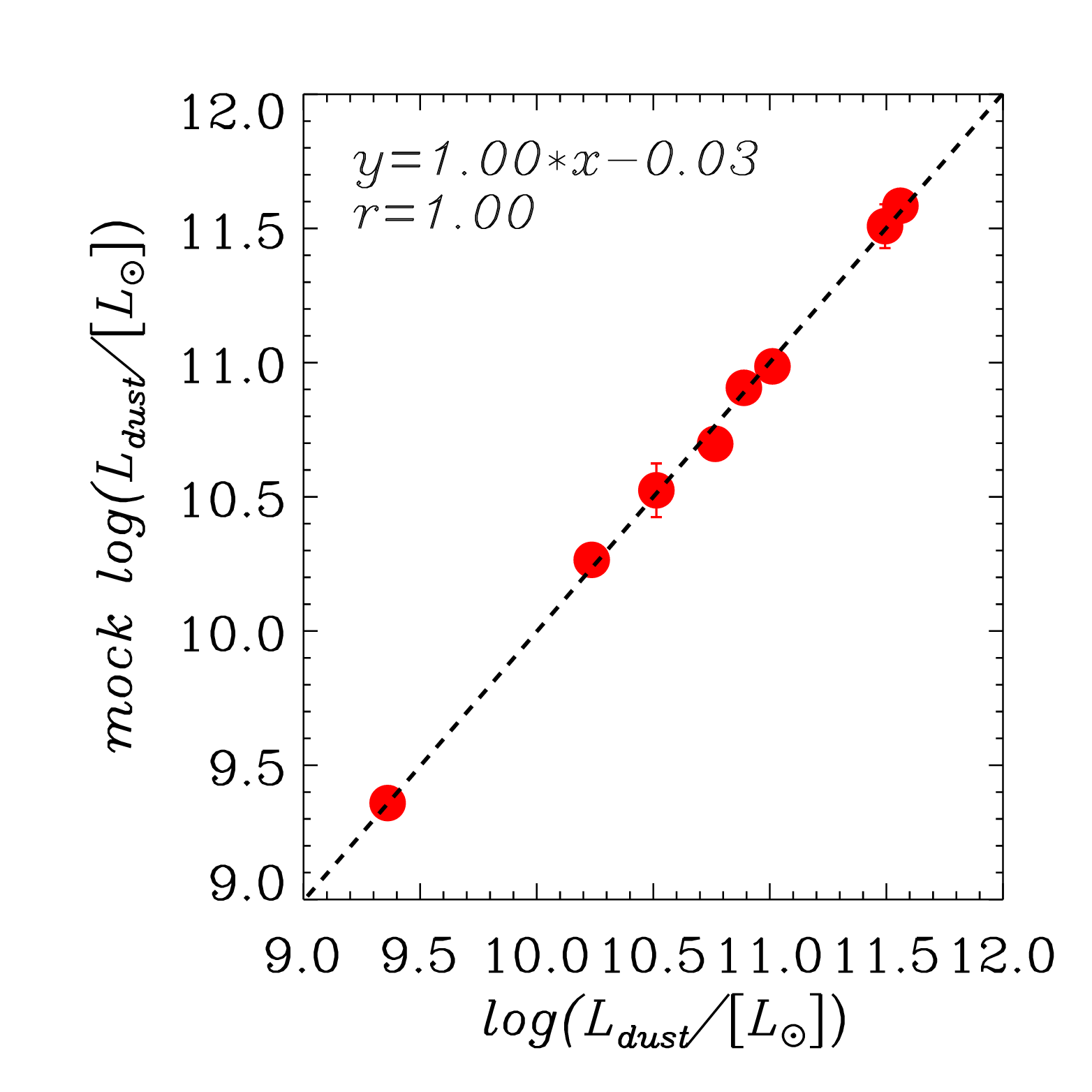}
  \includegraphics[width=.6\columnwidth]{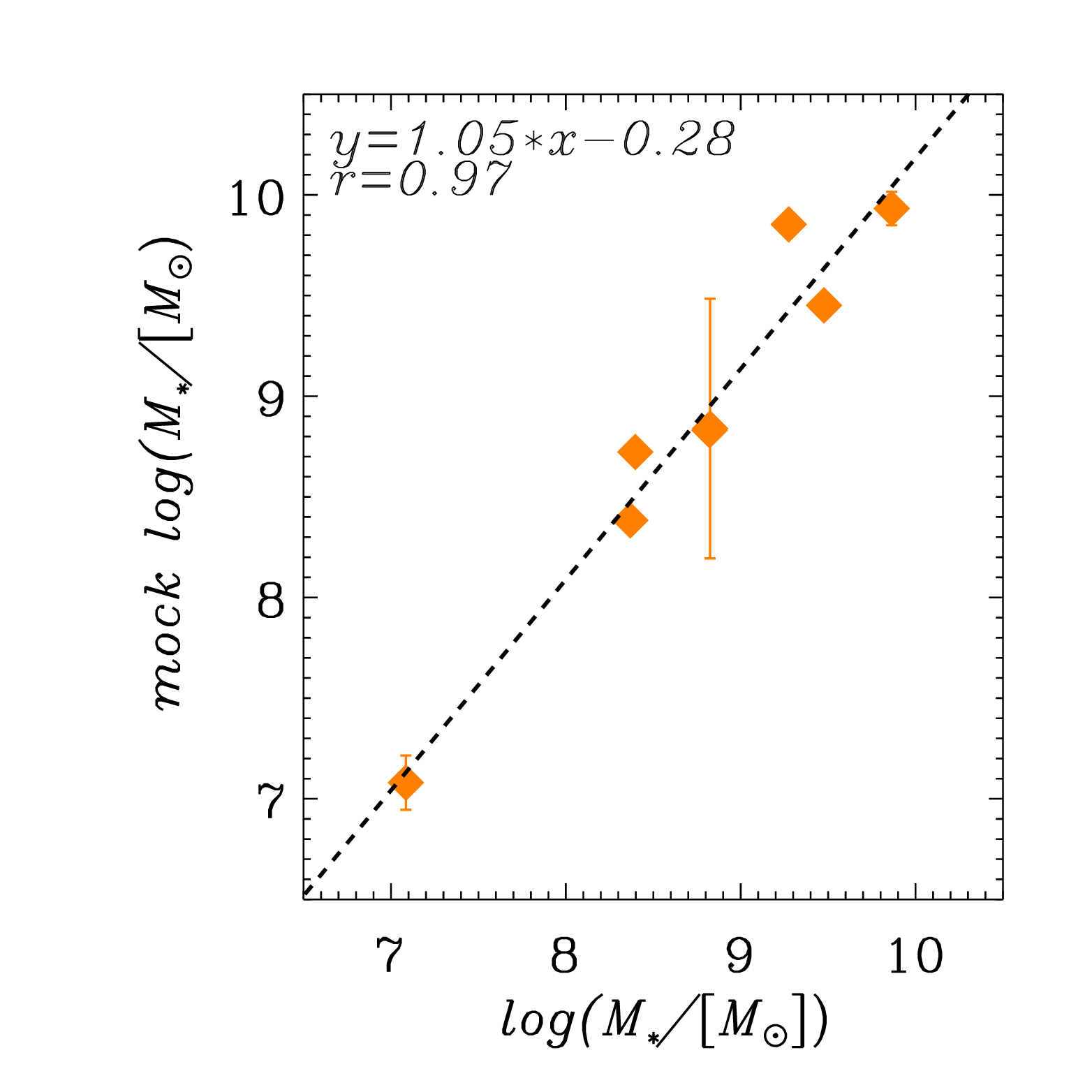}
  \includegraphics[width=.6\columnwidth]{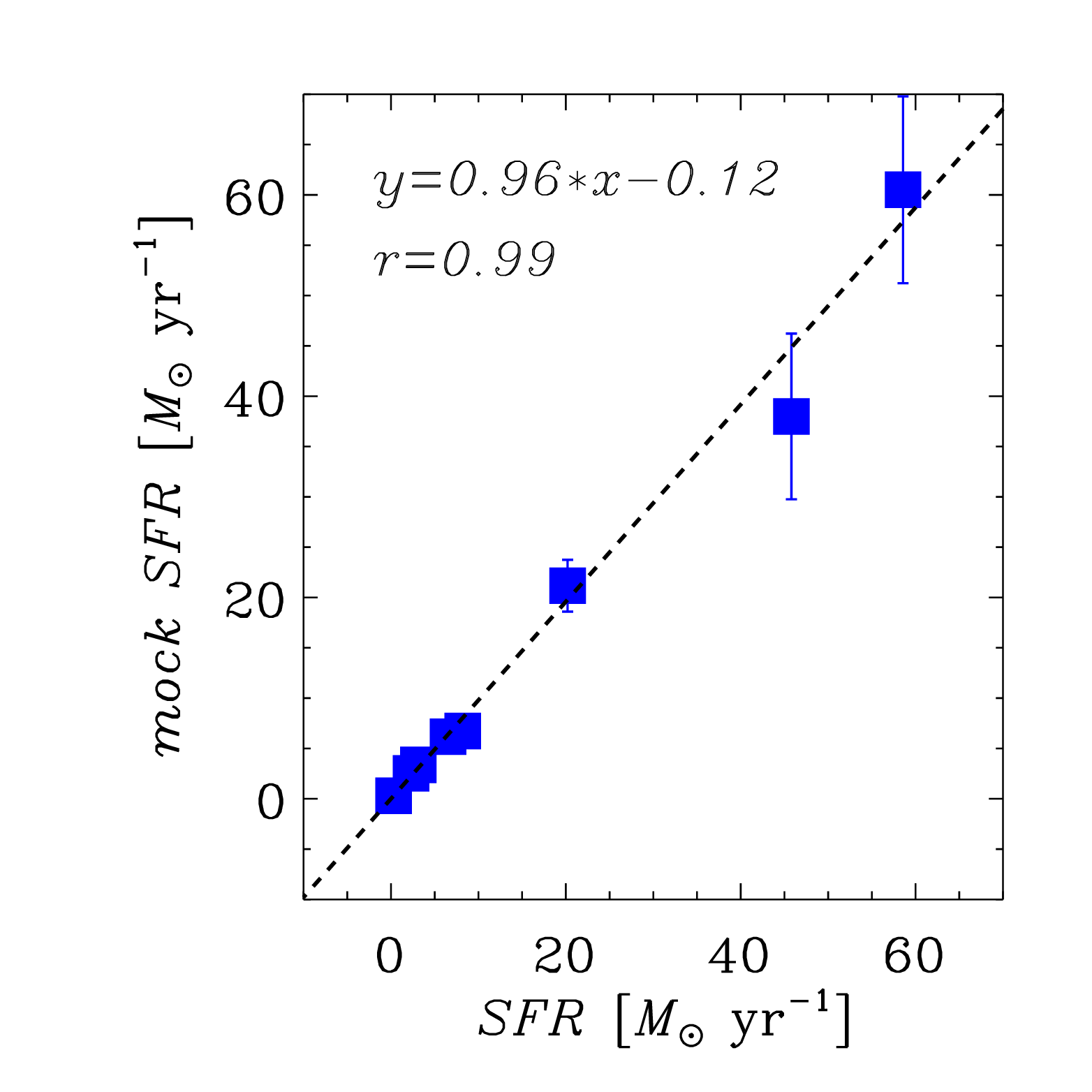}
   \includegraphics[width=.6\columnwidth]{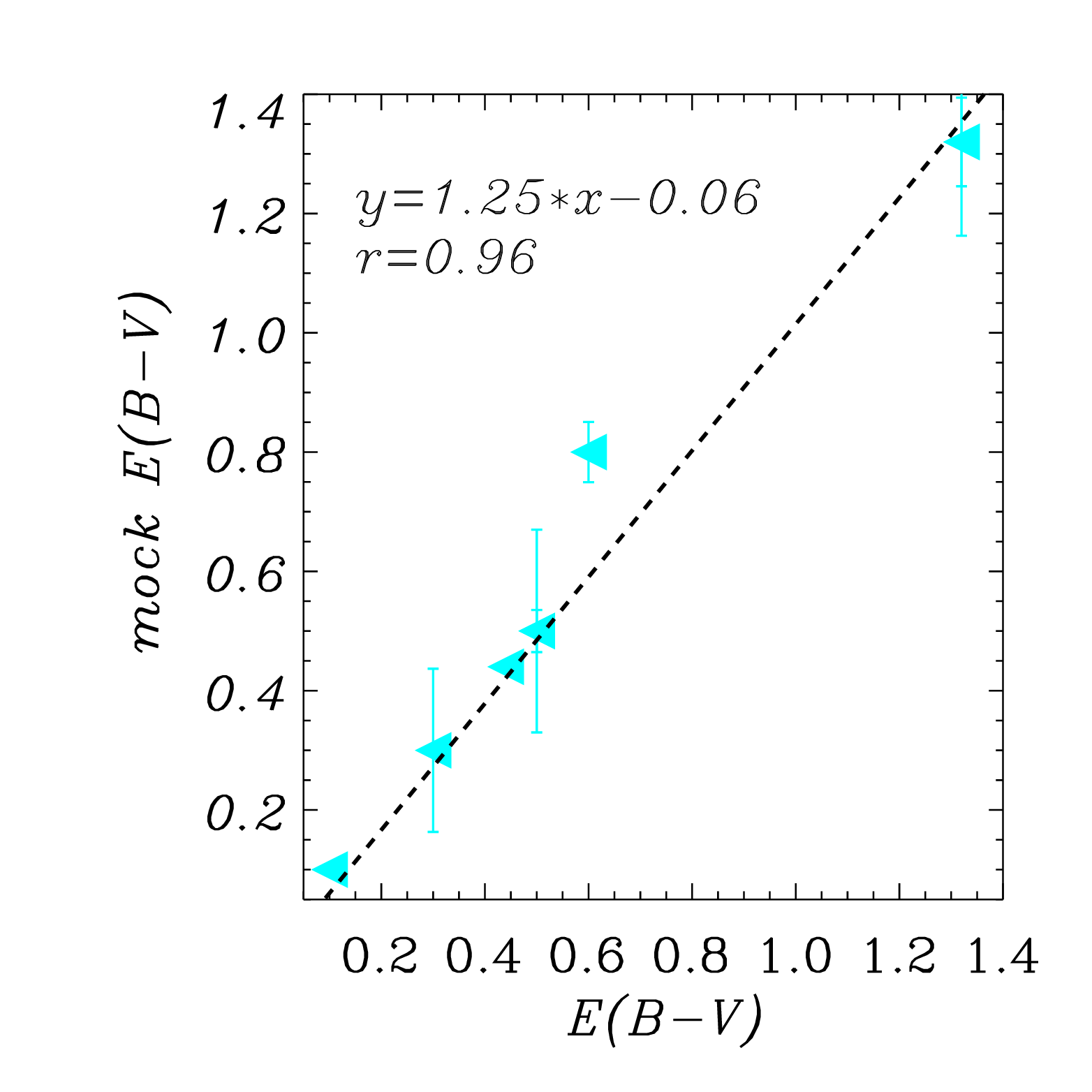}

\label{cdf}
\end{figure*}

Figure~\ref{cdf} shows the comparison between the value estimated by the code and true value of the output parameter provided by the  best-fit model for the mock catalogue. We estimate the reliability for the following parameters: AGN fraction, stellar mass, dust luminosity, attenuation, and SFR. The blue line represents the regression line with the equation given in the legend. The Pearson product-moment correlation coefficient is given as an {\it r} value. We find very good correlations ($r>0.95$) for all estimated parameters.

\begin{figure*}[ht]
\label{sedapp}
\centering
\subfloat{
\includegraphics[width=70mm]{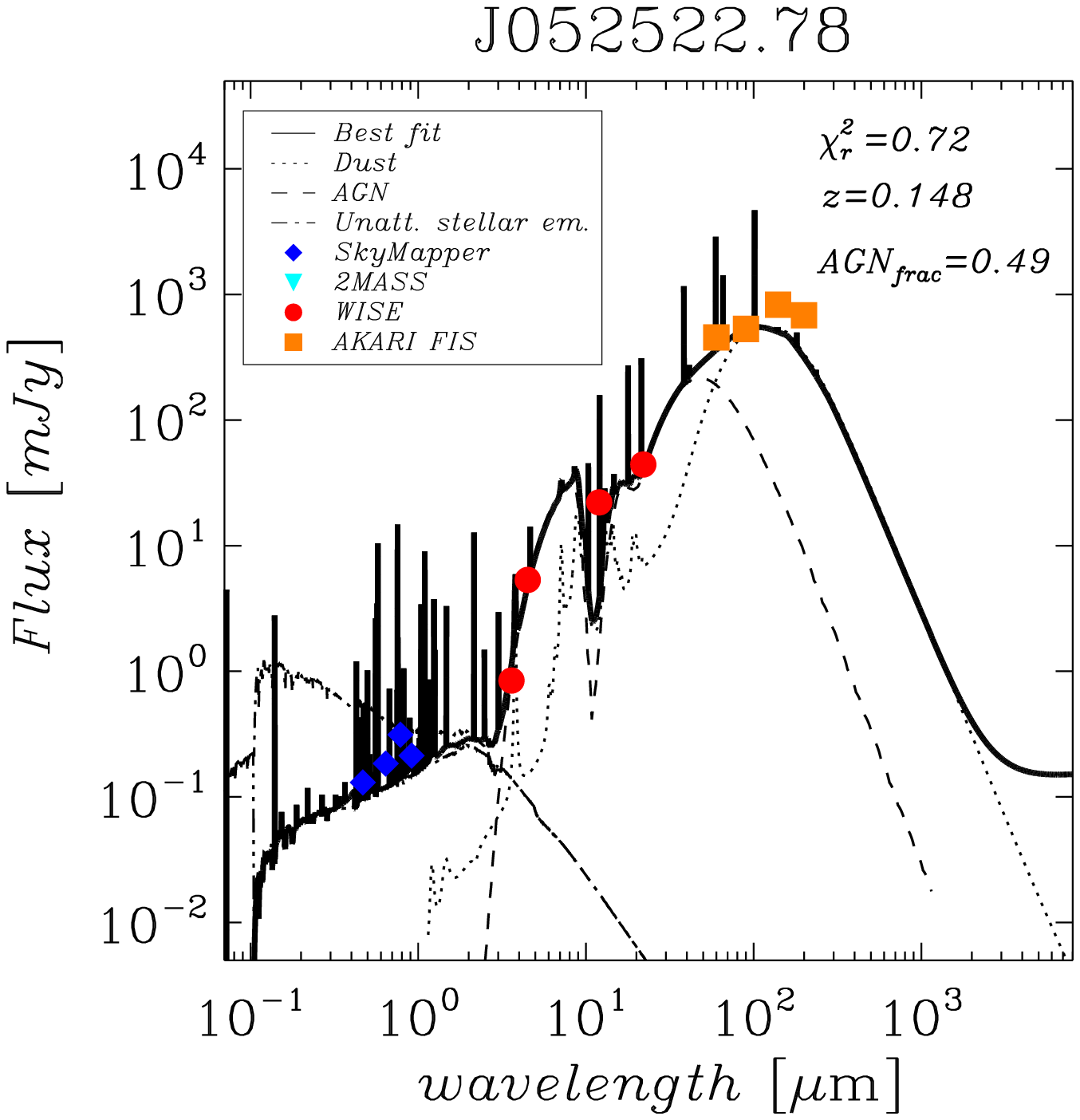}
}
\subfloat{
\includegraphics[width=70mm]{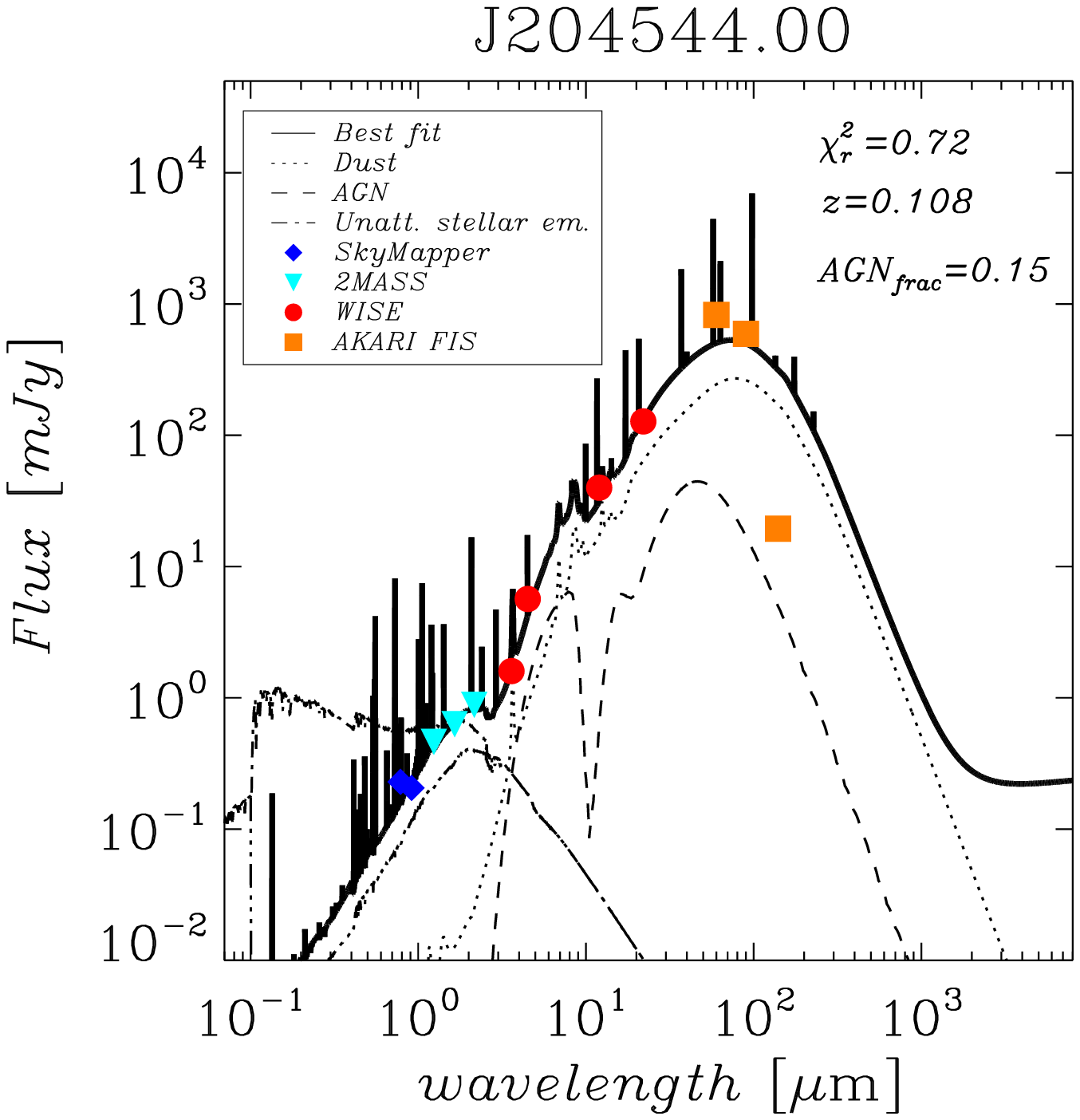}
}
\hspace{0mm}
\subfloat{
\includegraphics[width=70mm]{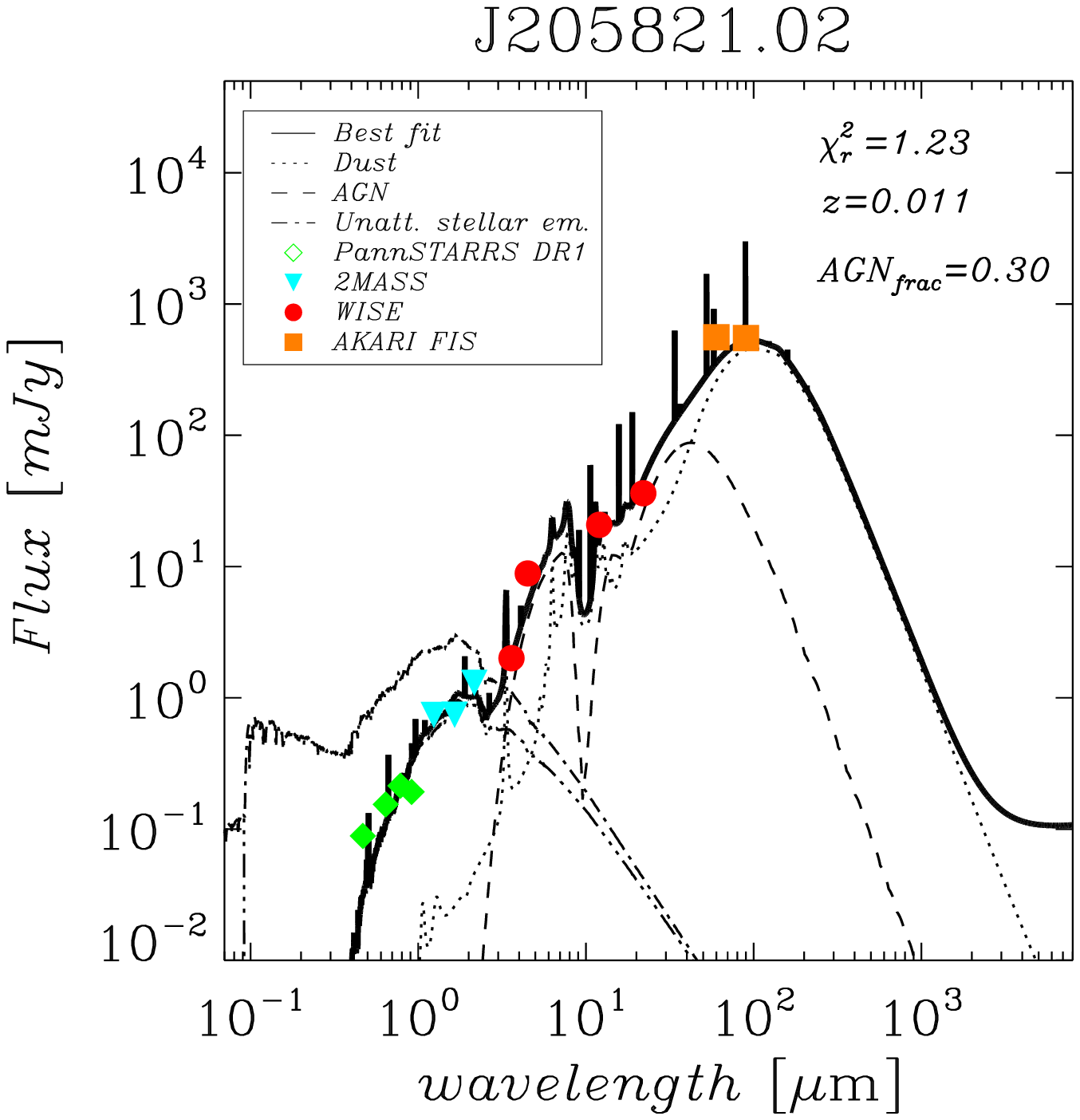}
}
\subfloat{
\includegraphics[width=70mm]{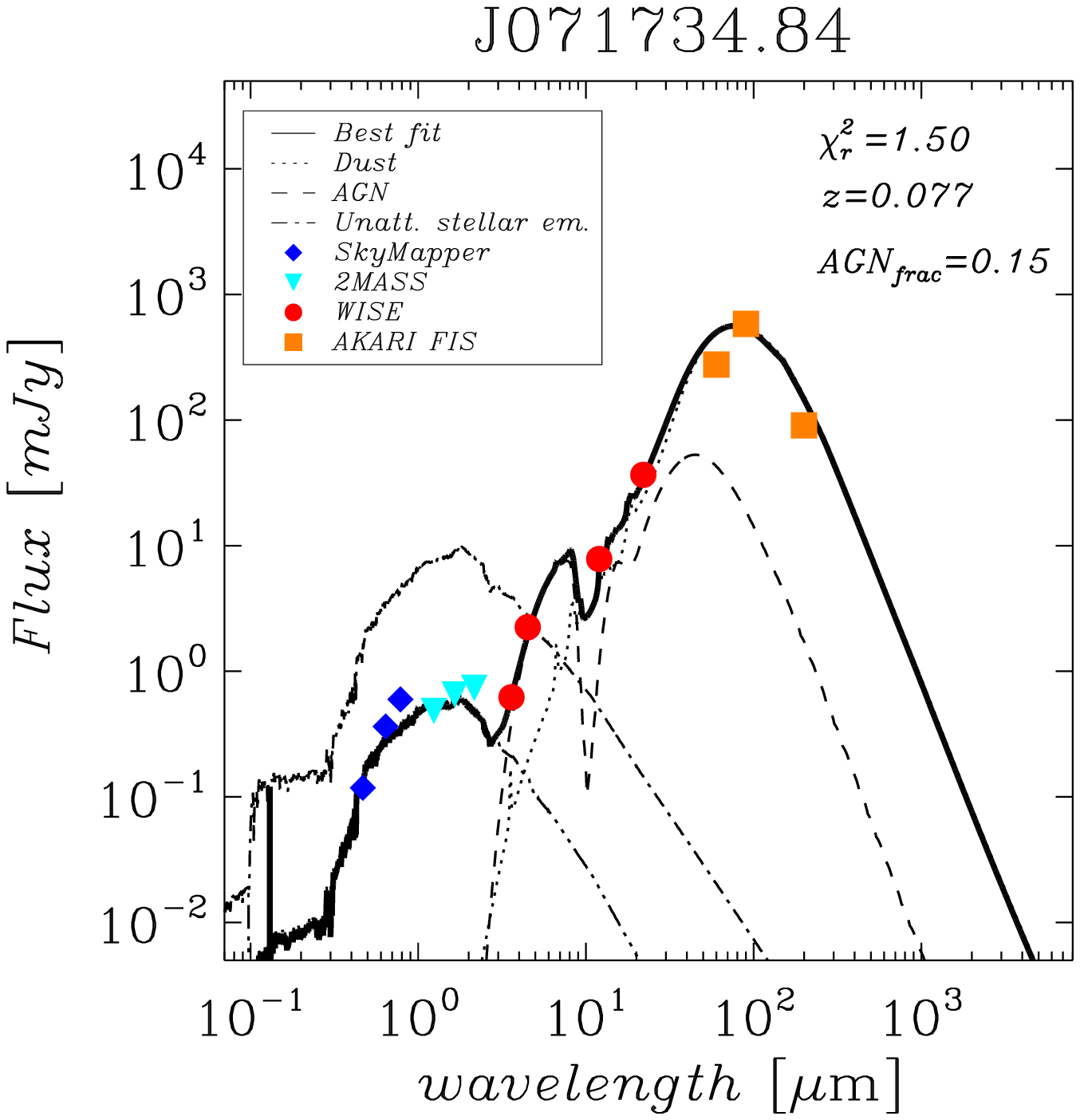}
}\\
\subfloat{
\includegraphics[width=70mm]{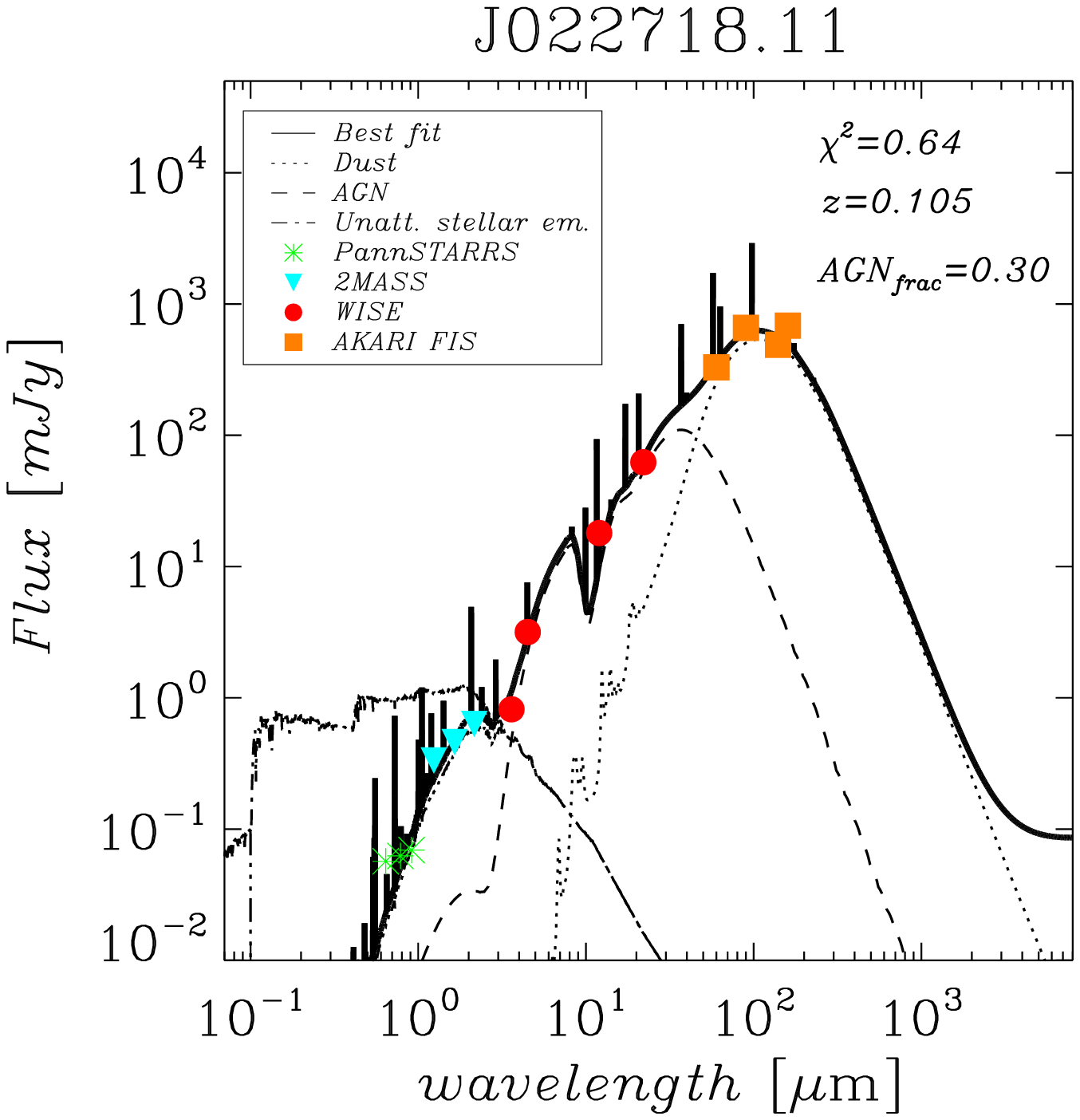}
}
\caption{SED fitting results for anomalies identified as local ($z$ $<$ $0.15$) galaxies. Errorbars are smaller than the point size used in the plot.}
\end{figure*}

\end{appendix}
\begin{acknowledgements}
We are grateful for the comments and suggestions by the  anonymous  referee,  which  helped  to  improve  the manuscript. 
This paper is based on observations made with ESO telescopes at the La Silla Paranal Observatory under program IDs 0101.A-0539 and 0102.A-0305.
AS  was  supported  by the ESO grant SSDF19/13 and  MNiSW grant 212727/E-78//M/2018.
MG is supported by the Polish NCN MAESTRO grant 2014/14/A/ST9/00121.
{\L}W acknowledges support from the Polish NCN HARMONIA grant 2018/06/M/ST9/00311.
Special thanks to Romain Thomas for the Photon \citep{photon} software and  Mark Taylor for the TOPCAT \citep{topcat} software.
CGD acknowledges the support of the 
Consejo de Investigaciones Cient\'ificas y T\'ecnicas (CONICET).
MPK acknowledges support from the First TEAM grant of the Foundation for Polish Science No. POIR.04.04.00-00-5D21/18-00.
\end{acknowledgements}

\bibliographystyle{aa.bst}
\bibliography{solarz_corr.bib}

\begin{thebibliography}{88}
\expandafter\ifx\csname natexlab\endcsname\relax\def\natexlab#1{#1}\fi

\bibitem[{{Albareti} {et~al.}(2017){Albareti}, {Allende Prieto}, {Almeida},
  {Anders}, {Anderson}, {Andrews}, {Arag{\'o}n-Salamanca},
  {Argudo-Fern{\'a}ndez}, {Armengaud}, {Aubourg}, {Avila-Reese}, {Badenes},
  {Bailey}, {Barbuy}, {Barger}, {Barrera-Ballesteros}, {Bartosz}, {Basu},
  {Bates}, {Battaglia}, {Baumgarten}, {Baur}, {Bautista}, {Beers}, {Belfiore},
  {Bershady}, {Bertran de Lis}, {Bird}, {Bizyaev}, {Blanc}, {Blanton},
  {Blomqvist}, {Bolton}, {Borissova}, {Bovy}, {Brand t}, {Brinkmann},
  {Brownstein}, {Bundy}, {Burtin}, {Busca}, {Orlando Camacho Chavez}, {Cano
  D{\'\i}az}, {Cappellari}, {Carrera}, {Chen}, {Cherinka}, {Cheung},
  {Chiappini}, {Chojnowski}, {Chuang}, {Chung}, {Cirolini}, {Clerc}, {Cohen},
  {Comerford}, {Comparat}, {Correa do Nascimento}, {Cousinou}, {Covey},
  {Crane}, {Croft}, {Cunha}, {Darling}, {Davidson}, {Dawson}, {Da Costa}, {Da
  Silva Ilha}, {Deconto Machado}, {Delubac}, {De Lee}, {De la Macorra}, {De la
  Torre}, {Diamond-Stanic}, {Donor}, {Downes}, {Drory}, {Du}, {Du Mas des
  Bourboux}, {Dwelly}, {Ebelke}, {Eigenbrot}, {Eisenstein}, {Elsworth},
  {Emsellem}, {Eracleous}, {Escoffier}, {Evans}, {Falc{\'o}n-Barroso}, {Fan},
  {Favole}, {Fernandez-Alvar}, {Fernand ez-Trincado}, {Feuillet}, {Fleming},
  {Font-Ribera}, {Freischlad}, {Frinchaboy}, {Fu}, {Gao}, {Garcia},
  {Garcia-Dias}, {Garcia-Hern{\'a}ndez}, {Garcia P{\'e}rez}, {Gaulme}, {Ge},
  {Geisler}, {Gillespie}, {Gil Marin}, {Girardi}, {Goddard}, {Gomez Maqueo
  Chew}, {Gonzalez-Perez}, {Grabowski}, {Green}, {Grier}, {Grier}, {Guo},
  {Guy}, {Hagen}, {Hall}, {Harding}, {Harley}, {Hasselquist}, {Hawley},
  {Hayes}, {Hearty}, {Hekker}, {Hernandez Toledo}, {Ho}, {Hogg},
  {Holley-Bockelmann}, {Holtzman}, {Holzer}, {Hu}, {Huber}, {Hutchinson},
  {Hwang}, {Ibarra-Medel}, {Ivans}, {Ivory}, {Jaehnig}, {Jensen}, {Johnson},
  {Jones}, {Jullo}, {Kallinger}, {Kinemuchi}, {Kirkby}, {Klaene}, {Kneib},
  {Kollmeier}, {Lacerna}, {Lane}, {Lang}, {Laurent}, {Law}, {Leauthaud}, {Le
  Goff}, {Li}, {Li}, {Li}, {Li}, {Liang}, {Liang}, {Lima}, {Lin}, {Lin}, {Lin},
  {Liu}, {Long}, {Lucatello}, {MacDonald}, {MacLeod}, {Mackereth}, {Mahadevan},
  {Maia}, {Maiolino}, {Majewski}, {Malanushenko}, {Malanushenko}, {Mallmann},
  {Manchado}, {Maraston}, {Marques-Chaves}, {Martinez Valpuesta}, {Masters},
  {Mathur}, {McGreer}, {Merloni}, {Merrifield}, {M{\'e}sz{\'a}ros}, {Meza},
  {Miglio}, {Minchev}, {Molaverdikhani}, {Montero-Dorta}, {Mosser}, {Muna},
  {Myers}, {Nair}, {Nandra}, {Ness}, {Newman}, {Nichol}, {Nidever},
  {Nitschelm}, {O'Connell}, {Oravetz}, {Oravetz}, {Pace}, {Padilla},
  {Palanque-Delabrouille}, {Pan}, {Parejko}, {Paris}, {Park}, {Peacock},
  {Peirani}, {Pellejero-Ibanez}, {Penny}, {Percival}, {Percival},
  {Perez-Fournon}, {Petitjean}, {Pieri}, {Pinsonneault}, {Pisani}, {Prada},
  {Prakash}, {Price-Jones}, {Raddick}, {Rahman}, {Raichoor}, {Barboza Rembold},
  {Reyna}, {Rich}, {Richstein}, {Ridl}, {Riffel}, {Riffel}, {Rix}, {Robin},
  {Rockosi}, {Rodr{\'\i}guez-Torres}, {Rodrigues}, {Roe}, {Roman Lopes},
  {Rom{\'a}n-Z{\'u}{\~n}iga}, {Ross}, {Rossi}, {Ruan}, {Ruggeri}, {Runnoe},
  {Salazar-Albornoz}, {Salvato}, {Sanchez}, {Sanchez}, {Sanchez-Gallego},
  {Santiago}, {Schiavon}, {Schimoia}, {Schlafly}, {Schlegel}, {Schneider},
  {Sch{\"o}nrich}, {Schultheis}, {Schwope}, {Seo}, {Serenelli}, {Sesar},
  {Shao}, {Shetrone}, {Shull}, {Silva Aguirre}, {Skrutskie}, {Slosar}, {Smith},
  {Smith}, {Sobeck}, {Somers}, {Souto}, {Stark}, {Stassun}, {Steinmetz},
  {Stello}, {Storchi Bergmann}, {Strauss}, {Streblyanska}, {Stringfellow},
  {Suarez}, {Sun}, {Taghizadeh-Popp}, {Tang}, {Tao}, {Tayar}, {Tembe},
  {Thomas}, {Tinker}, {Tojeiro}, {Tremonti}, {Troup}, {Trump}, {Unda-Sanzana},
  {Valenzuela}, {Van den Bosch}, {Vargas-Maga{\~n}a}, {Vazquez}, {Villanova},
  {Vivek}, {Vogt}, {Wake}, {Walterbos}, {Wang}, {Wang}, {Weaver}, {Weijmans},
  {Weinberg}, {Westfall}, {Whelan}, {Wilcots}, {Wild}, {Williams}, {Wilson},
  {Wood-Vasey}, {Wylezalek}, {Xiao}, {Yan}, {Yang}, {Ybarra}, {Yeche}, {Yuan},
  {Zakamska}, {Zamora}, {Zasowski}, {Zhang}, {Zhao}, {Zhao}, {Zheng}, {Zheng},
  {Zhou}, {Zhu}, {Zinn}, \& {Zou}}]{sdss13}
{Albareti}, F.~D., {Allende Prieto}, C., {Almeida}, A., {et~al.} 2017, \apjs,
  233, 25

\bibitem[{{Allen} {et~al.}(2011){Allen}, {Hewett}, {Maddox}, {Richards}, \&
  {Belokurov}}]{allen11}
{Allen}, J.~T., {Hewett}, P.~C., {Maddox}, N., {Richards}, G.~T., \&
  {Belokurov}, V. 2011, \mnras, 410, 860

\bibitem[{{Assef} {et~al.}(2010){Assef}, {Kochanek}, {Brodwin}, {Cool},
  {Forman}, {Gonzalez}, {Hickox}, {Jones}, {Le Floc'h}, {Moustakas}, {Murray},
  \& {Stern}}]{assef10}
{Assef}, R.~J., {Kochanek}, C.~S., {Brodwin}, M., {et~al.} 2010, \apj, 713, 970

\bibitem[{{Bailer-Jones} {et~al.}(2018){Bailer-Jones}, {Rybizki}, {Fouesneau},
  {Mantelet}, \& {Andrae}}]{bailer-jones18}
{Bailer-Jones}, C.~A.~L., {Rybizki}, J., {Fouesneau}, M., {Mantelet}, G., \&
  {Andrae}, R. 2018, \aj, 156, 58

\bibitem[{{Baldwin} {et~al.}(1981){Baldwin}, {Phillips}, \& {Terlevich}}]{BPT}
{Baldwin}, J.~A., {Phillips}, M.~M., \& {Terlevich}, R. 1981, \pasp, 93, 5

\bibitem[{{Baron} \& {Poznanski}(2017)}]{baron17}
{Baron}, D. \& {Poznanski}, D. 2017, \mnras, 465, 4530

\bibitem[{Baumgartner {et~al.}(2013)Baumgartner, Tueller, Markwardt, Skinner,
  Barthelmy, Mushotzky, Evans, \& Gehrels}]{swift}
Baumgartner, W.~H., Tueller, J., Markwardt, C.~B., {et~al.} 2013, The
  Astrophysical Journal Supplement Series, 207, 19

\bibitem[{{Becker} {et~al.}(1995){Becker}, {White}, \& {Helfand}}]{first}
{Becker}, R.~H., {White}, R.~L., \& {Helfand}, D.~J. 1995, \apj, 450, 559

\bibitem[{{Bilicki} {et~al.}(2014){Bilicki}, {Jarrett}, {Peacock}, {Cluver}, \&
  {Steward}}]{bilicki14}
{Bilicki}, M., {Jarrett}, T.~H., {Peacock}, J.~A., {Cluver}, M.~E., \&
  {Steward}, L. 2014, \apjs, 210, 9

\bibitem[{{Bilicki} {et~al.}(2016){Bilicki}, {Peacock}, {Jarrett}, {Cluver},
  {Maddox}, {Brown}, {Taylor}, {Hambly}, {Solarz}, {Holwerda}, {Baldry},
  {Loveday}, {Moffett}, {Hopkins}, {Driver}, {Alpaslan}, \&
  {Bland-Hawthorn}}]{bilicki16}
{Bilicki}, M., {Peacock}, J.~A., {Jarrett}, T.~H., {et~al.} 2016, \apjs, 225, 5

\bibitem[{Breiman(2001)}]{Breiman2001}
Breiman, L. 2001, Machine Learning, 45, 5

\bibitem[{{Bruzual} \& {Charlot}(2003)}]{bc03}
{Bruzual}, G. \& {Charlot}, S. 2003, \mnras, 344, 1000

\bibitem[{{Buzzoni} {et~al.}(1984){Buzzoni}, {Delabre}, {Dekker}, {Dodorico},
  {Enard}, {Focardi}, {Gustafsson}, {Nees}, {Paureau}, \& {Reiss}}]{efosc}
{Buzzoni}, B., {Delabre}, B., {Dekker}, H., {et~al.} 1984, The Messenger, 38, 9

\bibitem[{{Caccianiga} {et~al.}(2007){Caccianiga}, {Severgnini}, {Della Ceca},
  {Maccacaro}, {Carrera}, \& {Page}}]{caccianiga07}
{Caccianiga}, A., {Severgnini}, P., {Della Ceca}, R., {et~al.} 2007, \aap, 470,
  557

\bibitem[{{Calzetti} {et~al.}(2000){Calzetti}, {Armus}, {Bohlin}, {Kinney},
  {Koornneef}, \& {Storchi-Bergmann}}]{calzetti00}
{Calzetti}, D., {Armus}, L., {Bohlin}, R.~C., {et~al.} 2000, \apj, 533, 682

\bibitem[{{Chabrier}(2003)}]{chabrier03}
{Chabrier}, G. 2003, \pasp, 115, 763

\bibitem[{Chambers {et~al.}(2016)Chambers, Magnier, Metcalfe, Flewelling,
  Huber, Waters, Denneau, Draper, Farrow, Finkbeiner, Holmberg, Koppenhoefer,
  Price, Rest, Saglia, Schlafly, Smartt, Sweeney, Wainscoat, Burgett, Chastel,
  Grav, Heasley, Hodapp, Jedicke, Kaiser, Kudritzki, Luppino, Lupton, Monet,
  Morgan, Onaka, Shiao, Stubbs, Tonry, White, Bañados, Bell, Bender, Bernard,
  Boegner, Boffi, Botticella, Calamida, Casertano, Chen, Chen, Cole, Deacon,
  Frenk, Fitzsimmons, Gezari, Gibbs, Goessl, Goggia, Gourgue, Goldman, Grant,
  Grebel, Hambly, Hasinger, Heavens, Heckman, Henderson, Henning, Holman, Hopp,
  Ip, Isani, Jackson, Keyes, Koekemoer, Kotak, Le, Liska, Long, Lucey, Liu,
  Martin, Masci, McLean, Mindel, Misra, Morganson, Murphy, Obaika, Narayan,
  Nieto-Santisteban, Norberg, Peacock, Pier, Postman, Primak, Rae, Rai, Riess,
  Riffeser, Rix, Röser, Russel, Rutz, Schilbach, Schultz, Scolnic, Strolger,
  Szalay, Seitz, Small, Smith, Soderblom, Taylor, Thomson, Taylor, Thakar,
  Thiel, Thilker, Unger, Urata, Valenti, Wagner, Walder, Walter, Watters,
  Werner, Wood-Vasey, \& Wyse}]{chambers2016}
Chambers, K.~C., Magnier, E.~A., Metcalfe, N., {et~al.} 2016, The Pan-STARRS1
  Surveys

\bibitem[{{Ciesla} {et~al.}(2015){Ciesla}, {Charmandaris}, {Georgakakis},
  {Bernhard}, {Mitchell}, {Buat}, {Elbaz}, {LeFloc'h}, {Lacey}, {Magdis}, \&
  {Xilouris}}]{ciesla15}
{Ciesla}, L., {Charmandaris}, V., {Georgakakis}, A., {et~al.} 2015, \aap, 576,
  A10

\bibitem[{{Clough} {et~al.}(2005){Clough}, {Shephard}, {Mlawer}, {Delamere},
  {Iacono}, {Cady-Pereira}, {Boukabara}, \& {Brown}}]{clough05}
{Clough}, S.~A., {Shephard}, M.~W., {Mlawer}, E.~J., {et~al.} 2005, \jqsrt, 91,
  233

\bibitem[{{Colless}(1999)}]{2dfgrs}
{Colless}, M. 1999, in Large-Scale Structure in the Universe, ed.
  G.~{Efstathiou} \& {et al.}, Vol. 357, 105

\bibitem[{{Condon} {et~al.}(1998){Condon}, {Cotton}, {Greisen}, {Yin},
  {Perley}, {Taylor}, \& {Broderick}}]{condon98}
{Condon}, J.~J., {Cotton}, W.~D., {Greisen}, E.~W., {et~al.} 1998, \aj, 115,
  1693

\bibitem[{{Crowther} {et~al.}(2007){Crowther}, {Carpano}, {Hadfield}, \&
  {Pollock}}]{crowther07}
{Crowther}, P.~A., {Carpano}, S., {Hadfield}, L.~J., \& {Pollock}, A.~M.~T.
  2007, \aap, 469, L31

\bibitem[{{Cutri} {et~al.}(2013){Cutri}, {Wright}, {Conrow}, {Fowler},
  {Eisenhardt}, {Grillmair}, {Kirkpatrick}, {Masci}, {McCallon}, {Wheelock},
  {Fajardo-Acosta}, {Yan}, {Benford}, {Harbut}, {Jarrett}, {Lake}, {Leisawitz},
  {Ressler}, {Stanford}, {Tsai}, {Liu}, {Helou}, {Mainzer}, {Gettings},
  {Gonzalez}, {Hoffman}, {Marsh}, {Padgett}, {Skrutskie}, {Beck}, {Papin}, \&
  {Wittman}}]{cutri13}
{Cutri}, R.~M., {Wright}, E.~L., {Conrow}, T., {et~al.} 2013, {Explanatory
  Supplement to the AllWISE Data Release Products}, Tech. rep., Explanatory
  Supplement to the AllWISE Data Release Products, by R. M. Cutri et al.

\bibitem[{{Della Ceca} {et~al.}(2008){Della Ceca}, {Caccianiga}, {Severgnini},
  {Maccacaro}, {Brunner}, {Carrera}, {Cocchia}, {Mateos}, {Page}, \&
  {Tedds}}]{dc08}
{Della Ceca}, R., {Caccianiga}, A., {Severgnini}, P., {et~al.} 2008, \aap, 487,
  119

\bibitem[{{Dewdney} {et~al.}(2009){Dewdney}, {Hall}, {Schilizzi}, \&
  {Lazio}}]{dewdney09}
{Dewdney}, P.~E., {Hall}, P.~J., {Schilizzi}, R.~T., \& {Lazio}, T.~J.~L.~W.
  2009, IEEE Proceedings, 97, 1482

\bibitem[{{Drinkwater} {et~al.}(2010){Drinkwater}, {Jurek}, {Blake}, {Woods},
  {Pimbblet}, {Glazebrook}, {Sharp}, {Pracy}, {Brough}, {Colless}, {Couch},
  {Croom}, {Davis}, {Forbes}, {Forster}, {Gilbank}, {Gladders}, {Jelliffe},
  {Jones}, {Li}, {Madore}, {Martin}, {Poole}, {Small}, {Wisnioski}, {Wyder}, \&
  {Yee}}]{wigglez}
{Drinkwater}, M.~J., {Jurek}, R.~J., {Blake}, C., {et~al.} 2010, \mnras, 401,
  1429

\bibitem[{Evans {et~al.}(2010)Evans, Primini, Glotfelty, Anderson, Bonaventura,
  Chen, Davis, Doe, Evans, Fabbiano, Galle, Gibbs, Grier, Hain, Hall, Harbo,
  He, Houck, Karovska, Kashyap, Lauer, McCollough, McDowell, Miller, Mitschang,
  Morgan, Mossman, Nichols, Nowak, Plummer, Refsdal, Rots, Siemiginowska,
  Sundheim, Tibbetts, Stone, Winkelman, \& Zografou}]{chandra}
Evans, I.~N., Primini, F.~A., Glotfelty, K.~J., {et~al.} 2010, The
  Astrophysical Journal Supplement Series, 189, 37

\bibitem[{{Farnes} {et~al.}(2018){Farnes}, {Mort}, {Dulwich}, {Salvini}, \&
  {Armour}}]{ska}
{Farnes}, J., {Mort}, B., {Dulwich}, F., {Salvini}, S., \& {Armour}, W. 2018,
  arXiv e-prints

\bibitem[{{Fischer} {et~al.}(2016){Fischer}, {Padgett}, {Stapelfeldt}, \&
  {Sewi{\l}o}}]{fischer16}
{Fischer}, W.~J., {Padgett}, D.~L., {Stapelfeldt}, K.~L., \& {Sewi{\l}o}, M.
  2016, \apj, 827, 96

\bibitem[{{Fritz} {et~al.}(2006){Fritz}, {Franceschini}, \&
  {Hatziminaoglou}}]{fritz06}
{Fritz}, J., {Franceschini}, A., \& {Hatziminaoglou}, E. 2006, \mnras, 366, 767

\bibitem[{{Gagn{\'e}} {et~al.}(2018){Gagn{\'e}}, {Mamajek}, {Malo}, {Riedel},
  {Rodriguez}, {Lafreni{\`e}re}, {Faherty}, {Roy-Loubier}, {Pueyo}, {Robin}, \&
  {Doyon}}]{scorpius}
{Gagn{\'e}}, J., {Mamajek}, E.~E., {Malo}, L., {et~al.} 2018, \apj, 856, 23

\bibitem[{{Gaia Collaboration} {et~al.}(2018){Gaia Collaboration}, {Brown},
  {Vallenari}, {Prusti}, {de Bruijne}, {Babusiaux}, {Bailer-Jones}, {Biermann},
  {Evans}, {Eyer}, \& et~al.}]{gaia2}
{Gaia Collaboration}, {Brown}, A.~G.~A., {Vallenari}, A., {et~al.} 2018, \aap,
  616, A1

\bibitem[{{Gaia Collaboration} {et~al.}(2016){Gaia Collaboration}, {Prusti},
  {de Bruijne}, {Brown}, {Vallenari}, {Babusiaux}, {Bailer-Jones}, {Bastian},
  {Biermann}, {Evans}, \& et~al.}]{gaia1}
{Gaia Collaboration}, {Prusti}, T., {de Bruijne}, J.~H.~J., {et~al.} 2016,
  \aap, 595, A1

\bibitem[{{Gibson} {et~al.}(2009){Gibson}, {Jiang}, {Brandt}, {Hall}, {Shen},
  {Wu}, {Anderson}, {Schneider}, {Vand en Berk}, {Gallagher}, {Fan}, \&
  {York}}]{gibson09}
{Gibson}, R.~R., {Jiang}, L., {Brandt}, W.~N., {et~al.} 2009, \apj, 692, 758

\bibitem[{{Giovannoli} {et~al.}(2011){Giovannoli}, {Buat}, {Noll},
  {Burgarella}, \& {Magnelli}}]{giovannoli11}
{Giovannoli}, E., {Buat}, V., {Noll}, S., {Burgarella}, D., \& {Magnelli}, B.
  2011, \aap, 525, A150

\bibitem[{{Griffith} {et~al.}(2011){Griffith}, {Tsai}, {Stern}, {Blain},
  {Eisenhardt}, {Harrison}, {Jarrett}, {Madsen}, {Stanford}, {Wright}, {Wu},
  {Wu}, \& {Yan}}]{griffith11}
{Griffith}, R.~L., {Tsai}, C.-W., {Stern}, D., {et~al.} 2011, \apjl, 736, L22

\bibitem[{{Hainline} {et~al.}(2016){Hainline}, {Reines}, {Greene}, \&
  {Stern}}]{hainline16}
{Hainline}, K.~N., {Reines}, A.~E., {Greene}, J.~E., \& {Stern}, D. 2016, \apj,
  832, 119

\bibitem[{{Hall} {et~al.}(2002){Hall}, {Anderson}, {Strauss}, {York},
  {Richards}, {Fan}, {Knapp}, {Schneider}, {Vanden Berk}, {Geballe}, {Bauer},
  {Becker}, {Davis}, {Rix}, {Nichol}, {Bahcall}, {Brinkmann}, {Brunner},
  {Connolly}, {Csabai}, {Doi}, {Fukugita}, {Gunn}, {Haiman}, {Harvanek},
  {Heckman}, {Hennessy}, {Inada}, {Ivezi{\'c}}, {Johnston}, {Kleinman},
  {Krolik}, {Krzesinski}, {Kunszt}, {Lamb}, {Long}, {Lupton}, {Miknaitis},
  {Munn}, {Narayanan}, {Neilsen}, {Newman}, {Nitta}, {Okamura}, {Pentericci},
  {Pier}, {Schlegel}, {Snedden}, {Szalay}, {Thakar}, {Tsvetanov}, {White}, \&
  {Zheng}}]{hall02}
{Hall}, P.~B., {Anderson}, S.~F., {Strauss}, M.~A., {et~al.} 2002, \apjs, 141,
  267

\bibitem[{{Hambly} {et~al.}(2001){Hambly}, {MacGillivray}, {Read}, {Tritton},
  {Thomson}, {Kelly}, {Morgan}, {Smith}, {Driver}, {Williamson}, {Parker},
  {Hawkins}, {Williams}, \& {Lawrence}}]{supercosmos}
{Hambly}, N.~C., {MacGillivray}, H.~T., {Read}, M.~A., {et~al.} 2001, \mnras,
  326, 1279

\bibitem[{{Hatziminaoglou} {et~al.}(2008){Hatziminaoglou}, {Fritz},
  {Franceschini}, {Afonso-Luis}, {Hern{\'a}n-Caballero}, {P{\'e}rez-Fournon},
  {Serjeant}, {Lonsdale}, {Oliver}, {Rowan-Robinson}, {Shupe}, {Smith}, \&
  {Surace}}]{hatziminaoglou08}
{Hatziminaoglou}, E., {Fritz}, J., {Franceschini}, A., {et~al.} 2008, \mnras,
  386, 1252

\bibitem[{{Hewett} \& {Foltz}(2003)}]{hewett03}
{Hewett}, P.~C. \& {Foltz}, C.~B. 2003, \aj, 125, 1784

\bibitem[{{Hocking} {et~al.}(2018){Hocking}, {Geach}, {Sun}, \&
  {Davey}}]{hocking18}
{Hocking}, A., {Geach}, J.~E., {Sun}, Y., \& {Davey}, N. 2018, \mnras, 473,
  1108

\bibitem[{{Hoyle} {et~al.}(2015){Hoyle}, {Rau}, {Paech}, {Bonnett}, {Seitz}, \&
  {Weller}}]{hoyle15}
{Hoyle}, B., {Rau}, M.~M., {Paech}, K., {et~al.} 2015, \mnras, 452, 4183

\bibitem[{{Ivezic} {et~al.}(2008){Ivezic}, {Axelrod}, {Brandt}, {Burke},
  {Claver}, {Connolly}, {Cook}, {Gee}, {Gilmore}, {Jacoby}, {Jones}, {Kahn},
  {Kantor}, {Krabbendam}, {Lupton}, {Monet}, {Pinto}, {Saha}, {Schalk},
  {Schneider}, {Strauss}, {Stubbs}, {Sweeney}, {Szalay}, {Thaler}, {Tyson}, \&
  {LSST Collaboration}}]{ivezic08}
{Ivezic}, Z., {Axelrod}, T., {Brandt}, W.~N., {et~al.} 2008, Serbian
  Astronomical Journal, 176, 1

\bibitem[{{Izotov} {et~al.}(2011){Izotov}, {Guseva}, {Fricke}, \&
  {Henkel}}]{izotov11}
{Izotov}, Y.~I., {Guseva}, N.~G., {Fricke}, K.~J., \& {Henkel}, C. 2011, \aap,
  536, L7

\bibitem[{{Izotov} {et~al.}(2014){Izotov}, {Guseva}, {Fricke}, {Kr{\"u}gel}, \&
  {Henkel}}]{izotov14}
{Izotov}, Y.~I., {Guseva}, N.~G., {Fricke}, K.~J., {Kr{\"u}gel}, E., \&
  {Henkel}, C. 2014, \aap, 570, A97

\bibitem[{{Jarrett} {et~al.}(2017){Jarrett}, {Cluver}, {Magoulas}, {Bilicki},
  {Alpaslan}, {Bland-Hawthorn}, {Brough}, {Brown}, {Croom}, {Driver},
  {Holwerda}, {Hopkins}, {Loveday}, {Norberg}, {Peacock}, {Popescu}, {Sadler},
  {Taylor}, {Tuffs}, \& {Wang}}]{jarrett17}
{Jarrett}, T.~H., {Cluver}, M.~E., {Magoulas}, C., {et~al.} 2017, \apj, 836,
  182

\bibitem[{{Jones} {et~al.}(2004){Jones}, {Saunders}, {Colless}, {Read},
  {Parker}, {Watson}, {Campbell}, {Burkey}, {Mauch}, {Moore}, {Hartley},
  {Cass}, {James}, {Russell}, {Fiegert}, {Dawe}, {Huchra}, {Jarrett}, {Lahav},
  {Lucey}, {Mamon}, {Proust}, {Sadler}, \& {Wakamatsu}}]{6dfgs}
{Jones}, D.~H., {Saunders}, W., {Colless}, M., {et~al.} 2004, \mnras, 355, 747

\bibitem[{{Jones} {et~al.}(2017){Jones}, {Woods}, {Kemper}, {Kraemer}, {Sloan},
  {Srinivasan}, {Oliveira}, {van Loon}, {Boyer}, {Sargent}, {McDonald},
  {Meixner}, {Zijlstra}, {Ruffle}, {Lagadec}, {Pauly}, {Sewi{\l}o}, {Clayton},
  \& {Volk}}]{jones17}
{Jones}, O.~C., {Woods}, P.~M., {Kemper}, F., {et~al.} 2017, \mnras, 470, 3250

\bibitem[{{Kaiser} {et~al.}(2010){Kaiser}, {Burgett}, {Chambers}, {Denneau},
  {Heasley}, {Jedicke}, {Magnier}, {Morgan}, {Onaka}, \& {Tonry}}]{kaiser10}
{Kaiser}, N., {Burgett}, W., {Chambers}, K., {et~al.} 2010, in Society of
  Photo-Optical Instrumentation Engineers (SPIE) Conference Series, Vol. 7733,
  \procspie, 77330E

\bibitem[{{Kauffmann}(2018)}]{kauffmann18}
{Kauffmann}, G. 2018, \mnras, 473, 5210

\bibitem[{{Kauffmann} {et~al.}(2003){Kauffmann}, {Heckman}, {Tremonti},
  {Brinchmann}, {Charlot}, {White}, {Ridgway}, {Brinkmann}, {Fukugita}, {Hall},
  {Ivezi{\'c}}, {Richards}, \& {Schneider}}]{kauff03}
{Kauffmann}, G., {Heckman}, T.~M., {Tremonti}, C., {et~al.} 2003, \mnras, 346,
  1055

\bibitem[{{Kawada} {et~al.}(2007){Kawada}, {Baba}, {Barthel}, {Clements},
  {Cohen}, {Doi}, {Figueredo}, {Fujiwara}, {Goto}, {Hasegawa}, {Hibi}, {Hirao},
  {Hiromoto}, {Jeong}, {Kaneda}, {Kawai}, {Kawamura}, {Kester}, {Kii},
  {Kobayashi}, {Kwon}, {Lee}, {Makiuti}, {Matsuo}, {Matsuura}, {M{\"u}ller},
  {Murakami}, {Nagata}, {Nakagawa}, {Narita}, {Noda}, {Oh}, {Okada}, {Okuda},
  {Oliver}, {Ootsubo}, {Pak}, {Park}, {Pearson}, {Rowan-Robinson}, {Saito},
  {Salama}, {Sato}, {Savage}, {Serjeant}, {Shibai}, {Shirahata}, {Sohn},
  {Suzuki}, {Takagi}, {Takahashi}, {Thomson}, {Usui}, {Verdugo}, {Watabe},
  {White}, {Wang}, {Yamamura}, {Yamauchi}, \& {Yasuda}}]{kawada07}
{Kawada}, M., {Baba}, H., {Barthel}, P.~D., {et~al.} 2007, \pasj, 59, S389

\bibitem[{{Kewley} {et~al.}(2001){Kewley}, {Dopita}, {Sutherland}, {Heisler},
  \& {Trevena}}]{kewley01}
{Kewley}, L.~J., {Dopita}, M.~A., {Sutherland}, R.~S., {Heisler}, C.~A., \&
  {Trevena}, J. 2001, \apj, 556, 121

\bibitem[{{Kewley} \& {Ellison}(2008)}]{ke08}
{Kewley}, L.~J. \& {Ellison}, S.~L. 2008, \apj, 681, 1183

\bibitem[{{Kewley} {et~al.}(2006){Kewley}, {Groves}, {Kauffmann}, \&
  {Heckman}}]{kewley06}
{Kewley}, L.~J., {Groves}, B., {Kauffmann}, G., \& {Heckman}, T. 2006, \mnras,
  372, 961

\bibitem[{{Kobulnicky} \& {Kewley}(2004)}]{kk04}
{Kobulnicky}, H.~A. \& {Kewley}, L.~J. 2004, \apj, 617, 240

\bibitem[{Kohonen(1982)}]{som}
Kohonen, T. 1982, Biological Cybernetics, 43, 59

\bibitem[{{Kurcz} {et~al.}(2016){Kurcz}, {Bilicki}, {Solarz}, {Krupa}, {Pollo},
  \& {Ma{\l}ek}}]{kurcz2016}
{Kurcz}, A., {Bilicki}, M., {Solarz}, A., {et~al.} 2016, \aap, 592, A25

\bibitem[{{Liang} {et~al.}(2020){Liang}, {Li}, {Li}, {Yan}, {Mo}, {Zhang},
  {Machuca}, \& {Roman-Lopes}}]{liang20}
{Liang}, F.-H., {Li}, C., {Li}, N., {et~al.} 2020, arXiv e-prints,
  arXiv:2001.00431

\bibitem[{{Lo Faro} {et~al.}(2017){Lo Faro}, {Buat}, {Roehlly},
  {Alvarez-Marquez}, {Burgarella}, {Silva}, \& {Efstathiou}}]{lofaro17}
{Lo Faro}, B., {Buat}, V., {Roehlly}, Y., {et~al.} 2017, \mnras, 472, 1372

\bibitem[{{Meusinger} {et~al.}(2012){Meusinger}, {Schalldach}, {Scholz}, {in
  der Au}, {Newholm}, {de Hoon}, \& {Kaminsky}}]{unusualqso}
{Meusinger}, H., {Schalldach}, P., {Scholz}, R.~D., {et~al.} 2012, \aap, 541,
  A77

\bibitem[{{Moorwood} {et~al.}(1998){Moorwood}, {Cuby}, \& {Lidman}}]{sofi}
{Moorwood}, A., {Cuby}, J.-G., \& {Lidman}, C. 1998, The Messenger, 91, 9

\bibitem[{{Noll} {et~al.}(2009){Noll}, {Burgarella}, {Giovannoli}, {Buat},
  {Marcillac}, \& {Mu{\~n}oz-Mateos}}]{noll09}
{Noll}, S., {Burgarella}, D., {Giovannoli}, E., {et~al.} 2009, \aap, 507, 1793

\bibitem[{{O'Connor} {et~al.}(2016){O'Connor}, {Rosenberg}, {Satyapal}, \&
  {Secrest}}]{oconnor16}
{O'Connor}, J.~A., {Rosenberg}, J.~L., {Satyapal}, S., \& {Secrest}, N.~J.
  2016, \mnras, 463, 811

\bibitem[{{Pagel} {et~al.}(1979){Pagel}, {Edmunds}, {Blackwell}, {Chun}, \&
  {Smith}}]{pagel}
{Pagel}, B.~E.~J., {Edmunds}, M.~G., {Blackwell}, D.~E., {Chun}, M.~S., \&
  {Smith}, G. 1979, \mnras, 189, 95

\bibitem[{{Patat} {et~al.}(2011){Patat}, {Moehler}, {O'Brien}, {Pompei},
  {Bensby}, {Carraro}, {de Ugarte Postigo}, {Fox}, {Gavignaud}, {James},
  {Korhonen}, {Ledoux}, {Randall}, {Sana}, {Smoker}, {Stefl}, \&
  {Szeifert}}]{patat11}
{Patat}, F., {Moehler}, S., {O'Brien}, K., {et~al.} 2011, \aap, 527, A91

\bibitem[{{Pepiak} {et~al.}(2014){Pepiak}, {Pollo}, {Takeuchi}, {Solarz}, \&
  {Jurusik}}]{pepcus}
{Pepiak}, A., {Pollo}, A., {Takeuchi}, T.~T., {Solarz}, A., \& {Jurusik}, W.
  2014, \planss, 100, 12

\bibitem[{{Pilyugin} \& {Thuan}(2005)}]{pt05}
{Pilyugin}, L.~S. \& {Thuan}, T.~X. 2005, \apj, 631, 231

\bibitem[{{Reines} \& {Comastri}(2016)}]{reines16}
{Reines}, A.~E. \& {Comastri}, A. 2016, \pasa, 33, e054

\bibitem[{{Reis} {et~al.}(2018){Reis}, {Poznanski}, {Baron}, {Zasowski}, \&
  {Shahaf}}]{reis18}
{Reis}, I., {Poznanski}, D., {Baron}, D., {Zasowski}, G., \& {Shahaf}, S. 2018,
  \mnras, 476, 2117

\bibitem[{{Ricci} {et~al.}(2015){Ricci}, {Ueda}, {Koss}, {Trakhtenbrot},
  {Bauer}, \& {Gandhi}}]{ricci15}
{Ricci}, C., {Ueda}, Y., {Koss}, M.~J., {et~al.} 2015, \apjl, 815, L13

\bibitem[{{Rosen, S. R.} {et~al.}(2016){Rosen, S. R.}, {Webb, N. A.}, {Watson,
  M. G.}, {Ballet, J.}, {Barret, D.}, {Braito, V.}, {Carrera, F. J.},
  {Ceballos, M. T.}, {Coriat, M.}, {Della Ceca, R.}, {Denkinson, G.}, {Esquej,
  P.}, {Farrell, S. A.}, {Freyberg, M.}, {Gris\'e, F.}, {Guillout, P.}, {Heil,
  L.}, {Koliopanos, F.}, {Law-Green, D.}, {Lamer, G.}, {Lin, D.}, {Martino,
  R.}, {Michel, L.}, {Motch, C.}, {Nebot Gomez-Moran, A.}, {Page, C. G.},
  {Page, K.}, {Page, M.}, {Pakull, M. W.}, {Pye, J.}, {Read, A.}, {Rodriguez,
  P.}, {Sakano, M.}, {Saxton, R.}, {Schwope, A.}, {Scott, A. E.}, {Sturm, R.},
  {Traulsen, I.}, {Yershov, V.}, \& {Zolotukhin, I.}}]{xmm}
{Rosen, S. R.}, {Webb, N. A.}, {Watson, M. G.}, {et~al.} 2016, A\&A, 590, A1

\bibitem[{{Sartori} {et~al.}(2015){Sartori}, {Schawinski}, {Treister},
  {Trakhtenbrot}, {Koss}, {Shirazi}, \& {Oh}}]{sartori15}
{Sartori}, L.~F., {Schawinski}, K., {Treister}, E., {et~al.} 2015, \mnras, 454,
  3722

\bibitem[{{Satyapal} {et~al.}(2018){Satyapal}, {Abel}, \&
  {Secrest}}]{satyapal18}
{Satyapal}, S., {Abel}, N.~P., \& {Secrest}, N.~J. 2018, \apj, 858, 38

\bibitem[{Sch\"{o}lkopf {et~al.}(2000)Sch\"{o}lkopf, Smola, Williamson, \&
  Bartlett}]{ocsvm}
Sch\"{o}lkopf, B., Smola, A.~J., Williamson, R.~C., \& Bartlett, P.~L. 2000,
  Neural Comput., 12, 1207

\bibitem[{{Smartt} {et~al.}(2015){Smartt}, {Valenti}, {Fraser}, {Inserra},
  {Young}, {Sullivan}, {Pastorello}, {Benetti}, {Gal-Yam}, {Knapic},
  {Molinaro}, {Smareglia}, {Smith}, {Taubenberger}, {Yaron}, {Anderson},
  {Ashall}, {Balland}, {Baltay}, {Barbarino}, {Bauer}, {Baumont}, {Bersier},
  {Blagorodnova}, {Bongard}, {Botticella}, {Bufano}, {Bulla}, {Cappellaro},
  {Campbell}, {Cellier-Holzem}, {Chen}, {Childress}, {Clocchiatti},
  {Contreras}, {Dall'Ora}, {Danziger}, {de Jaeger}, {De Cia}, {Della Valle},
  {Dennefeld}, {Elias-Rosa}, {Elman}, {Feindt}, {Fleury}, {Gall},
  {Gonzalez-Gaitan}, {Galbany}, {Morales Garoffolo}, {Greggio}, {Guillou},
  {Hachinger}, {Hadjiyska}, {Hage}, {Hillebrandt}, {Hodgkin}, {Hsiao}, {James},
  {Jerkstrand}, {Kangas}, {Kankare}, {Kotak}, {Kromer}, {Kuncarayakti},
  {Leloudas}, {Lundqvist}, {Lyman}, {Hook}, {Maguire}, {Manulis}, {Margheim},
  {Mattila}, {Maund}, {Mazzali}, {McCrum}, {McKinnon}, {Moreno-Raya},
  {Nicholl}, {Nugent}, {Pain}, {Pignata}, {Phillips}, {Polshaw}, {Pumo},
  {Rabinowitz}, {Reilly}, {Romero-Ca{\~n}izales}, {Scalzo}, {Schmidt},
  {Schulze}, {Sim}, {Sollerman}, {Taddia}, {Tartaglia}, {Terreran},
  {Tomasella}, {Turatto}, {Walker}, {Walton}, {Wyrzykowski}, {Yuan}, \&
  {Zampieri}}]{epessto}
{Smartt}, S.~J., {Valenti}, S., {Fraser}, M., {et~al.} 2015, \aap, 579, A40

\bibitem[{{Smith} {et~al.}(2014){Smith}, {Koss}, \& {Mushotzky}}]{smith14}
{Smith}, K.~L., {Koss}, M., \& {Mushotzky}, R.~F. 2014, \apj, 794, 112

\bibitem[{{Solarz} {et~al.}(2017){Solarz}, {Bilicki}, {Gromadzki}, {Pollo},
  {Durkalec}, \& {Wypych}}]{solarz17}
{Solarz}, A., {Bilicki}, M., {Gromadzki}, M., {et~al.} 2017, \aap, 606, A39

\bibitem[{{Stern} {et~al.}(2012){Stern}, {Assef}, {Benford}, {Blain}, {Cutri},
  {Dey}, {Eisenhardt}, {Griffith}, {Jarrett}, {Lake}, {Masci}, {Petty},
  {Stanford}, {Tsai}, {Wright}, {Yan}, {Harrison}, \& {Madsen}}]{stern12}
{Stern}, D., {Assef}, R.~J., {Benford}, D.~J., {et~al.} 2012, \apj, 753, 30

\bibitem[{{Taylor}(2005)}]{topcat}
{Taylor}, M.~B. 2005, in Astronomical Society of the Pacific Conference Series,
  Vol. 347, Astronomical Data Analysis Software and Systems XIV, ed.
  P.~{Shopbell}, M.~{Britton}, \& R.~{Ebert}, 29

\bibitem[{{Thomas}(2019)}]{photon}
{Thomas}, R. 2019, {Astrophysics Source Code Library, [record ascl:1901.007]}

\bibitem[{{Trump} {et~al.}(2006){Trump}, {Hall}, {Reichard}, {Richards},
  {Schneider}, {Vand en Berk}, {Knapp}, {Anderson}, {Fan}, {Brinkman},
  {Kleinman}, \& {Nitta}}]{trump06}
{Trump}, J.~R., {Hall}, P.~B., {Reichard}, T.~A., {et~al.} 2006, \apjs, 165, 1

\bibitem[{{van der Marel} {et~al.}(2002){van der Marel}, {Alves}, {Hardy}, \&
  {Suntzeff}}]{lmc}
{van der Marel}, R.~P., {Alves}, D.~R., {Hardy}, E., \& {Suntzeff}, N.~B. 2002,
  \aj, 124, 2639

\bibitem[{{Winkler} {et~al.}(2003){Winkler}, {Courvoisier}, {Di Cocco},
  {Gehrels}, {Gim{\'e}nez}, {Grebenev}, {Hermsen}, {Mas-Hesse}, {Lebrun},
  {Lund}, {Palumbo}, {Paul}, {Roques}, {Schnopper}, {Sch{\"o}nfelder},
  {Sunyaev}, {Teegarden}, {Ubertini}, {Vedrenne}, \& {Dean}}]{integral}
{Winkler}, C., {Courvoisier}, T.~J.~L., {Di Cocco}, G., {et~al.} 2003, \aap,
  411, L1

\bibitem[{{Wolf} {et~al.}(2018){Wolf}, {Onken}, {Luvaul}, {Schmidt}, {Bessell},
  {Chang}, {Da Costa}, {Mackey}, {Martin-Jones}, {Murphy}, {Preston}, {Scalzo},
  {Shao}, {Smillie}, {Tisserand}, {White}, \& {Yuan}}]{skymapper}
{Wolf}, C., {Onken}, C.~A., {Luvaul}, L.~C., {et~al.} 2018, \pasa, 35, e010

\bibitem[{{Wright} {et~al.}(2010){Wright}, {Eisenhardt}, {Mainzer}, {Ressler},
  {Cutri}, {Jarrett}, {Kirkpatrick}, {Padgett}, {McMillan}, {Skrutskie},
  {Stanford}, {Cohen}, {Walker}, {Mather}, {Leisawitz}, {Gautier}, {McLean},
  {Benford}, {Lonsdale}, {Blain}, {Mendez}, {Irace}, {Duval}, {Liu}, {Royer},
  {Heinrichsen}, {Howard}, {Shannon}, {Kendall}, {Walsh}, {Larsen}, {Cardon},
  {Schick}, {Schwalm}, {Abid}, {Fabinsky}, {Naes}, \& {Tsai}}]{wright10}
{Wright}, E.~L., {Eisenhardt}, P.~R.~M., {Mainzer}, A.~K., {et~al.} 2010, \aj,
  140, 1868

\bibitem[{{York} {et~al.}(2000){York}, {Adelman}, {Anderson}, {Anderson},
  {Annis}, {Bahcall}, {Bakken}, {Barkhouser}, {Bastian}, {Berman}, {Boroski},
  {Bracker}, {Briegel}, {Briggs}, {Brinkmann}, {Brunner}, {Burles}, {Carey},
  {Carr}, {Castander}, {Chen}, {Colestock}, {Connolly}, {Crocker}, {Csabai},
  {Czarapata}, {Davis}, {Doi}, {Dombeck}, {Eisenstein}, {Ellman}, {Elms},
  {Evans}, {Fan}, {Federwitz}, {Fiscelli}, {Friedman}, {Frieman}, {Fukugita},
  {Gillespie}, {Gunn}, {Gurbani}, {de Haas}, {Haldeman}, {Harris}, {Hayes},
  {Heckman}, {Hennessy}, {Hindsley}, {Holm}, {Holmgren}, {Huang}, {Hull},
  {Husby}, {Ichikawa}, {Ichikawa}, {Ivezi{\'c}}, {Kent}, {Kim}, {Kinney},
  {Klaene}, {Kleinman}, {Kleinman}, {Knapp}, {Korienek}, {Kron}, {Kunszt},
  {Lamb}, {Lee}, {Leger}, {Limmongkol}, {Lindenmeyer}, {Long}, {Loomis},
  {Loveday}, {Lucinio}, {Lupton}, {MacKinnon}, {Mannery}, {Mantsch}, {Margon},
  {McGehee}, {McKay}, {Meiksin}, {Merelli}, {Monet}, {Munn}, {Narayanan},
  {Nash}, {Neilsen}, {Neswold}, {Newberg}, {Nichol}, {Nicinski}, {Nonino},
  {Okada}, {Okamura}, {Ostriker}, {Owen}, {Pauls}, {Peoples}, {Peterson},
  {Petravick}, {Pier}, {Pope}, {Pordes}, {Prosapio}, {Rechenmacher}, {Quinn},
  {Richards}, {Richmond}, {Rivetta}, {Rockosi}, {Ruthmansdorfer}, {Sandford},
  {Schlegel}, {Schneider}, {Sekiguchi}, {Sergey}, {Shimasaku}, {Siegmund},
  {Smee}, {Smith}, {Snedden}, {Stone}, {Stoughton}, {Strauss}, {Stubbs},
  {SubbaRao}, {Szalay}, {Szapudi}, {Szokoly}, {Thakar}, {Tremonti}, {Tucker},
  {Uomoto}, {Vanden Berk}, {Vogeley}, {Waddell}, {Wang}, {Watanabe},
  {Weinberg}, {Yanny}, {Yasuda}, \& {SDSS Collaboration}}]{york00}
{York}, D.~G., {Adelman}, J., {Anderson}, Jr., J.~E., {et~al.} 2000, \aj, 120,
  1579

\end{thebibliography}
\end{document}